\documentclass{article}

\usepackage{microtype}
\usepackage{graphicx}
\usepackage{booktabs} 
\usepackage[table]{xcolor}
\usepackage{colortbl}
\usepackage{array}
\usepackage[many]{tcolorbox}

\usepackage[backref=page]{hyperref}

\usepackage[accepted]{icml2026}

\usepackage{amsmath}
\usepackage{amssymb}
\usepackage{mathtools}
\usepackage{amsthm}

\usepackage[capitalize,noabbrev]{cleveref}

\theoremstyle{plain}

\theoremstyle{definition}

\theoremstyle{remark}

\usepackage[textsize=tiny]{todonotes}
\usepackage{xspace}
\usepackage[dvipsnames, table]{xcolor}
\usepackage{bbm}
\usepackage{bbold}
\usepackage{enumitem}
\usepackage{tabularx}
\usepackage{multirow}
\usepackage{multicol}
\usepackage{caption}
\usepackage{subcaption}
\usepackage{stackengine,graphicx}
\usepackage{stfloats}

\icmltitlerunning{\projectname: Runtime-Efficient Multi-Agent Embodied Planning}

\newcommand\drsh{\mathop{\ensurestackMath{%
  \stackengine{-1.2pt}{\rightarrow}{\scalebox{1}[.35]{%
    $\mkern-1.3mu\vert$}}{O}{l}{F}{F}{S}}}}

\newcommand{\projectname}[0]{\textsc{Mosaic}\xspace}

\newcommand{\positive}[1]{\textcolor{Green}{{(#1)}}}
\newcommand{\negative}[1]{\textcolor{red}{{(#1)}}}
\definecolor{successgreen}{RGB}{230,245,230}
\definecolor{failred}{RGB}{245,230,230}
\newcommand{\SuccessAction}[1]{%
  {\setlength{\fboxsep}{0.5pt}%
   \setlength{\fboxrule}{0.8pt}%
   \fbox{\colorbox{successgreen}{\strut\,\textsf{#1}\,}}}%
}
\newcommand{\FailureAction}[1]{%
  {\setlength{\fboxsep}{0.5pt}%
   \setlength{\fboxrule}{0.8pt}%
   \fbox{\colorbox{failred}{\strut\,\textsf{#1}\,}}}%
}
\tcbuselibrary{breakable}
\newtcolorbox{boxJ}{
    breakable,
    sharpish corners, 
    title = Prompt Enhancements for Constraint-Aware and Runtime-Efficient Planning,
    colback = sub, 
    colframe = main, 
    boxrule = 0pt, 
    toprule = 4.5pt, 
    enhanced,
    fuzzy shadow = {0pt}{-2pt}{-0.5pt}{0.5pt}{black!35},
}
\newtcolorbox{boxK}{
    breakable,
    sharpish corners, 
    title = Prompt related to the use of Agent-centric Spatial Memory,
    colback = sub, 
    colframe = main, 
    boxrule = 0pt, 
    toprule = 4.5pt, 
    enhanced,
    fuzzy shadow = {0pt}{-2pt}{-0.5pt}{0.5pt}{black!35},
}
\definecolor{main}{HTML}{5989cf}    
\definecolor{sub}{HTML}{f3f2ee}     

\begin{document}

\twocolumn[
\icmltitle{\projectname: Runtime-Efficient Multi-Agent Embodied Planning}



\icmlsetsymbol{equal}{*}

\begin{icmlauthorlist}
\icmlauthor{Kunjal Panchal}{equal,umass}
\icmlauthor{Saayan Mitra}{adobe}
\icmlauthor{Sunav Choudhary}{adobe}
\icmlauthor{Victor Bursztyn}{adobe}
\icmlauthor{Somdeb Sarkhel}{adobe}
\icmlauthor{Hui Guan}{umass}
\end{icmlauthorlist}

\icmlaffiliation{umass}{College of Information and Computer Sciences, University of Massachusetts, Amherst}
\icmlaffiliation{adobe}{Adobe, San Jose}

\icmlcorrespondingauthor{Kunjal Panchal}{kpanchal@umass.edu}

\icmlkeywords{Machine Learning, ICML}

\vskip 0.3in
]



\printAffiliationsAndNotice{$^*$Partial work completed during an internship at Adobe.}  

\begin{abstract}

LLM-based multi-agent embodied planning remains impractical due to prohibitively high execution latency. 
We identify failed actions as the dominant bottleneck, stemming from two core challenges: 
inaccurate state tracking under partial observability and inefficient coordination that produces redundant or conflicting actions. 
We introduce \projectname, a runtime-efficient multi-agent planning framework that addresses both challenges. 
\projectname maintains accurate yet lightweight state tracking through agent-centric semantic memory that stores objects in relative coordinates, enabling geometric transformations and coordination. 
It ensures efficient coordination through Integer Linear Programming that allocates actions at every planning step, enforcing physical feasibility and inter-agent coordination constraints. 
Across AI2-THOR and search-and-rescue benchmarks, \projectname achieves 27–32\% faster execution, 30–33\% fewer LLM calls, 25–31\% fewer steps, and 4–10\% points higher success rates. 
These results demonstrate that efficient memory and constraint-guided coordination are critical for scalable, low-latency multi-agent planning.
\end{abstract}

\section{Introduction}
\label{sec:introduction}

Many real-world embodied tasks such as collaborative search and rescue, household rearrangement, and environmental exploration require multiple agents (physical or simulated entities executing coordinated plans or policies) operating simultaneously in shared spaces~\cite{liu2024hetero, qian2024scalingllm, skrynnik2024decentralized, chen-etal-2024-llmarena}.
Leveraging multiple agents offers clear advantages: 
they can parallelize subtasks, cover larger areas, and recover from local failures~\cite{qian2024scalingllm, nayak2024llamar}, leading to faster and more robust task completion compared to single-agent systems. 
Recent works have explored the use of Large Language Models (LLMs) as planners in such multi-agent environments, demonstrating impressive inference-time generalization across unstructured tasks and domains~\cite{zhang2025lamma, bai2025twostepmultiagenttaskplanning}. 
In this paradigm, LLMs generate action sequences for each agent, while the agents execute these planned actions in the environment.

However, practical deployment of LLM-driven multi-agent planning remains limited due to prohibitively high latency, which include both physical action execution time and LLM inference overhead.
For example, running a state-of-the-art multi-agent system~\cite{nayak2024llamar} to solve a simple embodied task such as ``Turn off the faucet and light'' 
requires approximately 7.2 minutes to complete. 
More complex tasks such as ``Rescue two people from a fire-affected area'' in a simplified simulation environment require 10.5 minutes to complete.

\begin{figure}[t]
    \centering
    \includegraphics[width=0.95\linewidth]{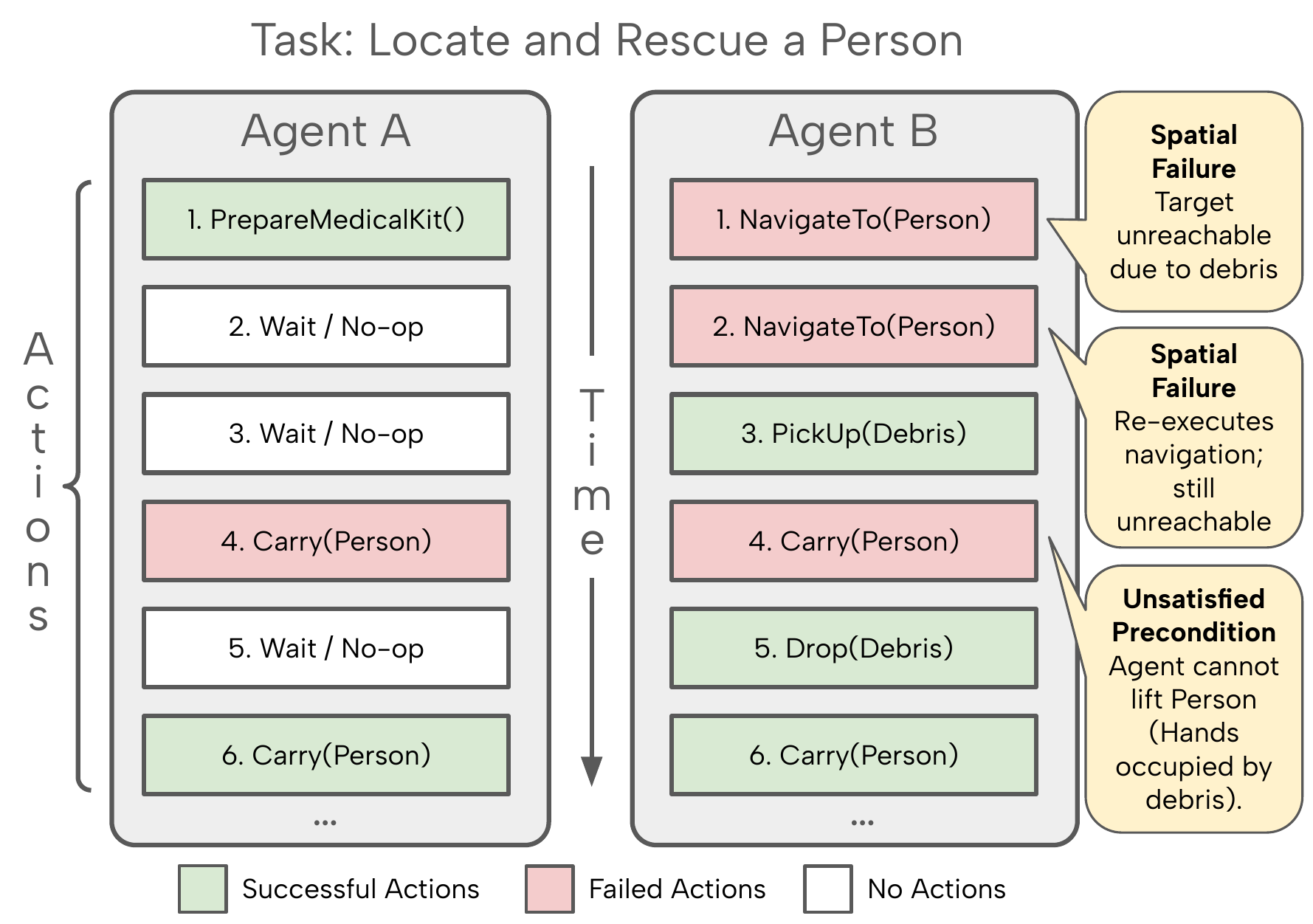}
    \caption{Multi-agent task execution showing failures driven by spatial and coordination inefficiencies.
    These inefficiencies also induce agent underutilization during stalled periods.
    }
    \label{fig:failed-actions}
    \vspace{-0.3cm}
\end{figure}

We find that a substantial portion of the runtime is spent on failed actions during execution, which trigger replanning, recovery, or redundant exploration. Figure~\ref{fig:failed-actions} illustrates a sequence of such failed actions in a rescue scenario: Agent B repeatedly fails to navigate to a person due to spatial reasoning errors (steps 1--2), then violates action preconditions by attempting to carry the person while holding debris (step 4), while Agent A remains idle waiting for coordination (steps 2--3, 5). 
In our preliminary experiments on state-of-the-art LLM-based multi-agent systems for embodied planning~\cite{nayak2024llamar, zhang2023coela}, agents spent up to 16--51\% of planning steps on failed actions, frequently attempting unreachable goals due to incorrect spatial estimates or poor coordination. Failed actions increase runtime latency in two critical ways: 
(a)~they consume execution time without contributing to task completion, as agents must retry or replan after each failure; and 
(b)~they cause cascading delays, forcing other agents to stall or execute no-ops while waiting for failed actions to resolve, leading to severe underutilization (as shown in Agent A's idle periods in Figure~\ref{fig:failed-actions}). 
As a result, current systems remain impractical for resource-constrained or time-critical applications, requiring up to 9 minutes per episode for tasks such as search and rescue, and 15.6 minutes for fast-paced household activities.

These failures stem from two open challenges in LLM-based multi-agent planning. 
First, \textit{state tracking under partial observability}: LLMs must maintain accurate beliefs about object locations as agents move and fields of view change, yet without effective tracking mechanisms, agents navigate to outdated locations or re-explore visited areas. Prior work on state tracking faces critical tradeoffs: temporal memories~\cite{nayak2024llamar, yao2022react, lin2025stopwastingtokensefficient} are lightweight but force error-prone spatial inference from text, while spatial and spatio-temporal memories~\cite{fang2019scenememory, hu2025dllmmem, mao2025metamemory, he2025mem4nav} encode geometry explicitly at high computational cost. 

Second, \textit{multi-agent coordination}: inadequate task allocation causes redundant behaviors, motion interference, and unmet action preconditions. 
Existing coordination approaches~\cite{choe2025askreasonassist, brawer2023towards, obata2024lip, wu2024hierarchical} perform coarse-grained target-agent allocation 
or checking for task conflicts through target object dependencies~\cite{choe2025askreasonassist, brawer2023towards, obata2024lip, wu2024hierarchical}. 
These assignments assume fixed responsibilities and do not adapt to changing environments or resolve fine-grained, action-level decisions where action failures occur. 

In this work, we introduce \projectname, a runtime-efficient multi-agent planning framework that addresses both challenges through two novel designs. 
\textbf{(a)~Agent-centric Semantic Memory} addresses the state tracking challenge by storing task-relevant objects as structured entries in relative coordinates from each agent's perspective. 
Our approach enables geometric transformations through explicit relative positioning while remaining lightweight and scalable: When objects leave an agent's field of view, their locations are inferred from motion history; cross-agent reasoning is enabled by transforming between agent coordinate frames. 
\textbf{(b)~Action Selection via Constraint Optimization} addresses the coordination challenge by selecting one action per agent from LLM-generated candidates at each timestep via an Integer Linear Programming (ILP) framework. 
While LLMs can propose contextually plausible actions, they struggle to reason reliably about fine-grained interactions among multiple agents; by formalizing these interactions as symbolic constraints, the ILP efficiently filters infeasible combinations and produces globally consistent action assignments.
The constraints capture collision avoidance, blocked-path prevention, and workload balancing, while a cost function penalizes actions with high failure likelihood and workload imbalance.
Unlike prior approaches that perform coarse-grained target object allocation over entire episodes, our ILP-based allocation operates at the action level at every planning step, continuously adapting to environmental changes captured by the agent-centric semantic memory.

Empirically, across the AI2-THOR~\cite{ai2thor} household environment and the SAR~\cite{nayak2024llamar} search-and-rescue setting, \projectname outperforms the strongest baseline in terms of success rate, delivering 27--32\% faster execution and 30--33\% fewer LLM calls.
Under this comparison, agents with \projectname execute 25–31\% fewer physical steps, driven by a 7–12\% point reduction in failed actions.
This failure reduction also yields a 4–10\% points improvement in task success for \projectname under a fixed planning-step budget.
Even relative to its single-agent variant, \projectname demonstrates substantial gains: 
by leveraging coordinated action assignments across agents, it reduces redundant exploration and failed attempts, resulting in 1.25--1.33$\times$ shorter episode duration.
The advantages persist at scale: with five agents, we still observe 1.25–1.30$\times$ end-to-end speedups.
Under per-agent sensor noise reflecting real-world conditions, \projectname maintains a $\sim$1.25$\times$ multi-agent speedup, with task performance remaining within 70–90\% of the noiseless setting.
Taken together, these findings indicate that \projectname delivers runtime efficiency and coordination benefits beyond single-agent deployment, without compromising planning quality.

We summarize the contributions as follows: 
\begin{enumerate}[nosep, leftmargin=0.7cm]
    \item \textbf{Agent-centric Semantic Memory}: 
    A lightweight memory that tracks task-relevant objects and destinations across planning steps, reducing failed actions by up to 12\%.
    
    \item \textbf{Action selection via constraint optimization}: 
    A constraint optimization ILP framework that converts LLM action proposals into globally consistent multi-agent actions at every planning step. 
    A cost function that reduces unnecessary navigation and high-risk actions, achieving up to 32\% faster execution and 33\% fewer LLM calls.
    
    \item \textbf{Scalable low-latency execution}: 
    End-to-end speedups of up to 1.30$\times$ with five agents, while improving task success by up to 10\%.
\end{enumerate}

\section{Related Work}
\label{sec:related-work}

We review prior work on memory representation for embodied planning and agent coordination. 
Further discussion is deferred to Appendix~\ref{adx:extended-related-work}. 

\textbf{Memory Representations for Embodied Agents.} 
Work on memory for embodied agents can be categorized into (a)~temporal, (b)~spatial, and (c)~spatio-temporal memory. 
\textit{Temporal memory} retains past observations without explicitly storing geometry, including short-term memory over limited context windows~\cite{ yao2022react, lin2025stopwastingtokensefficient}, long-term episodic memory across extended horizons~\cite{nayak2024llamar, fang2019scenememory, gupta2025memo, lei2025robomemory, zhang2025ella, yadla2025temporal}, and hierarchical memory organizing information at multiple abstraction levels~\cite{wang2025karma, hu2024hiagenthierarchicalworkingmemory, lei2025stma, zhang2025gmemorytracinghierarchicalmemory, han2025llmmultiagentsystemschallenges}. 
While effective for retaining history, temporal memories often rely on unstructured textual logs~\cite{pan2025whymultiagentfail, zhang2025agentcausestaskfailures, li2024survey}, resulting in error-prone spatial inference from text. 
\textit{Spatial memory} explicitly stores environmental layout independent of time, including topology-based navigation~\cite{gupta2017cognitive, chaplot2020learning} and dense 3D representations such as \textsc{VLMaps}~\cite{huang23vlmaps}, \textsc{ConceptGraphs}~\cite{gu2024conceptgraphs}, \textsc{Mem4Nav}~\cite{he2025mem4nav}, TME~\cite{ye2025taskmemoryengine}, and \textsc{Mem2Ego}~\cite{zhang2025memego}, which incur significant overhead for high-fidelity geometry. 
\textit{Spatio-temporal memory} combines both dimensions, maintaining spatial representations that evolve over time~\cite{fang2019scenememory, yang20253dmem, lei2025stma, mao2025metamemory, anwar2025remembr, hu2025dllmmem}, typically requiring accumulation over long horizons with high computational cost. 
In contrast, our agent-centric semantic memory maintains a structured textual representation of task-relevant objects in relative coordinates, updated at every planning step to track distances and goals even when objects fall outside the field of view, enabling efficient spatial grounding without dense reconstruction overhead.

\textbf{Multi-Agent Coordination.} Coordination methods differ in how task/action allocation is produced and failures are handled. \textit{Neural methods} propose candidate actions from learned experience but are largely reactive in failure handling, correcting errors only after execution~\cite{yao2022react, kannan2024smart, zhang2023coela, wang2023voyager, ahn2022saycan, ji2024testing, huang2022inner}. 
\textit{Symbolic approaches} leverage preconditions and logical rules to proactively reason about feasibility~\cite{zheng2024evaluating, cornelio2024recover}, commonly through Task and Motion Planning (TAMP) frameworks~\cite{dantam2018incremental, kaelbling2011hierarchical, lozano2014constraint, garrett2021integrated, faroni2024tamp}, but lack the flexible reasoning capabilities of LLMs. 
\textit{Neuro-symbolic coordination} uses LLMs to propose actions and symbolic optimization to enforce action feasibility.
\citet{choe2025askreasonassist} uses symbolic constraints to manage failures reactively via help request actions, whereas \projectname proactively coordinates all possible actions.
\citet{brawer2023towards} proposes LLMs to convert user-defined constraints into ILP feasibility checks, but does not address runtime optimization. 
\textsc{LiP-LLM}~\cite{obata2024lip} and \citet{wu2024hierarchical} assign tasks to agents for the entire episode, without reasoning about action-level conflicts at each step.
In contrast, \projectname performs joint action-level assignments at every timestep, avoiding failures through fine-grained, LLM-guided feasibility reasoning.

\section{Methodology}
\label{sec:methodology}

We first formulate the problem setting and introduce a general multi-agent planning architecture to contextualize \projectname, and then elaborate on \projectname's two key designs.

\subsection{Preliminaries}
\label{subsec:preliminaries}

\textbf{Problem Setting: Multi-Agent POMDP.}
The multi-agent embodied planning problem is typically formulated as a Partially Observable Markov Decision Process (POMDP) without access to explicit transition dynamics or reward functions. 
The environment supports a joint action space
$\mathcal{X} = \mathcal{X}_1 \times \dots \times \mathcal{X}_N$,
where $\mathcal{X}_i$ denotes the discrete action space of agent $a_i$.
The goal in this work is to satisfy a high-level natural language instruction $\mathsf{I}$ by producing a sequence of joint actions
$\mathbf{x}_t = \{x_{1, t}, \dots, x_{N, t}\}$, $x_{i, t} \in \mathcal{X}_i$, whose execution leads to fulfilling the semantic requirements of $\mathsf{I}$, while minimizing runtime overhead. 

Partial observability implies that agents operate without complete state information and act based on local observations accumulated over time, including past actions, execution outcomes, and internal memory. 
In this setting, effective action planning requires 
(a)~maintaining and updating internal state representations (analogous to a belief state) rather than relying solely on instantaneous observations, as agents must reason about objects and locations beyond their current field of view, and 
(b)~selecting joint actions via a constrained policy (conditioned on the state and the task), avoiding conflicts under physical constraints, satisfying action preconditions, and preventing redundant exploration (implicitly accounting for transition uncertainty).

Our framework captures the functional roles of a POMDP solver: 
it maintains an internal state representation that aggregates partial observations (analogous to a belief state) and selects joint actions via a constrained policy conditioned on this state and the task. 
Enforcing feasibility under physical and resource constraints implicitly accounts for transition uncertainty, approximating the role of a transition model.

\textbf{Multi-Agent Planning Architecture.}
\projectname is designed to be compatible with general LLM-based planning frameworks. 
Here, we describe its instantiation within a state-of-the-art centralized planning architecture. 
This architecture uses distinct LLM calls for: 
(a) decomposing high-level task instructions into subtasks, 
(b) proposing actions for each subtask, and 
(c) verifying subtask completion. 
This role separation, adopted from prior LLM-based embodied planning systems~\cite{nayak2024llamar, zhang2023coela, kannan2024smart}, decouples high-level reasoning from low-level action selection to improve execution reliability. 
Appendix~\ref{adx:multi-agent-planning-architecture} provides a procedural description, while here we summarize the core conceptual steps.

{(a)~High-level Task Decomposition:}
Given an instruction $\mathsf{I}$ and shared memory $\mathsf{M}$, the \textsc{LLM-Planner} decomposes the task into a set of open subtasks.
The framework maintains an interaction history $\mathsf{H}$ and a set of completed subtasks $\mathsf{S}_c$, which together provide persistent context.
Execution proceeds iteratively until all subtasks are completed.\\
{(b)~Action Proposals:} 
At each planning step, a centralized \textsc{LLM-Actor} proposes joint actions for all agents based on the currently open subtasks.
Actions are executed in parallel, producing observations that update $\mathsf{H}$ and $\mathsf{M}$.\\
{(c)~Verification of Subtask Completion:}
After execution, \textsc{LLM-Verifier} determines which subtasks have been completed.
This step accounts for execution errors and newly revealed dependencies.
Then \textsc{LLM-Planner} is re-invoked to revise remaining subtasks using the updated context.

\begin{figure*}[t]
    \centering
    \includegraphics[width=\linewidth]{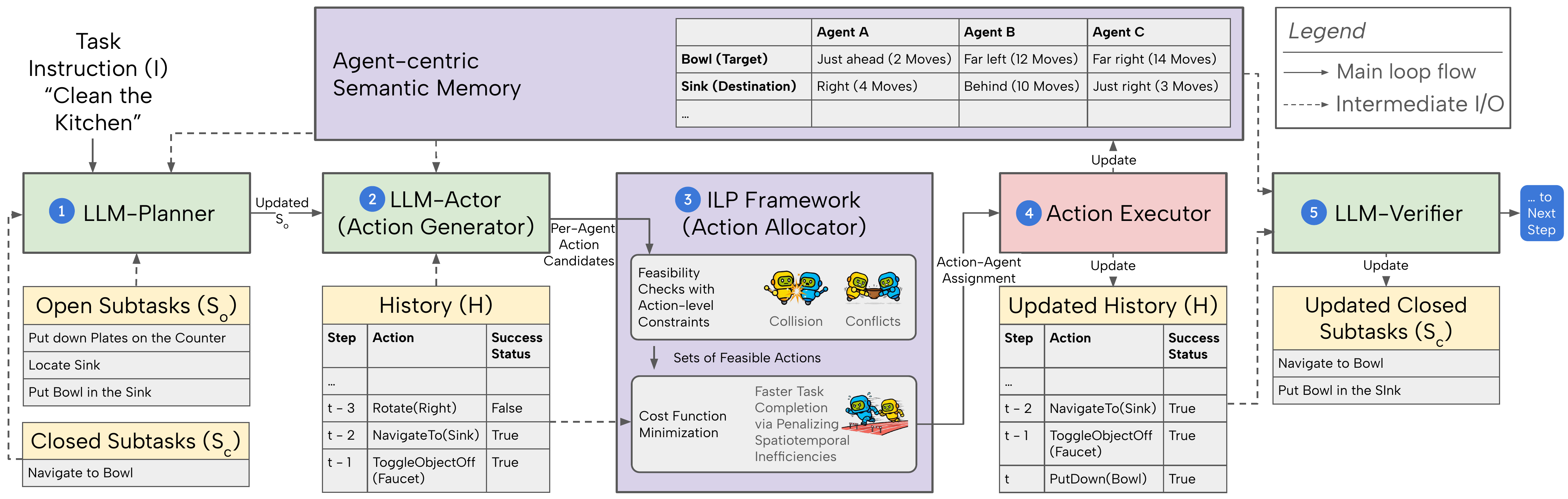}
    \caption{Overview of \projectname. 
    \textbf{Agent-centric semantic memory} (ASM) maintains task-relevant spatial context across timesteps under partial observability, improving action generation.
    The \textsc{LLM-Actor} proposes action candidates based on ASM, which an \textbf{integer linear programming (ILP) framework} resolves into globally consistent joint actions through \textbf{feasibility constraints}.
    A \textbf{cost function} guides the ILP to balance agent workloads and reduce navigation and temporal overhead.
    Together, these components reduce execution failures and redundant actions, enabling low-latency, fine-grained multi-agent coordination. 
    }
    \label{fig:mosaic-overview}
    \vspace{-0.3cm}
\end{figure*}

The above LLM-based multi-agent planning framework exhibits two key limitations that lead to long time to task completion.
First, inefficient state tracking under partial observability degrades the quality of action generation.
Second, insufficient coordination of generated actions across agents results in frequent execution failures.
We address these issues with \projectname, a runtime-efficient planning system composed of two novel designs (Figure~\ref{fig:mosaic-overview}).
First, \projectname maintains an \textbf{agent-centric semantic memory} (ASM) that tracks task-relevant objects and destinations for each agent (\S~\ref{subsec:egocentric-memory}).
Second, it introduces action-level coordination through an \textbf{integer linear programming (ILP)} framework.
This framework resolves \textsc{LLM-Actor} action proposals into globally consistent, feasible, and efficient assignments via action constraints and a cost function (\S~\ref{subsec:action-level-constraints}). 

\subsection{Agent-centric Semantic Memory}
\label{subsec:egocentric-memory}

Agent-centric Semantic Memory (ASM) maintains task-relevant objects and destinations using \textit{agent-centric relative distances}, updated at each planning step based on observations and agent motion. The design is guided by two principles: 
(a)~Lightweight spatial reasoning:
tracking actionable spatial relations in relative coordinates is more computationally efficient than constructing global maps or dense 3D representations. 
(b)~Cross-agent transferability: The relative coordinate representation can be translated across agents' perspectives through simple geometric transformations. 
Once an agent moves or interacts with an object, others can update their relative distances from that object accordingly.
This transfer avoids repeated exploration attempts, motion collisions, and interference.

For each agent ($a \in \mathsf{A}$) and task-relevant object or destination ($o \in \mathcal{O}$), we maintain agent-centric distance and orientation:
$r_{a,o} = \frac{|x_o - x_a| + |z_o - z_a|}{u}$, and
$\phi_{a,o} = \mathrm{atan2}(x_o - x_a, z_o - z_a) - \theta_a$,
where $(x_a, z_a)$ and $\theta_a$ denote the agent’s location and orientation, $(x_o, z_o)$ the object or destination’s global position, and $u$ a unit of motion corresponding to one discrete step.
We note that even for 3D simulators, locomotion is effectively planar in $(x,z)$; following prior work, spatial failures arise only from horizontal movement, as vertical actions do not fail and crouch/stand is not required.
These coordinates are continuously updated as the agent moves. 
I.e., Forward motion decreases the relative distance ($r_{a,o}^{\text{new}} = r_{a,o}^{\text{old}} - d$), and changes in heading adjust the relative orientation ($\phi_{a,o}^{\text{new}} = (\phi_{a,o}^{\text{old}} - \Delta \theta_a) \bmod 360^\circ$).
Lateral movements are updated similarly.
Additionally, re-observations implicitly relocalize objects by updating their coordinates, correcting accumulated sensor noise.

To encode spatial information semantically, distances $r_{a,o}$ are converted into semantic tiers: 
an object 1–2 steps ahead is labeled `just ahead’, while $>$10 steps away is labeled `far’. 
Orientations $\phi_{a,o}$ are similarly mapped to semantic directions, 
e.g., `left’, `right’, or `ahead’.
These underlying continuous metrics are only discretized within the \textsc{LLM-Actor} prompt to facilitate reasoning, preventing the accumulation of discretization errors over long horizons.
This discretization is applied only at the interface level, while the underlying memory retains full continuous geometry, ensuring numerical precision while providing a stable and compact interface for LLM reasoning.
Figure~\ref{fig:mosaic-overview} illustrates these semantic tiers in the Agent-centric Semantic Memory: 
Agent A sees the Bowl `just ahead (2 moves)’ while Agent B perceives it `far left (12 moves)’.

The memory also stores relative distances between agents, enabling cross-agent transferability of information.
Let  $(r_{a,b}, \phi_{a,b})$ be the relative distance and orientation between agents $a$ and $b$, and $(r_{a,o}, \phi_{a,o})$ between agent $a$ and object $o$.  
We define the relative displacement in global coordinates:
\begin{align}
\Delta x &= r_{a,o} \sin(\phi_{a,o} + \theta_a) - r_{a,b} \sin(\phi_{a,b} + \theta_a), \nonumber \\
\Delta z &= r_{a,o} \cos(\phi_{a,o} + \theta_a) - r_{a,b} \cos(\phi_{a,b} + \theta_a). \nonumber
\end{align}
Then, agent $b$'s location and orientation with respect to the object $o$ would be:
\begin{align}
r_{b,o} = \frac{|\Delta x| + |\Delta z|}{u}, \text{and }
\phi_{b,o} = \mathrm{atan2}(\Delta x, \Delta z) - \theta_b. \nonumber
\end{align}

\subsection{Action Selection via Constraint Optimization}
\label{subsec:action-level-constraints}

Next, \projectname introduces an Integer Linear Programming (ILP) framework that selects one action per agent from \textsc{LLM-Actor}-generated candidates. 
It addresses a limitation of prior works: 
while LLMs excel at proposing contextually relevant actions, they cannot guarantee constraint satisfaction across multiple agents. 
Conversely, constraint optimization methods can enforce feasibility guarantees but lack the semantic understanding to generate meaningful action proposals. 
\projectname combines both strengths, leveraging LLMs for candidate action generation and ILP for constraint enforcement.

\textbf{Action-Level Feasibility Constraints.} At each planning step, the \textsc{LLM-Actor} generates a small set of plausible action candidates for each agent $a_i \in \mathsf{A}$:
$\mathsf{X}_{i,t} = \{x_{i,t}^{(1)}, x_{i,t}^{(2)}, \dots\}$.
The \textsc{LLM-Actor} generates action candidates \emph{per agent}, respecting each agent’s local context to avoid irrelevant proposals.
It produces \emph{multiple candidates} per agent, allowing the ILP framework to resolve conflicts and trade-offs and select a feasible and conflict-free joint action.
The ILP framework selects exactly one action per agent at each planning step through binary  variables
\[
z_{i,t}^{(k)} =
\begin{cases}
1 & \text{if agent } a_i \text{ executes candidate } x_{i,t}^{(k)} \text{ at step } t, \\
0 & \text{otherwise}.
\end{cases}
\]

\begin{table}[t]
\centering
\renewcommand{\arraystretch}{1.2}
\setlength{\tabcolsep}{3.3pt}
\footnotesize
\caption{Action-level constraints enforced by the ILP framework.}
\label{tab:ilp-constraints}
\vspace{-0.2cm}
\begin{tabular}{p{1.7cm}>{\raggedright\arraybackslash}p{2.8cm}p{3cm}} 
\toprule
\textbf{Constraint} & \textbf{Description} & \textbf{Formula / Key Info} \\
\midrule
Eligibility & Prevent physically impossible actions & $z_{i,t}^{(k)} = 0$ if $x_{i,t}^{(k)}$ is infeasible (invalid preconditions, blocked navigation.) \\
\rowcolor{gray!10} Per-Agent Action Count & Each agent executes exactly one action per planning step & $\sum_k z_{i,t}^{(k)} = 1 \;\; \forall i$ (use \texttt{no-op} if necessary) \\
Multi-Agent Joint-Action & Ensure subtasks requiring multiple agents are fully staffed & $\sum_i \sum_{k \in \mathsf{X}_{i,t}^{(s)}} z_{i,t}^{(k)} = k_s$ \\
\rowcolor{gray!10} {Resource \newline Exclusivity} & Prevent simultaneous use of exclusive resources & $\sum_i \sum_{k \in \mathsf{X}_{i,t}^{(r)}} z_{i,t}^{(k)} \le 1$ \\
Collision /\newline Interference & Avoid spatial or kinematic conflicts between agents & $z_{i,t}^{(k)} + z_{i',t}^{(k')} \le 1$ for incompatible candidate pairs \\
\bottomrule
\end{tabular}
\vspace{-0.5cm}
\end{table}

The executed action for agent $a_i$ at planning step $t$ is given by
$x_{i,t} = \sum_k z_{i,t}^{(k)} x_{i,t}^{(k)}$,
with constraints enforcing $\sum_k z_{i,t}^{(k)} = 1$ for each $(i,t)$.
Table~\ref{tab:ilp-constraints} summarizes the action-level feasibility constraints enforced by the ILP framework.
These constraints are described in detail in Appendix~\ref{adx:action-level-constraints}.
By enforcing feasibility before execution, ILP proactively prevents conflicts and invalid actions, rather than reacting to failures after they occur.
The hyperparameter selection of action candidate count is discussed in Appendix~\ref{subsec:hyperparameter-action-candidates}.

\textit{A Note on the ILP's Scalability.} 
\projectname mitigates ILP search-space explosion via \emph{action pruning}, limiting the \textsc{LLM-Actor} to $K$ candidates per agent; experiments (Appendix~\ref{subsec:hyperparameter-action-candidates}) show the trade-offs related to the choice of $K$. 
For larger systems, \emph{agent pruning} via hierarchical coordination or spatial partitioning can further cluster agents into local groups, keeping each ILP instance tractable.

\label{subsec:cost-function-for-ilp}

\textbf{Cost Function.} Once feasibility constraints are satisfied, multiple joint actions may still be possible, differing in execution efficiency.
Hence, the ILP framework uses a cost function to identify the suitable action per agent that minimizes the runtime cost of execution.
The cost function considers spatial efficiency, temporal coherence, and workload balance by incorporating two weighted components:
\vspace{-0.6cm}
\begin{align}
\mathcal{J}(z, \mathsf{H}) &= 
\sum_{i=1}^{N} \sum_{k=1}^{K_i} 
z_{i,t}^{(k)}\Big( 
C_{\textsf{penal}}(a_i, x_{i,t}^{(k)}, \mathsf{H}) \nonumber \\  
& \qquad \qquad +
\lambda_{\textsf{load}}\,C_{\textsf{load}}(a_i, x_{i,t}^{(k)}, \mathsf{H})
\Big). \label{eq:cost-function}
\end{align}
The scaling coefficient $\lambda_{\textsf{load}}$ balances the relative importance of load distribution versus penalties on failures, with a default value of 1.

\textit{Spatial and Temporal Penalties.} $C_{\textsf{penal}}$ captures common spatial and temporal inefficiencies, including cyclic behavior (repeated action sequences), repeated failures for a specific action, oscillatory back-and-forth spatial movements, backtracking to previously visited locations, and stagnation due to idle or no-progress steps.
Using the agent’s action-success history $\mathsf{H}$, these inefficiencies are incorporated as:
\begin{align}
C_{\textsf{penal}}(a_i, x_{i,t}^{(k)}, \mathsf{H}) & = 
\sum_{p \in \mathsf{P}} \lambda_p \cdot \textsf{Penalty}_p(a_i, x_{i,t}^{(k)}, \mathsf{H}), \label{eq:penal-cost-function}
\end{align}
Each pattern $p \in \mathsf{P}$ is scored using $\textsf{Penalty}_p(a_i, x_{i,t}^{(k)}, \mathsf{H}_t)$ and weighted by $\lambda_p$.
Detailed definitions of all penalty patterns and weights are provided in Appendix~\ref{subsec:hyperparameter-sensitivity} (Table~\ref{tab:temporal-patterns}).

\textit{Load Imbalance Penalty} $C_{\textsf{load}}$ promotes balanced workload by penalizing idle or underutilized agents: 
\begin{align}
C_{\textsf{load}}&(a_i, x_{i,t}^{(k)}, \mathsf{H}) \nonumber \\
&=  
\frac{\frac{1}{|\mathsf{A}|} \sum_{i' \in \mathsf{A}} \textsc{ActionCount}(a_{i'}, \mathsf{H})}
{\textsc{ActionCount}(a_i, \mathsf{H}) + \mathbb{1}[x_{i,t}^{(k)} \neq \texttt{no-op}]}, \nonumber
\end{align}
where $\textsc{ActionCount}(a_i, \mathsf{H})$ counts all non-idle actions executed by $a_i$ up to step $t$.
The numerator computes the average number of non-idle actions per agent, capturing overall team workload, while the denominator reflects agent $a_i$’s cumulative workload, including the current action if non-idle.
If $a_i$ selects a \texttt{no-op}, the denominator remains unchanged, yielding a higher penalty relative to active agents. 
To handle the only possible zero-denominator case (an agent with no prior actions selecting \texttt{no-op}), we cap $C_{\textsf{load}}$ at 3. 
This avoids unbounded penalties while preserving a strong incentive for agent utilization.

\section{Experiments and Results}
\label{sec:experiments-results}

In this section, we first describe the experimental setup, followed by results demonstrating \projectname's efficiency, effectiveness, and scalability.

\subsection{Experimental Setup}
\label{subsec:experimental-setup}

\textbf{Environments.}
We evaluate our framework in two environments: \textbf{AI2-THOR} (Apache 2.0) and \textbf{Search and Rescue (SAR)} (MIT) \cite{nayak2024llamar}.
AI2-THOR is a photorealistic 3D simulator for 36 multi-agent household tasks, supporting up to 5 agents operating across kitchen, bedroom, living room, and bathroom. 
Agents perceive objects within a 1.5m range and a 90$^\circ$ field of view under partial observability, with optional noise in position, rotation, and detection sensors. 

The SAR environment is a discrete, text-based grid simulator for emergency response missions. 
Agents navigate 2D grids to extinguish fires and rescue lost persons, managing limited visibility and dynamic conditions. 
Fires vary by intensity (low/medium/high) and type (chemical/non-chemical), requiring agents to collect appropriate resources (water or sand) from reservoirs. 
Rescue actions demand cooperative effort proportional to each person's load, enforcing multi-agent coordination for resource allocation, fire control, and rescue under time constraints.

Consistent with the prior work, we assume reliable localization, as handling raw sensor drift lies outside our high-level planning scope and is typically addressed at the control layer. 
Nevertheless, we include a spatial noise sensitivity analysis in Appendix~\ref{adx:performance-under-observation-noise} to assess robustness.

\textbf{Models.}
\projectname uses one LLM to handle all \textsc{LLM-Planner}, \textsc{LLM-Actor}, and \textsc{LLM-Verifier} calls; this LLM is selected from \textbf{GPT-4o}~\cite{hurst2024gpt4o}, \textbf{Claude Sonnet 4.5}~\cite{anthropic2025claudesonnet}, or \textbf{Gemini 3 Flash}~\cite{deepmind2025gemini3flash}\footnote{All trademarks and logos are the property of their respective owners and are used here for identification and descriptive purposes only. No affiliation, sponsorship, or endorsement is implied.}.
These hosted commercial models are used in accordance with their respective Terms of Service and exclusively for research and technical documentation purposes.
For AI2-THOR, \textbf{CLIP ViT-B/32}~\cite{radford2021cliplearning}, released under the MIT License and following the implementation used in \textbf{LLaMAR}, is additionally employed to compute visual-text similarity between agent camera observations and subtask descriptions, guiding agents toward task-relevant objects through optimal rotation or movement decisions.

\textbf{Baselines.}
We categorize baselines by their per-planning-step complexity:
\textbf{(a)~Single-call per step baselines}: a single LLM invocation generates actions for all agents at each planning step.
\textbf{(a1)~\textsc{Act}:} Direct action prediction, where a single LLM observes the environment and directly outputs actions for all agents at each planning step, without explicit reasoning or planning modules.
\textbf{(a2)~Chain-of-Thought (\textsc{CoT})}~\cite{wei2022cot}: Prompts the LLM to reason step-by-step before outputting coordinated actions for all agents.
\textbf{(a3)~\textsc{ReAct}}~\cite{yao2022react}: A reasoning-and-acting approach in which the LLM explicitly outputs both a reasoning trace (Think) and agent actions (Action) at each step.
\textbf{(b)~Multi-call per step baselines}: employ multiple LLM invocations per planning step.
\textbf{(b1)~\textsc{SmartLLM}}~\cite{kannan2024smart}: A hierarchical planning approach that decomposes tasks into subtasks, assigns them to agents based on their skills and generates executable Python code.
\textbf{(b2)~\textsc{CoELA}}~\cite{zhang2023coela}: A communication-enabled decentralized system where each agent independently generates messages and actions via separate LLM calls, coordinating through shared dialogue history.
\textbf{(b3)~\textsc{LLaMAR}}~\cite{nayak2024llamar}: A Plan–Act–Verify architecture with explicit modules for planning, execution, failure correction, and progress verification.
For the best-performing baselines, we evaluate variants that incorporate:
(i) \projectname's agent-centric semantic memory (\S~\ref{subsec:egocentric-memory}), and 
(ii) \projectname's ILP framework (\S~\ref{subsec:action-level-constraints}).

\textbf{Metrics.}
We evaluate \projectname along two dimensions: 
planning effectiveness (the first five metrics) and execution efficiency (the remaining four). 
Planning effectiveness metrics are bounded in $[0,1]$.
\textbf{(a)~Success Rate}: fraction of episodes in which all tasks are completed.
\textbf{(b)~Transport Rate}: per-episode fraction of completed subtasks.
\textbf{(c)~Coverage}: fraction of successful interactions with task-relevant objects.
\textbf{(d)~Balance}: workload balance ($B$) across agents, \(B = \min_i s_i / (\max_j s_j + \epsilon)\), where \(s_i\) is the number of successful actions by agent \(i\), $\epsilon$ is $10^{-4}$, and $i,j\in [n]$. 
\(B=1\) denotes perfect balance, \(B \approx 0\) indicates at least one inactive agent.
\textbf{(e)~Failure Rate}: fraction of actions which could not be successfully executed, 
lower rate indicates better coordination among agents and conflict avoidance.
\textbf{(f)~Agent Steps}: total \# of physical steps and primitive movements (e.g., pickup, putdown) across all agents per episode. 
A single symbolic action (e.g., moving between two distant locations) may consist of multiple agent steps.
\textbf{(g)~Runtime}: total elapsed wall-clock time per episode from start to termination, capturing overall execution latency. 
\textbf{(h)~LLM Calls}: \# of LLM invocations per episode (e.g., API calls or prompts), quantifying reasoning overhead independent of step duration.
\textbf{(i)~Prompt Token Count}: \# of input tokens supplied to the LLM per episode, reflecting the communication overhead and memory load required for planning.

Results for all metrics are averaged over three episodes for all the tasks. 
Each episode is capped at 100 planning steps.
Furthermore, agents execute actions synchronously at each step, and the system proceeds only after all agents complete their actions, ensuring coordinated updates.

\subsection{Planning Effectiveness}
\label{subsubsec:planning-effectiveness}

Table~\ref{tab:main-results-ai2thor} reports the planning effectiveness of \projectname and its baseline. 
For both AI2-THOR and SAR, \projectname consistently achieves the highest success and transport rates, maximum coverage and balanced workload, through maintaining the lowest failure rate.

\begin{table*}[t]
\setlength{\tabcolsep}{3pt} 
\caption{Comparative evaluation of \projectname against baselines across effectiveness and efficiency metrics in \textbf{AI2-THOR} and \textbf{SAR} environments.
\projectname integrates \emph{Agent-centric Semantic Memory (ASM)} and an \emph{Integer Linear Programming (ILP)} framework.
Arrows~($\uparrow$/$\downarrow$) indicate whether higher or lower values are better. 
The numbers in subscript indicate the change relative to \textsc{ReAct} or \textsc{LLaMaR}. 
For the first five columns (effectiveness metrics), subscript is the absolute difference, while for the last four columns (efficiency metrics) subscript is percent difference. 
Positive improvements are highlighted in \textcolor{Green}{green}, and drops are highlighted in \textcolor{red}{red}.
{Best results are highlighted in bold} for both single-call and multi-call method classes.
Standard deviation and variance are reported in Appendix~\ref{adx:variance}.}
\label{tab:main-results-ai2thor}
\scriptsize
\begin{tabular}{llccccccccc}
\toprule \multicolumn{2}{c}{AI2-THOR Environment} &  
\multicolumn{5}{c}{Effectiveness Metrics} & \multicolumn{4}{c}{Efficiency Metrics} \\
\cmidrule(lr){1-2} \cmidrule(lr){3-7} \cmidrule(lr){8-11}
\textbf{\begin{tabular}[c]{@{}c@{}}Method\\Class\end{tabular}} & 
\textbf{Method} & 
\textbf{\begin{tabular}[c]{@{}c@{}}Success\\ Rate ($\uparrow$)\end{tabular}} & 
\textbf{\begin{tabular}[c]{@{}c@{}}Transport\\ Rate ($\uparrow$)\end{tabular}} & 
\textbf{\begin{tabular}[c]{@{}c@{}}Coverage\\ ($\uparrow$)\end{tabular}} & 
\textbf{\begin{tabular}[c]{@{}c@{}}Balance\\ ($\uparrow$)\end{tabular}} & 
\textbf{\begin{tabular}[c]{@{}c@{}}Failure\\ Rate ($\downarrow$)\end{tabular}} & 
\textbf{\begin{tabular}[c]{@{}c@{}} Runtime\\ (Seconds, $\downarrow$)\end{tabular}} & 
\textbf{\begin{tabular}[c]{@{}c@{}}\# of LLM\\ Calls ($\downarrow$)\end{tabular}} & 
\textbf{\begin{tabular}[c]{@{}c@{}}\# of Agent\\ Steps ($\downarrow$)\end{tabular}} & 
\textbf{\begin{tabular}[c]{@{}c@{}}\# of Tokens\\ ($\times 10^3$, $\downarrow$)\end{tabular}} \\
\midrule

\multirow{6}{*}{\begin{tabular}[c]{@{}l@{}}
Single\\LLM call\\ per\\planning\\step
\end{tabular}}
& \textsc{Act}  & 0.42 & 0.63 & 0.82 & 0.70 & 0.42 & 398.0 & 156.2 & 251.1 & 1152 \\
& \textsc{CoT}   & 0.08 & 0.29 & 0.43 & 0.62 & 0.47 & \textbf{229.1} & \phantom{0}\textbf{62.4} & 346.2 & \phantom{0}\textbf{501} \\
& \textsc{ReAct} & {0.44} & {0.68} & {0.86} & {0.71} & {0.39} & 367.5 & 135.4 & {235.4} & \phantom{0}649 \\
& $\drsh$ + ASM
& 0.47$_{\positive{+0.03}}$
& 0.72$_{\positive{+0.04}}$
& \textbf{0.90}$_{\positive{+0.04}}$
& 0.73$_{\positive{+0.02}}$
& 0.37$_{\positive{-0.02}}$
& 371.0$_{\negative{+0.9\%}}$
& 130.6$_{\positive{-\phantom{0}3.5\%}}$
& 230.3$_{\positive{-\phantom{0}2.2\%}}$
& \phantom{0}722$_{\negative{+11.3\%}}$ \\
& $\drsh$ + ILP
& 0.49$_{\positive{+0.05}}$
& 0.71$_{\positive{+0.03}}$
& \textbf{0.90}$_{\positive{+0.04}}$
& 0.74$_{\positive{+0.03}}$
& 0.37$_{\positive{-0.02}}$
& 355.8$_{\positive{-3.2\%}}$
& 124.8$_{\positive{-\phantom{0}7.8\%}}$
& \textbf{211.3}$_{\positive{-10.3\%}}$
& \phantom{0}729$_{\negative{+12.3\%}}$ \\
& $\drsh$ + \projectname
& \textbf{0.51}$_{\positive{+0.07}}$
& \textbf{0.73}$_{\positive{+0.05}}$
& \textbf{0.90}$_{\positive{+0.04}}$
& \textbf{0.76}$_{\positive{+0.05}}$
& \textbf{0.36}$_{\positive{-0.03}}$
& 342.0$_{\positive{-7.0\%}}$
& 121.7$_{\positive{-10.1\%}}$
& 216.1$_{\positive{-\phantom{0}8.3\%}}$
& \phantom{0}724$_{\negative{+11.5\%}}$ \\

\midrule

\multirow{6}{*}{\begin{tabular}[c]{@{}l@{}} 
Multiple\\LLM calls\\per \\ planning \\ step
\end{tabular}}
& \textsc{SmartLLM}   & 0.06 & 0.19 & 0.38 & 0.56 & 0.51 & 935.9 & \phantom{0}\textbf{89.3} & 461.6 & \phantom{0}\textbf{595} \\
& \textsc{CoELA}      & 0.11 & 0.34 & 0.63 & 0.65 & 0.46 & 889.4 & 386.1 & 422.2 & 2953 \\
& \textsc{LLaMaR}     & 0.59 & 0.85 & 0.90 & 0.79 & 0.32 & 757.9 & 346.8 & 295.8 & 1472 \\
& $\drsh$ + ASM & 0.61$_{\positive{+0.02}}$ & 0.87$_{\positive{+0.02}}$ & 0.91$_{\positive{+0.01}}$ & 0.82$_{\positive{+0.03}}$ & 0.29$_{\positive{-0.03}}$ &
723.8$_{\positive{-\phantom{0}4.5\%}}$ 
& 310.2$_{\positive{-10.6\%}}$ 
& 284.2$_{\positive{-\phantom{0}3.9\%}}$ 
& 1562$_{\negative{+6.1\%}}$\\
& $\drsh$ + ILP & 0.67$_{\positive{+0.08}}$ & 0.90$_{\positive{+0.05}}$ & 0.93$_{\positive{+0.03}}$ & 0.85$_{\positive{+0.06}}$ & 0.24$_{\positive{-0.08}}$ & 
628.1$_{\positive{-17.1\%}}$ 
& 258.4$_{\positive{-25.5\%}}$ 
& 240.3$_{\positive{-18.8\%}}$ 
& 1420$_{\positive{-3.5\%}}$\\
& $\drsh$ + \textsc{\projectname} & \textbf{0.69}$_{\positive{+0.10}}$ & \textbf{0.92}$_{\positive{+0.07}}$ & \textbf{0.95}$_{\positive{+0.05}}$ & \textbf{0.87}$_{\positive{+0.08}}$ & \textbf{0.20}$_{\positive{-0.12}}$ & \textbf{554.1}$_{\positive{-26.9\%}}$ 
& {231.5}$_{\positive{-33.3\%}}$ 
& \textbf{204.1}$_{\positive{-31.0\%}}$ 
& {1364}$_{\positive{-7.3\%}}$ \\
\bottomrule
\end{tabular}

\begin{tabular}{llccccccccc}
\toprule 
\multicolumn{2}{c}{SAR Environment} & 
\multicolumn{5}{c}{Effectiveness Metrics} & \multicolumn{4}{c}{Efficiency Metrics} \\
\cmidrule(lr){1-2} \cmidrule(lr){3-7} \cmidrule(lr){8-11}
\textbf{\begin{tabular}[c]{@{}c@{}}Method\\Class\end{tabular}} & 
\textbf{Method} & 
\textbf{\begin{tabular}[c]{@{}c@{}}Success\\ Rate ($\uparrow$)\end{tabular}} & 
\textbf{\begin{tabular}[c]{@{}c@{}}Transport\\ Rate ($\uparrow$)\end{tabular}} & 
\textbf{\begin{tabular}[c]{@{}c@{}}Coverage\\ ($\uparrow$)\end{tabular}} & 
\textbf{\begin{tabular}[c]{@{}c@{}}Balance\\ ($\uparrow$)\end{tabular}} & 
\textbf{\begin{tabular}[c]{@{}c@{}}Failure\\ Rate ($\downarrow$)\end{tabular}} & 
\textbf{\begin{tabular}[c]{@{}c@{}} Runtime\\ (Seconds, $\downarrow$)\end{tabular}} & 
\textbf{\begin{tabular}[c]{@{}c@{}}\# of LLM\\ Calls ($\downarrow$)\end{tabular}} & 
\textbf{\begin{tabular}[c]{@{}c@{}}\# of Agent\\ Steps ($\downarrow$)\end{tabular}} & 
\textbf{\begin{tabular}[c]{@{}c@{}}\# of Tokens\\ ($\times 10^3$, $\downarrow$)\end{tabular}} \\
\midrule

\multirow{6}{*}{\begin{tabular}[c]{@{}l@{}}
Only one\\ LLM call\\per\\planning\\step
\end{tabular}}
& \textsc{Act}   & 0.08 & 0.15 & 0.43 & 0.79 & 0.25 & \phantom{0}51.9 & \phantom{0}84.4 & 163.3 & \phantom{0}437 \\
& \textsc{CoT}   & 0.15 & 0.41 & 0.75 & 0.86 & 0.25 & \phantom{0}66.1 & \phantom{0}81.5 & 162.9 & \phantom{0}393 \\
& \textsc{ReAct} & {0.55} & {0.76} & {0.89} & {0.87} & {0.19} & \phantom{0}{47.4} & \phantom{0}{76.3} & \phantom{0}{95.3} & \phantom{0}\textbf{321} \\
& $\drsh$ + ASM & 0.58$_{\positive{+0.03}}$ & 0.80$_{\positive{+0.04}}$ & 0.95$_{\positive{+0.06}}$ & 0.90$_{\positive{+0.03}}$ & 0.18$_{\positive{-0.01}}$ & \phantom{0}47.5$_{\positive{-0.04\%}}$ & \phantom{0}74.0$_{\positive{-\phantom{0}2.9\%}}$ & \phantom{0}89.6$_{\positive{-\phantom{0}6.0\%}}$ & \phantom{0}371$_{\negative{+15.6\%}}$ 
\\
& $\drsh$ + ILP & 0.62$_{\positive{+0.07}}$ & 0.81$_{\positive{+0.05}}$ & 0.95$_{\positive{+0.06}}$ & 0.93$_{\positive{+0.06}}$ & 0.18$_{\positive{-0.01}}$ & \phantom{0}44.9$_{\positive{-5.3\%}}$ & \phantom{0}71.0$_{\positive{-\phantom{0}6.9\%}}$ & \phantom{0}86.1$_{\positive{-\phantom{0}9.7\%}}$ & \phantom{0}366$_{\negative{+14.0\%}}$ \\
& $\drsh$ + \projectname & \textbf{0.63}$_{\positive{+0.08}}$ & \textbf{0.85}$_{\positive{+0.09}}$ & \textbf{0.95}$_{\positive{+0.06}}$ & \textbf{0.95}$_{\positive{+0.08}}$ & \textbf{0.16}$_{\positive{-0.03}}$ & \phantom{0}\textbf{44.2}$_{\positive{-6.9\%}}$ & \phantom{0}\textbf{65.7}$_{\positive{-13.9\%}}$ & \phantom{0}\textbf{85.5}$_{\positive{-10.3\%}}$ & \phantom{0}364$_{\negative{+13.4\%}}$ \\

\midrule

\multirow{6}{*}{\begin{tabular}[c]{@{}l@{}} 
Multiple\\LLM calls\\per \\ planning \\ step
\end{tabular}}
& \textsc{SmartLLM}   & 0.12 & 0.34 & 0.51 & 0.82 & 0.23 & 513.3 & \textbf{101.6} & 146.5 & \phantom{0}\textbf{486} \\
& \textsc{CoELA}      & 0.09 & 0.32 & 0.49 & 0.82 & 0.23 & 545.3 & 383.7 & 191.9 & 2145 \\
& \textsc{LLaMaR}     & 0.55 & 0.76 & 0.89 & 0.89 & 0.16 & 333.8 & 256.5 & 107.6 & \phantom{0}857 \\
& $\drsh$ + ASM & 0.55$_{(-)\phantom{000}}$ & 0.77$_{\positive{+0.01}}$ & 0.90$_{\positive{+0.01}}$ & 0.90$_{\positive{+0.01}}$ & 0.14$_{\positive{-0.02}}$ & 310.0$_{\positive{-\phantom{0}7.1\%}}$ & 239.3$_{\positive{-\phantom{0}6.7\%}}$ & 105.9$_{\positive{-\phantom{0}1.6\%}}$ & \phantom{0}825$_{\positive{-\phantom{0}3.6\%}}$ \\
& $\drsh$ + ILP & 0.58$_{\positive{+0.03}}$ & 0.78$_{\positive{+0.02}}$ & 0.91$_{\positive{+0.02}}$ & 0.92$_{\positive{+0.03}}$ & 0.10$_{\positive{-0.06}}$ & 250.4$_{\positive{-25.0\%}}$ & 199.7$_{\positive{-22.2\%}}$ & \phantom{0}88.9$_{\positive{-17.4\%}}$ & \phantom{0}730$_{\positive{-14.7\%}}$\\
& $\drsh$ + \textsc{\projectname} & \textbf{0.59}$_{\positive{+0.04}}$ & \textbf{0.79}$_{\positive{+0.03}}$ & \textbf{0.92}$_{\positive{+0.03}}$ & \textbf{0.93}$_{\positive{+0.04}}$ & \textbf{0.09}$_{\positive{-0.07}}$ & \textbf{227.4}$_{\positive{-31.9\%}}$ & 179.2$_{\positive{-30.2\%}}$ & \phantom{0}\textbf{80.6}$_{\positive{-25.1\%}}$ & \phantom{0}683$_{\positive{-20.3\%}}$ \\

\bottomrule
\end{tabular}
\vspace{-0.4cm}
\end{table*}

\textbf{\projectname achieves consistently the lowest failure rates.}
Baseline single-call planners such as \textsc{Act}, \textsc{CoT}, and \textsc{ReAct} rely on short-term textual memory limited to the current object or destination. 
As a result, previously relevant but incomplete information is discarded,
leading to redundant exploration and infeasible actions, with failure rates of 0.39--0.47 in AI2-THOR and 0.19--0.25 in SAR.
Replacing short-term memory with ASM improves success rates by 1--3\% points, but performance remains lower than multi-call \textsc{LLaMaR}, highlighting that task decomposition across multiple planning calls provides the largest boost in success.
In the multi-call setting, \textsc{SmartLLM} suffers from frequent errors in Python-based plan generation, while \textsc{CoELA}’s decentralized design causes context loss and inconsistent action selection, resulting in high failure rates of 0.46--0.51 in AI2-THOR and 0.23 in SAR.
Stronger multi-call planners such as \textsc{LLaMaR} retain longer-term context as textual logs, but this memory remains unstructured due to absolute coordinates and orientations. 
This limits spatial reasoning during LLM inference, resulting in comparatively high failure rates (0.32 in AI2-THOR, 0.16 in SAR).

In contrast, \projectname's ASM tracks all task-relevant objects and destinations, including those out of view, with the ILP framework enforcing feasibility constraints and penalizes conflicting or redundant actions, thereby reducing failures.
When combined, ASM and ILP in \projectname reduce failures by 3\% points for \textsc{ReAct} and 7--12\% points for \textsc{LLaMaR}.
The improvements are particularly pronounced in the multi-call setting of \textsc{LLaMaR}, where task decomposition breaks complex goals into smaller subtasks, enabling more informed action selection and allocation.

\textbf{\projectname achieves the highest task success rate because of reduced failures.}
Action execution failures can lead agents into patterns from which recovery is difficult.
Such failures can prevent task completion and consequently reduce overall success rates. 
This effect is particularly pronounced for Python-based \textsc{SmartLLM}, decentralized \textsc{CoELA}, and single-call method \textsc{CoT}, which exhibit higher failure rates (0.46--0.51 in AI2-THOR and 0.23--0.25 in SAR), corresponding to lower success rates (0.06--0.11 in AI2-THOR and 0.09--0.15 in SAR).
\textsc{CoT} struggles on longer planning horizons, as it cannot maintain coherence across extended sequences, consistent with~\cite{nayak2024llamar, stechly2024chain}, which find it effective only with very specific prompts.

By reducing execution failures, \projectname enables agents to follow more feasible action trajectories, resulting in higher success rates across both environments under a fixed planning-step budget. 
In AI2-THOR, \projectname improves success from 0.44 to 0.51 (+7\% points) in the single-call \textsc{ReAct} setting and from 0.59 to 0.69 (+10\% points) in the multi-call \textsc{LLaMaR} setting.
For SAR, the improvements in success rate are from 0.55 to 0.63 (+8\% points) for \textsc{ReAct} and from 0.55 to 0.59 (+4\% points) for \textsc{LLaMaR}.
The larger gains for \textsc{ReAct} in SAR are likely due to its shorter-horizon and less complex environment, where structured memory and coordination allow immediate corrections and more effective action selection. 
Moreover, these improvements are not due to aggressive exploration or increased LLM usage, but rather by preventing execution failures.

\subsection{Execution Efficiency}
\label{subsubsec:runtime-and-steps}

Table~\ref{tab:main-results-ai2thor} 
additionally shows that \projectname reduces execution costs.
Appendix~\ref{adx:ai2thor-sar-effectiveness} further discusses effectiveness as a function of agent steps.

\textbf{\projectname reduces runtime through lower failure rates.}
We observe that execution failures are the major contributor to high latency in embodied multi-agent planning. 
Each failed action triggers replanning, redundant steps, and extra LLM calls, compounding runtime. 
Consequently, methods with higher failure rates (\textsc{Act}, \textsc{CoT}, \textsc{SmartLLM}, \textsc{CoELA}) exhibit longer execution times within a fixed planning-step budget, increasing LLM usage without advancing task completion.

By reducing failure rates, \projectname directly limits inefficiencies such as redundant actions and additional LLM calls.
In AI2-THOR and SAR, \projectname reduces the failure rate of \textsc{ReAct} by 3\% points, which corresponds to a runtime reduction of $\sim$7.0\%, fewer LLM calls by 10.1--13.9\%, and fewer agent steps 8.3--10.3\%. 
In the multi-call \textsc{LLaMaR} setting, failure rates drop by 7--12\% points, yielding a much larger runtime reduction 26.9--31.9\%, alongside decreases in LLM calls 30.2--33.3\% and agent steps 25.1--31.0\%.

\textbf{\projectname achieves the best tradeoff between efficiency and effectiveness.}
Single-call planners attain low runtime primarily by minimizing LLM calls, but this efficiency comes at the cost of poor action quality, and low task success. 
In AI2-THOR, \textsc{CoT} achieves the lowest runtime (229.1s) and fewest LLM calls (62.4), yet requires 346.2 agent steps and succeeds in only 8\% of tasks, whereas \projectname reduces agent steps to 204.1 (-41\%) and increases success to 0.69 (+0.61), despite higher runtime (554.1s) and more LLM calls (231.5). 
A similar pattern appears in SAR, where \textsc{Act} and \textsc{CoT} execute quickly (51.9s and 66.1s) but attain low success (0.08--0.15) and high failure rates (0.25), while \projectname on \textsc{LLaMaR} reaches 0.59 success with substantially fewer redundant actions. 

Beyond outperforming single-call methods in effectiveness, \projectname also improves efficiency relative to strong multi-call baselines: 
compared to \textsc{LLaMaR}, across both environments, it reduces runtime by 26.9--31.9\%, LLM calls by 30.2--33.3\%, and agent steps by 25.0--31.0\%, and achieves consistent gains over \textsc{ReAct} as well. 
While ASM increases token usage by 11.5–13.4\% for \textsc{ReAct}+\projectname, this modest overhead is outweighed by improvements in success rate (7--8\% points), runtime reduction (7\%), and decreased agent steps (8.3--10.3\%).

\subsection{Computation Scalability of ILP}
\label{subsec:ilp-runtime}
\begin{table*}[!ht]
\caption{Performance and ILP overhead in the SAR environment across varying team sizes (3--8 agents). 
The ILP solver accounts for only 1–2\% of total runtime and scales modestly, with negligible impact on overall latency.}
\vspace{-0.2cm}
\centering
\renewcommand{\arraystretch}{1.2}
\footnotesize
\label{tab:ilp-runtime}
\begin{tabular}{cclccccc}
\toprule
\begin{tabular}[c]{@{}c@{}}\textbf{Agent}\\ \textbf{Count}\end{tabular} & \begin{tabular}[c]{@{}c@{}}\textbf{Total Action-Agent}\\ \textbf{Combinations}\end{tabular} & \textbf{Method} & \begin{tabular}[c]{@{}c@{}}\textbf{Success}\\ \textbf{Rate ($\uparrow$)}\end{tabular} & \begin{tabular}[c]{@{}c@{}}\textbf{Failure}\\ \textbf{Rate ($\downarrow$)}\end{tabular} & \begin{tabular}[c]{@{}c@{}}\textbf{Total}\\ \textbf{Runtime ($\downarrow$, s)}\end{tabular} & \begin{tabular}[c]{@{}c@{}}\textbf{Total ILP Solver} \\ \textbf{Runtime ($\downarrow$, s)}\end{tabular} & \begin{tabular}[c]{@{}c@{}}\textbf{\% of Total}\\ \textbf{Runtime}\end{tabular} \\
\midrule
\rowcolor{gray!10} &  &  \textsc{LLaMaR} & 0.63 & 0.14 & 264.3 & -- & -- \\
\rowcolor{gray!10} \multirow{-2}{*}{3} & \multirow{-2}{*}{9} & $\drsh$ + \projectname & 0.67 & 0.07 & 176.6 & 2.22 & 1.3 \\ 
\multirow{2}{*}{4} & \multirow{2}{*}{12} & \textsc{LLaMaR} & 0.65 & 0.12 & 269.9 & -- & -- \\
 &  & $\drsh$ + \projectname & 0.68 & 0.05 & 188.5 & 2.66 & 1.4 \\
\rowcolor{gray!10} & & \textsc{LLaMaR} & 0.64 & 0.12 & 272.8 & -- & -- \\
\rowcolor{gray!10} \multirow{-2}{*}{5} & \multirow{-2}{*}{15} & $\drsh$ + \projectname & 0.66 & 0.05 & 188.8 & 2.49 & 1.3 \\
\multirow{2}{*}{6} & \multirow{2}{*}{18} & \textsc{LLaMaR} & 0.62 & 0.16 & 280.3 & -- & -- \\
&  & $\drsh$ + \projectname & 0.65 & 0.08 & 199.1 & 3.34 & 1.7 \\
\rowcolor{gray!10} & & \textsc{LLaMaR} & 0.59 & 0.21 & 298.3 & -- & -- \\
\rowcolor{gray!10} \multirow{-2}{*}{7} & \multirow{-2}{*}{21} & $\drsh$ + \projectname & 0.62 & 0.16 & 234.2 & 4.38 & 1.9 \\
\multirow{2}{*}{8} & \multirow{2}{*}{24} & \textsc{LLaMaR} & 0.51 & 0.26 & 319.1 & -- & -- \\
 &  & $\drsh$ + \projectname & 0.55 & 0.19 & 301.2 & 5.87 & 1.9 \\
 \bottomrule
\end{tabular}
\vspace{-0.4cm}
\end{table*}
Table~\ref{tab:ilp-runtime} shows that ILP overhead remains consistently low, contributing only 1.3--1.9\% of total runtime as agent count increases from 3 to 8. 
Although the number of action-agent combinations grows from 9 to 24, ILP solver runtime increases only modestly from 2.22s to 5.87s, indicating favorable scaling behavior. 
Across all team sizes, \projectname{} improves success rates by 2--4 points while reducing failure rates by up to 9 points relative to \textsc{LLaMaR}. 
Notably, despite introducing an additional optimization stage, \projectname{} also reduces overall runtime by 33--88 seconds in most settings. 
This efficiency is enabled by action pruning with a fixed $K{=}3$, with further scalability achievable through agent pruning in larger systems.

\subsection{Additional Experiments}
\label{subsec:additional-experiments-main}

We provide detailed analyses of \projectname's robustness, efficiency, coordination, and ablations in the appendices.

\textbf{(a)~Effects of agent counts.} \projectname scales effectively with team size. In AI2-THOR with 3–5 agents, 
\textsc{ReAct}+\projectname improves success by 15--20\% points over \textsc{ReAct} while reducing runtime by 15--25\%, and 
\textsc{LLaMaR}+\projectname reaches up to 0.79 success rate with 20--30\% faster execution compared to \textsc{LLaMaR}, yielding consistently better success–runtime tradeoffs.
\textbf{(b)~Effects of feasibility constraints and cost function.} 
Appendix~\ref{subsec:ablation} highlights ILP constraints improve coordination (4--5\% points success increase) and the cost function boosts efficiency (12--22\% fewer LLM calls, 12\% faster runtime). 
\textbf{(c)~LLM model choices.}
Appendix~\ref{subsubsec:models} shows that \projectname improves success across GPT-4o, Claude Sonnet 4.5, and Gemini 3 Flash (4--7\% points for \textsc{ReAct}, 10--11\% points for \textsc{LLaMaR}) while reducing runtime by 27\%, showing model-agnostic robustness.
\textbf{(d)~Effects of sensor noises.}
In Appendix~\ref{adx:performance-under-observation-noise}, planning with \projectname remains effective under severe observation noise, with runtime at 88\% of the noiseless setting.
\textbf{(e)~Effect of planning step budgets.}
Appendix~\ref{subsec:planning-step-budget} shows that increasing the planning budget from 50 to 100 steps raises success by 0.21--0.25, while further increases to 150--200 steps yield minimal gains ($\leq$0.05) with higher runtime.
\textbf{(f)~Qualitative analysis.}
Appendix~\ref{adx:qualitative-analysis} presents planning trajectories: \textsc{LLaMaR} exhibits cycles, repeated failures, and stagnation, whereas \projectname ensures fast, balanced, and consistently progressive execution.

\vspace{-0.1cm}
\section{Conclusion}
\label{sec:conclusion}

\projectname demonstrates that runtime-efficient multi-agent planning can be achieved through lightweight agent-centric state tracking and constraint-guided coordination.
Its core contributions, Agent-centric Semantic Memory (ASM) and Integer Linear Programming (ILP), address inefficiencies in state tracking and multi-agent execution.
ASM enables compact spatial reasoning under partial observability, reducing failed actions by up to 12\%, while ILP enforces step-level feasibility constraints over LLM proposals, achieving up to 32\% faster execution, 33\% fewer LLM calls, and 4--10\% higher success rates.
Across both AI2-THOR and search-and-rescue environments, these modules consistently reduce redundant actions and execution overhead, improving latency and task success.
Overall, \projectname shows that structured memory and explicit feasibility constraints are effective plug-and-play mechanisms for scalable multi-agent embodied planning.

\clearpage
\section*{Acknowledgement}
\label{sec:acknowledgement}

This material is based upon work supported by the National Science Foundation under Grant No. CNS-2312396, CNS-2338512, IIS-2435822, and CCF-2449995.
Any opinions, findings, and conclusions or recommendations expressed in this material are those of the author(s) and do not necessarily reflect the views of the National Science Foundation.

\section*{Impact Statement}
\label{sec:impact-statement}

\projectname advances the practical deployment of LLM-based multi-agent systems in embodied environments by addressing the primary sources of inefficiency: failed actions, redundant exploration, and coordination breakdowns.
By integrating lightweight, spatially grounded agent-centric memory with action-level joint optimization, the framework reduces execution latency and inference cost while maintaining high task success.
Improved coordination efficiency limits unnecessary actions and resource waste, helping mitigate unsafe or unpredictable behavior and reducing environmental impact in shared environments.
As more reliable multi-agent systems become viable for public-facing domains, these gains also highlight the growing need for transparent constraints and oversight as such systems increasingly shape real-world decisions.

Together, these advances enable operation under realistic time and resource constraints, narrowing the gap between research prototypes and deployable systems.
The resulting improvements are especially impactful for safety- and time-critical applications such as search and rescue, household robotics, and long-horizon exploration, and demonstrate how principled integration of symbolic optimization with neural reasoning can unlock scalable, real-world multi-agent intelligence.

\bibliography
{bibliography}
\bibliographystyle{icml2026}

\newpage
\appendix
\onecolumn

\section{Extended Related Work}
\label{adx:extended-related-work}

We review prior work on memory representations for embodied planning, as well as approaches for coordinating actions among multiple agents.

\paragraph{Memory Representations for Embodied Agents.}
Work on memory for embodied agents can be broadly categorized into \textbf{(a) temporal, (b) spatial, and (c) spatio-temporal memory}.

\textbf{(a) Temporal memory} retains past observations over time without explicitly storing object locations or geometry. It includes \textit{(i) short-term memory}, which stores the most recent observations or latent states over a few timesteps via limited context windows~\cite{ yao2022react, lin2025stopwastingtokensefficient}; \textit{(ii) long-term or episodic memory}, which preserves information across extended horizons such as full task episodes or multiple trajectories using persistent memory modules or databases~\cite{nayak2024llamar, fang2019scenememory, gupta2025memo, lei2025robomemory, zhang2025ella, yadla2025temporal}; and 
\textit{(iii) hierarchical memory}, which organizes temporal information at multiple abstraction levels (e.g., low-level observations, subgoals, high-level plans)~\cite{wang2025karma, hu2024hiagenthierarchicalworkingmemory, lei2025stma, zhang2025gmemorytracinghierarchicalmemory, han2025llmmultiagentsystemschallenges}. 
While effective for retaining history, temporal-only memories often rely on unstructured textual logs~\cite{pan2025whymultiagentfail, zhang2025agentcausestaskfailures, li2024survey}, requiring the LLM to infer object relationships and positions from text alone, which is error-prone for spatial reasoning.

\textbf{(b) Spatial memory} explicitly stores the layout of the environment (such as object locations, or free space) independent of time, enabling direct spatial reasoning; representative approaches include classical topology-based or sensor-based navigation systems~\cite{gupta2017cognitive, chaplot2020learning} and recent dense 3D representations such as \textsc{VLMaps}~\cite{huang23vlmaps}, \textsc{ConceptGraphs}~\cite{gu2024conceptgraphs}, \textsc{Mem4Nav}~\cite{he2025mem4nav}, TME~\cite{ye2025taskmemoryengine}, and \textsc{Mem2Ego}~\cite{zhang2025memego}. 
In contrast to these dense spatial representations, which often incur significant overhead to store high-fidelity geometry, our memory focuses on an agent-centric abstraction of task-relevant objects and destinations, prioritizing runtime-efficient tracking over complete environmental reconstruction. 
\textbf{(c) Spatio-temporal memory} combines both dimensions by maintaining spatial representations that evolve over time, allowing LLMs to reason about dynamic environments in which objects move, appear, or disappear, as explored in \textsc{Scene Memory Transformer}~\cite{fang2019scenememory}, \textsc{3D-Mem}~\cite{yang20253dmem}, STMA~\cite{lei2025stma}, \textsc{Meta-Memory}~\cite{mao2025metamemory}, \textsc{Remembr}~\cite{anwar2025remembr}, and \textsc{3DLLM-Mem}~\cite{hu2025dllmmem}. 
Constructing and updating spatio-temporal memories typically requires accumulating observations over long horizons and synchronizing rich representations, which introduces significant computational and latency overhead. 
In contrast, our approach adopts an agent-centric spatio-temporal memory 
in which each agent maintains a structured textual representation of task-relevant objects and destinations relative to itself.
This memory is updated at every planning step to track relative distances and goals (even when objects fall outside the agent’s current field of view) enabling efficient spatial grounding in dynamic and partially-observable environments.

\paragraph{Multi-Agent Coordination.}
Multi-agent coordination methods can be characterized along two key dimensions: 
(i)~\textbf{Coordination Topology}, distinguishing centralized approaches from decentralized, message-passing ones, and 
(ii)~\textbf{Coordination Mechanism}, which determines how task and action allocation decisions are produced, e.g., via symbolic planners or neural policies.
\textbf{Failure handling} is closely tied to the coordination mechanism: symbolic methods can proactively anticipate conflicts and infeasible plans, while neural approaches are largely reactive, correcting errors only after execution.

\textbf{(a) Coordination Topology: Centralized vs. Decentralized Message Passing.}
Coordination can be \emph{decentralized}, where agents exchange information through peer-to-peer communication under bandwidth or observability constraints~\cite{zhang2023building, ying2024goma, wu2024camon, nomura2025decentralized, owerko2025mast, pesce2020improving, gupta2025hammer, han2025sparse}, or \emph{centralized}, where a shared entity aggregates information to maintain consistent knowledge of the environment and agent actions~\cite{simoes2020multi, tan2025roboos, mandi2024roco, hill2025communicating}.
We adopt centralized message passing, as we focus on collaborative settings (such as search-and-rescue and household tasks) where inter-agent privacy concerns are limited and maximizing coordination is critical for task success.

\textbf{(b) Coordination Mechanism: Task/Action Allocation and Failure Management.}
Coordination mechanisms differ in how task and action allocation decisions are produced, ranging from symbolic planners, which rely on explicit constraints, to neural policies, which learn to suggest actions from data.
Alongside allocation, these mechanisms differ in how they handle execution failures: 
some methods are reactive, correcting conflicts or infeasible actions only after they occur, while others are proactive, anticipating and preventing failures before execution.

Neural methods typically propose candidate actions or task assignments based on learned experience, often requiring additional mechanisms to ensure feasibility and resolve failures. 
They are largely reactive, correcting errors only after the actions are executed~\cite{yao2022react, kannan2024smart, nayak2024llamar, zhang2023coela, wang2023voyager, ahn2022saycan, ji2024testing, huang2022inner}.
Symbolic approaches leverage preconditions and logical rules to proactively reason about feasibility and reduce execution failures~\cite{zheng2024evaluating, cornelio2024recover}, and are commonly instantiated through Task and Motion Planning (TAMP) frameworks or LLM-assisted optimization layers~\cite{faroni2024tamp, kaelbling2011hierarchical, garrett2021integrated, lozano2014constraint, dantam2018incremental}.
Existing neuro-symbolic coordination mechanisms have limited scope.
\citealp{choe2025askreasonassist} handles failures reactively, allowing warehouse agents to request or offer help via symbolic constraints. 
\citealp{brawer2023towards} checks feasibility among user-specified actions but does not aim for runtime optimization. 
\textsc{LiP-LLM}~\cite{obata2024lip} and \citealt{wu2024hierarchical} assign tasks to agents while ensuring no conflicts over target objects using symbolic constraints.
However, these methods do not perform action-level assignments, which are necessary to reason about conflicts beyond target object allocation at each planning step.

In contrast, \projectname uses a neuro-symbolic, \textbf{runtime optimization-based coordination mechanism} that performs \textbf{joint action-level assignments} at every timestep, guided by LLM-generated action suggestions.
This enables \textbf{proactive failure avoidance} through fine-grained action feasibility reasoning.

\section{Multi-Agent Planning Architecture}
\label{adx:multi-agent-planning-architecture}

\begin{algorithm}[!th]
\caption{General Architecture Multi-Agent Embodied Planning}
\label{alg:multi_agent_planning}
\begin{algorithmic}[1]

\Require Task instruction $\mathsf{I}$, Action constraints $\mathsf{C}$, Agents $\mathsf{A} = \{a_1, \dots, a_N\}$, 
Architecture mode $\mathsf{Mode} \in \{\text{Centralized}, \text{Decentralized}\}$
\Ensure Executed multi-agent plan

\If{$\mathsf{Mode} = \text{Centralized}$} \label{line:init_start}
    \State \textbf{initialize} global memory $\mathsf{M} \gets \emptyset$
\Else
    \For{each agent $a_i \in \mathsf{A}$}
        \State \textbf{initialize} local memory $\mathsf{M}_i \gets \emptyset$
    \EndFor
\EndIf \label{line:init_end}

\State \textbf{initialize} $\mathsf{H} \gets \emptyset$ 
\Comment{Per-agent action-success history} \label{line:vars_start}

\State \textbf{initialize} $\mathsf{S}_o \gets \emptyset$  \Comment{Open subtasks}
\State \textbf{initialize} $\mathsf{S}_c \gets \emptyset$  \Comment{Completed subtasks} \label{line:vars_end}

\State $\mathsf{S}_o \gets$ \textsc{LLM-Planner}($\mathsf{I}$, $\mathsf{S}_o$, $\mathsf{S}_c$, 
\textsf{M} if Centralized else $\{\mathsf{M}_i\}_{i=1}^N$) 
\Comment{Initial decomposition into subtasks using appropriate memory} \label{line:initial_decomp}

\While{$\mathsf{S}_o \neq \emptyset$} \label{line:loop_start}

    \If{$\mathsf{Mode} = \text{Centralized}$} \label{line:actor_centralized_if}
        \State $\{x_i\}_{i=1}^N \gets$ \textsc{LLM-Actor}($\mathsf{A}$, $\mathsf{C}$, $\mathsf{S}_o$, $\mathsf{H}$, $\mathsf{M}$) \label{line:actor_centralized_call}
        \Comment{Single LLM call produces actions for all agents} \label{line:actor_centralized_comment}
    \Else \label{line:actor_decentralized_else}
        \For{each agent $a_i \in \mathsf{A}$} \label{line:actor_decentralized_for_start}
            \State $x_i \gets$ \textsc{LLM-Actor}($a_i$, $\mathsf{C}$, $\mathsf{S}_o$, $\mathsf{H}$, $\mathsf{M}_i$) \label{line:actor_decentralized_call}
            \Comment{Each agent calls its own LLM} \label{line:actor_decentralized_comment}
        \EndFor \label{line:actor_decentralized_for_end}
    \EndIf \label{line:actor_end_if}

    \State $\mathsf{H}, \mathsf{M} \gets$ \textsc{ExecuteInParallel}($\{(a_i, x_i)\}_{i=1}^N$)
    \Comment{Execute actions and update history/memory} \label{line:execute}

    \State $\mathsf{S}_o, \mathsf{S}_c \gets$ \textsc{LLM-Verifier}($\mathsf{S}_o$, $\mathsf{S}_c$, $\mathsf{H}$, 
    \textsf{M} if Centralized else $\{\mathsf{M}_i\}_{i=1}^N$)
    \Comment{Check which subtasks are completed} \label{line:verify}

    \State $\mathsf{S}_o \gets$ \textsc{LLM-Planner}(
    $\mathsf{I}$, $\mathsf{S}_o$, $\mathsf{S}_c$, 
    \textsf{M} if Centralized else $\{\mathsf{M}_i\}_{i=1}^N$) 
    \Comment{Optional re-planning based on updated memory} \label{line:replan}
\EndWhile \label{line:loop_end}
\end{algorithmic}
\end{algorithm} 
Algorithm~\ref{alg:multi_agent_planning} describes a general planning loop for both centralized and decentralized multi-agent execution, in which different LLMs are used to 
(a) decompose high-level instructions into subtasks, 
(b) propose actions for each of the subtasks, and 
(c) verify subtask completion.
This separation of roles follows prior LLM-based embodied planning systems~\cite{nayak2024llamar, zhang2023coela, kannan2024smart}, which decouple \emph{high-level reasoning} (task decomposition and progress assessment) from \emph{low-level decision making} (action selection) to improve execution reliability and task success.
The algorithm begins by initializing memory according to the selected mode (lines~\ref{line:init_start}–\ref{line:init_end}).
In centralized mode, a single global memory $\mathsf{M}$ is maintained and shared across all agents, whereas in decentralized mode, each agent maintains its own local memory $\mathsf{M}_i$.
This ensures that LLMs can reason using either a shared or independent knowledge base depending on the architecture.
Simultaneously, the interaction history $\mathsf{H}$ and bookkeeping variables for open and completed subtasks, $\mathsf{S}_o$ and $\mathsf{S}_c$, are initialized (lines~\ref{line:vars_start}–\ref{line:vars_end}) to record prior actions, execution outcomes, and task progress, providing context for planning and verification.

Given the high-level instruction $\mathsf{I}$, the \textsc{LLM-Planner} produces an initial decomposition into subtasks, conditioned on the available memory (line~\ref{line:initial_decomp}), allowing the system to convert natural language goals into trackable objectives.
The algorithm then proceeds iteratively until all subtasks are completed (line~\ref{line:loop_start}), with each iteration corresponding to one execution step.
In each step, agent actions are generated via \textsc{LLM-Actor} (lines~\ref{line:actor_centralized_if}–\ref{line:actor_end_if}), which is responsible for translating the current task context into concrete, executable actions.
In centralized mode, a single LLM call produces actions for all agents jointly, enabling globally consistent coordination.
In decentralized mode, each agent calls its own LLM to propose an action based on its local memory.
In both cases, action generation considers the current set of open subtasks, the accumulated execution history, and action constraints, allowing the LLM to reason at the action level for fine-grained coordination.

Once proposed, the actions are executed in parallel, and the resulting outcomes are used to update both the interaction history and the relevant memory representations (line~\ref{line:execute}).
The \textsc{LLM-Verifier} then evaluates which subtasks have been successfully completed and updates $\mathsf{S}_o$ and $\mathsf{S}_c$ accordingly (line~\ref{line:verify}).
This verification is critical because completing a subtask may change which subsequent actions or subtasks are relevant, depending on how the environment responded to the agents’ actions.
The \textsc{LLM-Planner} may then be re-invoked to revise the remaining subtasks in light of this updated context (line~\ref{line:replan}), ensuring that planning remains adaptive to partial observability, execution uncertainty, and dynamic environments while maintaining a consistent framework across both centralized and decentralized architectures.

\section{Action-level Feasibility Constraints}
\label{adx:action-level-constraints}

We detail the ILP’s action-level feasibility constraints that regulate candidate actions to prevent conflicts, infeasible assignments, and execution failures.

\textbf{Eligibility Constraints.}
If an action candidate $x_{i,t}^{(k)}$ is physically infeasible for agent $a_i$ at step $t$
(e.g., attempting to place or manipulate an object that is not currently held, navigating into an occupied or non-traversable region such as a wall, or violating basic action preconditions),
we enforce \fbox{$z_{i,t}^{(k)} = 0$}.
This prevents the solver from considering impossible action assignments.

\textbf{Per-Agent Action Selection.}
At each planning step $t$, each agent is required to execute exactly one action, enforced by
\fbox{$\sum_k z_{i,t}^{(k)} = 1 \;\; \forall i$}.
If no meaningful action is available, the candidate set $\mathsf{X}_{i,t}$ includes a special
\texttt{no-op} or \texttt{done} action, ensuring the constraint remains satisfiable.
This guarantees that the joint action at each step is well-defined and avoids repeated re-planning or inconsistent partial assignments.

\textbf{Multi-Agent Joint-Action Requirements.}
Some subtasks require multiple agents to act simultaneously (e.g., carrying a heavy object or assisting a person).
Let $s$ denote such a subtask that requires $k_s$ agents, and let $\mathsf{X}_{i,t}^{(s)} \subseteq \mathsf{X}_{i,t}$ denote the subset of action candidates for agent $a_i$ at step $t$ that correspond to subtask $s$.
We enforce
\fbox{$\sum_i \sum_{k \in \mathsf{X}_{i,t}^{(s)}} z_{i,t}^{(k)} = k_s$}.
This ensures that joint subtasks are neither under-staffed, which would lead to execution failure, nor over-staffed, which would waste actions and increase runtime.

\textbf{Resource Exclusivity.}
When multiple action candidates require exclusive access to the same resource (e.g., an object, tool, or environment element), we impose
\fbox{$\sum_{i} \sum_{k \in \mathsf{X}_{i,t}^{(r)}} z_{i,t}^{(k)} \le 1$},
where $\mathsf{X}_{i,t}^{(r)} \subseteq \mathsf{X}_{i,t}$ denotes the set of action candidates for agent $a_i$ at step $t$ that require resource $r$.
This prevents conflicts such as simultaneous grasps of the same object, concurrent attempts to open the same door, or contention over shared tools.

\textbf{Collision and Interference Avoidance.}
Certain action candidates may be mutually incompatible due to spatial or kinematic constraints.
For any pair of incompatible action candidates
$(x_{i,t}^{(k)}, x_{i',t}^{(k')})$,
we enforce
\fbox{$z_{i,t}^{(k)} + z_{i',t}^{(k')} \le 1$}.
This prevents collisions, overlapping target locations, and other mutually exclusive spatial configurations across agents at the same step.

\section{Additional Results}
\label{adx:additional-results}

This section presents additional experiments that further analyze the behavior, robustness, and generality of our approach. 
We examine how performance scales with the number of agents, characterize effectiveness trends across planning steps, and evaluate generalization across different models. 
We additionally study robustness under observation noise, isolate the contributions of individual components of the ILP framework through ablations, analyze the impact of the planning step budget, and report performance variance to assess stability.

\subsection{Scalability of \projectname}
\label{adx:scalability}

\begin{figure*}
     \centering
     \begin{subfigure}[b]{0.49\textwidth}
         \centering
         \includegraphics[width=\textwidth]{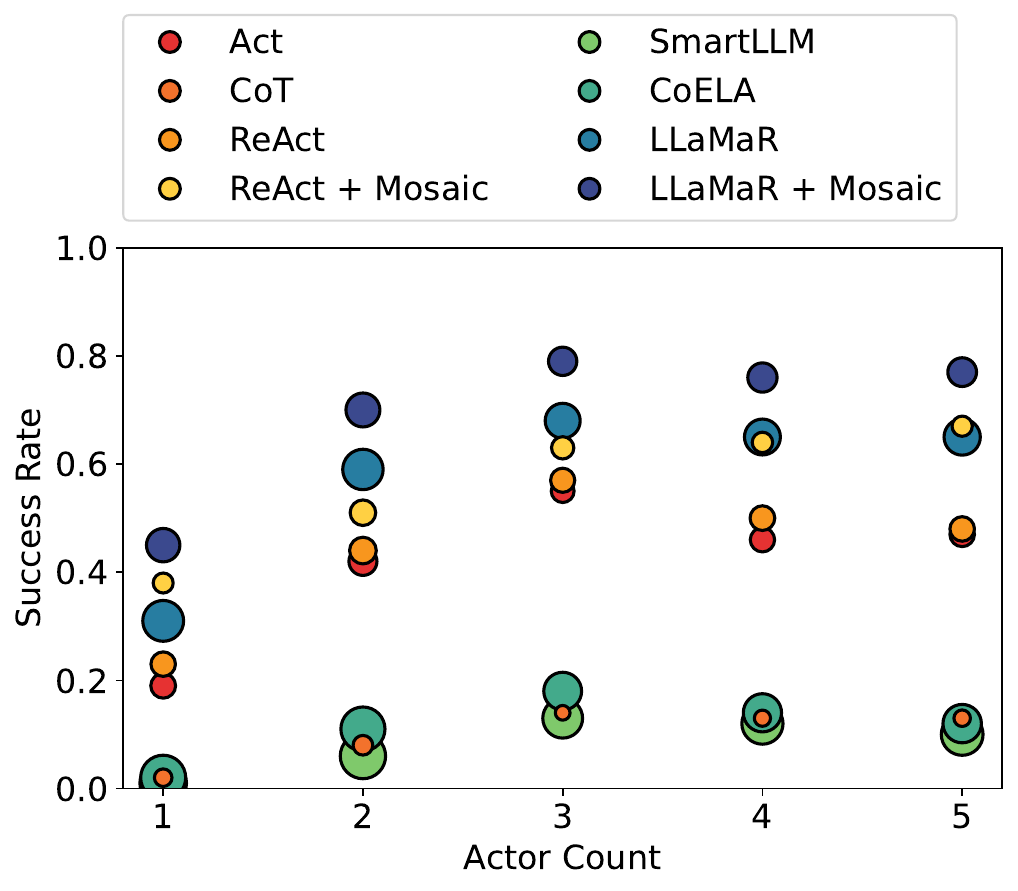}
         \caption{AI2-THOR}
         \label{fig:ai2thor-scalability}
     \end{subfigure}
     \hfill
     \begin{subfigure}[b]{0.49\textwidth}
         \centering
         \includegraphics[width=\textwidth]{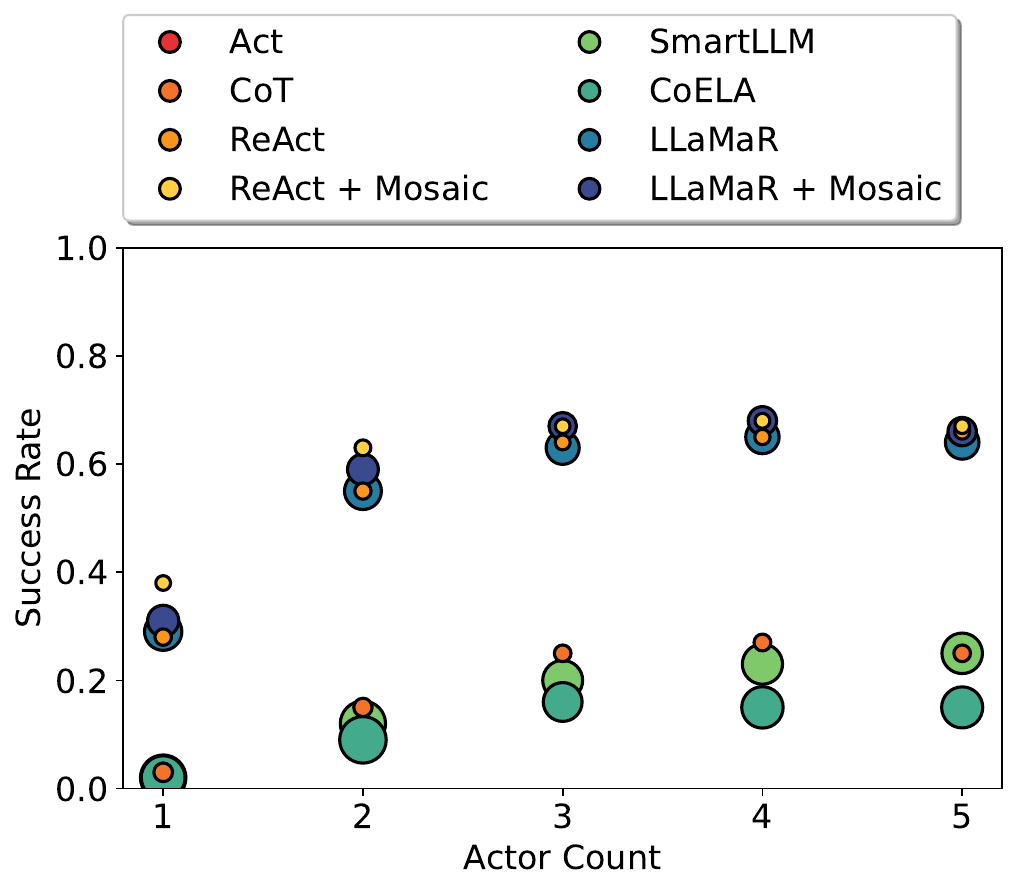}
         \caption{SAR}
         \label{fig:sar-scalability}
     \end{subfigure}
        \caption{A three-way comparison among success rate (y-axis), actor count (x-axis), and runtime (circle size) for \projectname and its baselines. 
        Smaller circles indicate lower runtime. 
        Across actor counts, our method consistently achieves the highest success rate while incurring  the lowest runtime among approaches that rely on multiple LLM calls. 
        Moreover, increasing the number of actors yields diminishing returns, reflecting saturation in the number of actors required by the environment.}
        \label{fig:scalability}
\end{figure*}

We evaluate scalability by varying agent count (1--5) and measuring success rate and runtime in AI2-THOR (Figure~\ref{fig:ai2thor-scalability}) and SAR (Figure~\ref{fig:sar-scalability}).
The corresponding numerical results, along with failure rates, are listed in Table~\ref{tab:scalability-summary}.

\textbf{\projectname improves efficiency and effectiveness as agent count increases to 3.}
Across both environments, augmenting existing planners with \projectname increases success rates while reducing execution time, with gains preserved as the number of agents grows. 
In AI2-THOR, \textsc{ReAct}+\projectname improves success over \textsc{ReAct} by 15--20\% points at higher agent counts (3--5), while \textsc{LLaMaR}+\projectname gains 10--14\% points and achieves the highest overall success (up to 0.79). 
In SAR, \projectname demonstrates similar benefits: integrating \projectname with \textsc{ReAct} raises success from 28\% to 38\% for a single agent (+10\% points) and from 65\% to 68\% for four agents (+3\% points). 
For \textsc{LLaMaR}, success improves from 29\% to 31\% (+2\% points) for one agent and from 65\% to 68\% (+3\% points) for four agents.

These improvements are accompanied by consistent runtime reductions. 
In AI2-THOR, \textsc{ReAct}+\projectname reduces runtime by 15--25\% across agent counts, while \textsc{LLaMaR}+\projectname achieves 20--30\% reductions. 
In SAR, \textsc{ReAct}+\projectname completes tasks in 35--44s vs 37--49s for the baseline (10--30\% faster), and \textsc{LLaMaR}+\projectname executes in 177--189s compared to 264--343s for \textsc{LLaMaR} alone (30--32\% faster). 
Gains are most pronounced for intermediate team sizes (2--3 agents), indicating that \projectname effectively reduces coordination bottlenecks as team size increases.

\textbf{Effectiveness saturates gracefully with 4--5 agents.}
Increasing agent count yields large initial gains in success rate, especially for \projectname-based methods. 
From 1 to 3 agents, \textsc{ReAct}+\projectname improves success by 25\% points in AI2-THOR (0.38$\rightarrow$0.63) and 29\% points in SAR (0.38$\rightarrow$0.67), while \textsc{LLaMaR}+\projectname gains 34\% points in AI2-THOR (0.45$\rightarrow$0.79) and 36\% points in SAR (0.31$\rightarrow$0.67). 
Beyond this range, performance saturates: from 3 to 5 agents, success varies by at most 2--3 points in AI2-THOR and 1--2 points in SAR. 
Notably, \projectname saturates at higher asymptotic success (0.76--0.79 in AI2-THOR and 0.66--0.68 in SAR) than non-\projectname baselines (0.63--0.68 and 0.63--0.65). 
This saturation likely reflects the spatial scale of the evaluated environments; in larger or more complex settings, additional agents are expected to yield more sustained, near-linear gains.

\begin{table}[h]
\centering
\setlength{\tabcolsep}{4pt} 

\caption{Scalability analysis of \projectname across AI2-THOR for multiple agent counts. Subtables show (a) Success Rate, (b) Failure Rate, and (c) Runtime, demonstrating improvements in task performance and efficiency when \projectname is integrated with baseline planners.}
\label{tab:scalability-summary}

\begin{subtable}[t]{0.49\linewidth}
\centering
\caption{Success Rate as a function of Agent Count.}
\label{tab:scalability-success-rate-ai2thor}
\begin{tabular}{lccccc}
\toprule
\textbf{Agent Count} $\rightarrow$ & 1 & 2 & 3 & 4 & 5 \\
\midrule
\textsc{Act} & 0.19 & 0.42 & 0.55 & 0.46 & 0.47 \\
\textsc{CoT} & 0.02 & 0.08 & 0.14 & 0.13 & 0.13 \\
\textsc{ReAct} & 0.23 & 0.44 & 0.57 & 0.50 & 0.48 \\
\textsc{ReAct}+\projectname & 0.38 & 0.51 & 0.63 & 0.64 & 0.67 \\
\midrule
\textsc{SmartLLM} & 0.01 & 0.06 & 0.13 & 0.12 & 0.10 \\
\textsc{CoELA} & 0.02 & 0.11 & 0.18 & 0.14 & 0.12 \\
\textsc{LLaMaR} & 0.31 & 0.59 & 0.68 & 0.65 & 0.65 \\
\textsc{LLaMaR}+\projectname & 0.45 & 0.70 & 0.79 & 0.76 & 0.76 \\
\bottomrule
\end{tabular}
\end{subtable}
\hfill
\begin{subtable}[t]{0.49\linewidth}
\centering
\caption{Failure Rate as a function of Agent Count.}
\label{tab:scalability-failure-rate-ai2thor}
\begin{tabular}{lccccc}
\toprule
\textbf{Agent Count} $\rightarrow$ & 1 & 2 & 3 & 4 & 5 \\
\midrule
\textsc{Act} & 0.55 & 0.42 & 0.38 & 0.44 & 0.45 \\
\textsc{CoT} & 0.52 & 0.47 & 0.45 & 0.48 & 0.48 \\
\textsc{ReAct} & 0.48 & 0.39 & 0.35 & 0.38 & 0.40 \\
\textsc{ReAct}+\projectname & 0.42 & 0.36 & 0.28 & 0.25 & 0.22 \\
\midrule
\textsc{SmartLLM} & 0.58 & 0.51 & 0.48 & 0.50 & 0.52 \\
\textsc{CoELA} & 0.54 & 0.46 & 0.42 & 0.45 & 0.47 \\
\textsc{LLaMaR} & 0.40 & 0.32 & 0.25 & 0.28 & 0.28 \\
\textsc{LLaMaR}+\projectname & 0.28 & 0.20 & 0.12 & 0.10 & 0.08 \\
\bottomrule
\end{tabular}
\end{subtable}

\vspace{0.5cm}

\begin{subtable}[t]{0.49\linewidth}
\centering
\caption{Runtime as a function of Agent Count.}
\label{tab:scalability-runtime-ai2thor}
\begin{tabular}{lccccc}
\toprule
\textbf{Actor Count} $\rightarrow$ & 1 & 2 & 3 & 4 & 5 \\
\midrule
\textsc{Act} & 326.2 & 398.1 & 286.5 & 316.4 & 320.9 \\
\textsc{CoT} & 196.1 & 229.1 & 156.8 & 176.1 & 180.3 \\
\textsc{ReAct} & 319.6 & 367.5 & 312.5 & 324.1 & 320.6 \\
\textsc{ReAct}+\projectname & 234.5 & 342.0 & 275.9 & 241.7 & 238.4 \\ 
\midrule
\textsc{SmartLLM} & 984.8 & 935.9 & 736.3 & 783.3 & 803.9 \\
\textsc{CoELA} & 923.3 & 889.4 & 671.1 & 683.5 & 690.6 \\
\textsc{LLaMaR} & 772.2 & 757.9 & 584.5 & 612.1 & 622.6 \\
\textsc{LLaMaR} + \projectname & 538.6 & 554.1 & 424.9 & 448.7 & 455.1 \\
\bottomrule
\end{tabular}
\end{subtable}

\end{table}

Overall, \projectname ensures favorable scaling behavior: higher success rates, reduced execution time, and graceful performance saturation as agent count increases, demonstrating that coordinated planning with ASM and ILP supports both effective and runtime-efficient multi-agent operation.

\subsection{Effectiveness Trends}
\label{adx:ai2thor-sar-effectiveness}

\begin{figure*}[h]
     \centering
     \begin{subfigure}[b]{0.3\textwidth}
         \centering
         \includegraphics[width=\textwidth]{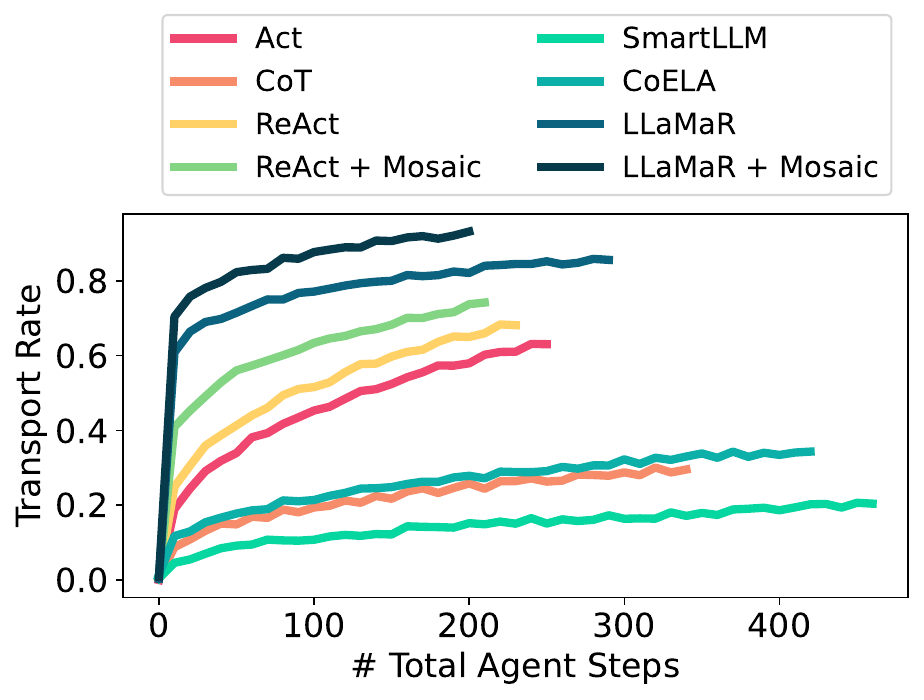}
         \caption{Transport Rate}
         \label{fig:ai2thor-transport-rate}
     \end{subfigure}
     \hfill
     \begin{subfigure}[b]{0.3\textwidth}
         \centering
         \includegraphics[width=\textwidth]{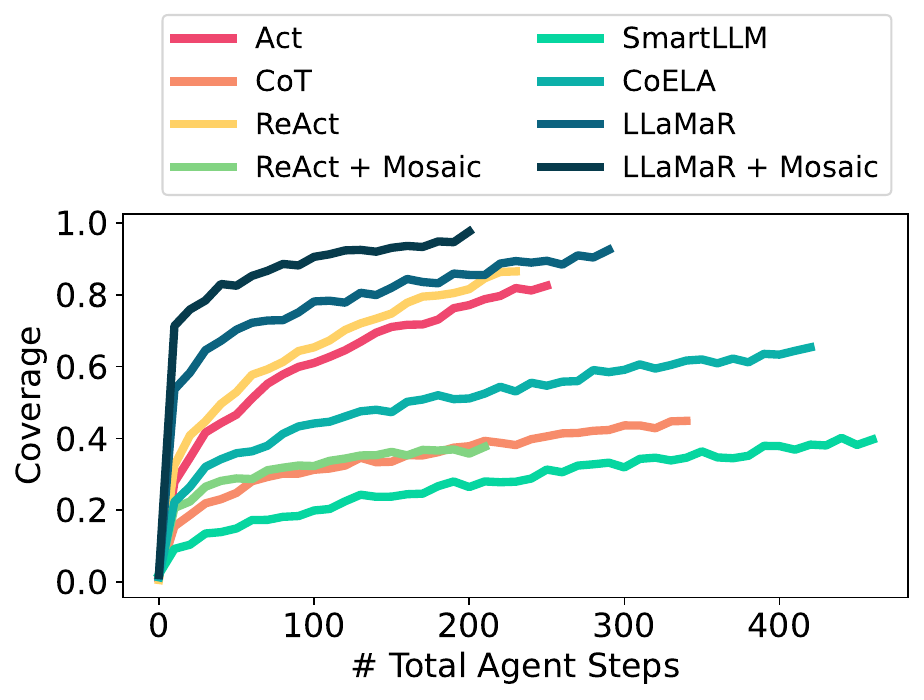}
         \caption{Coverage}
         \label{fig:ai2thor-coverage}
     \end{subfigure}
     \hfill
     \begin{subfigure}[b]{0.3\textwidth}
         \centering
         \includegraphics[width=\textwidth]{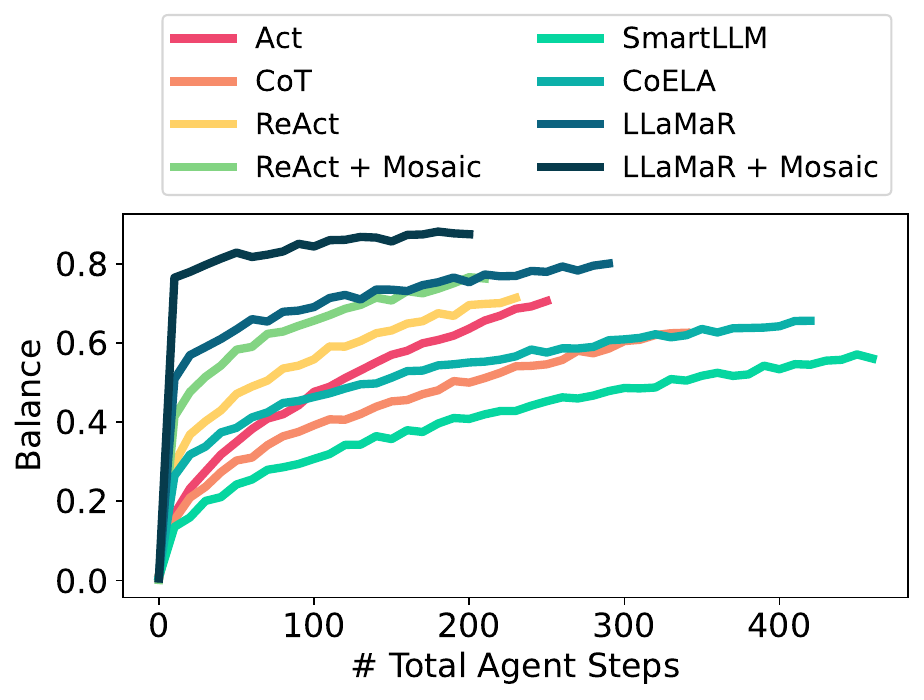}
         \caption{Balance}
         \label{fig:ai2thor-balance}
     \end{subfigure}
        \caption{Transport Rate, Coverage, and Balance vs. Number of Agent Steps in AI2-THOR.}
        \label{fig:ai2thor-plots}
\end{figure*}

\begin{figure*}[h]
     \centering
     \begin{subfigure}[b]{0.3\textwidth}
         \centering
         \includegraphics[width=\textwidth]{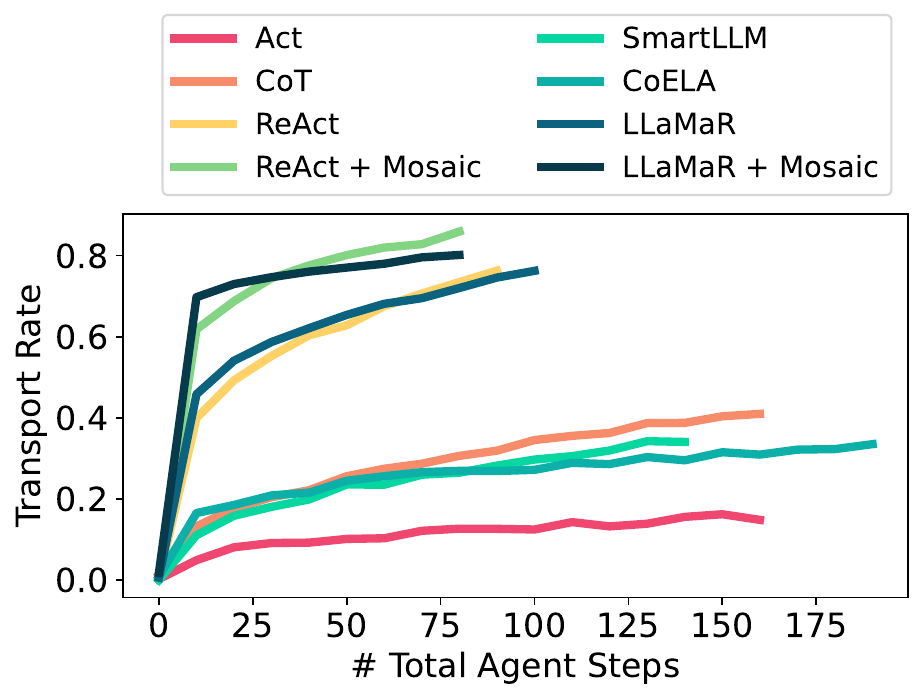}
         \caption{Transport Rate}
         \label{fig:sar-transport-rate}
     \end{subfigure}
     \hfill
     \begin{subfigure}[b]{0.3\textwidth}
         \centering
         \includegraphics[width=\textwidth]{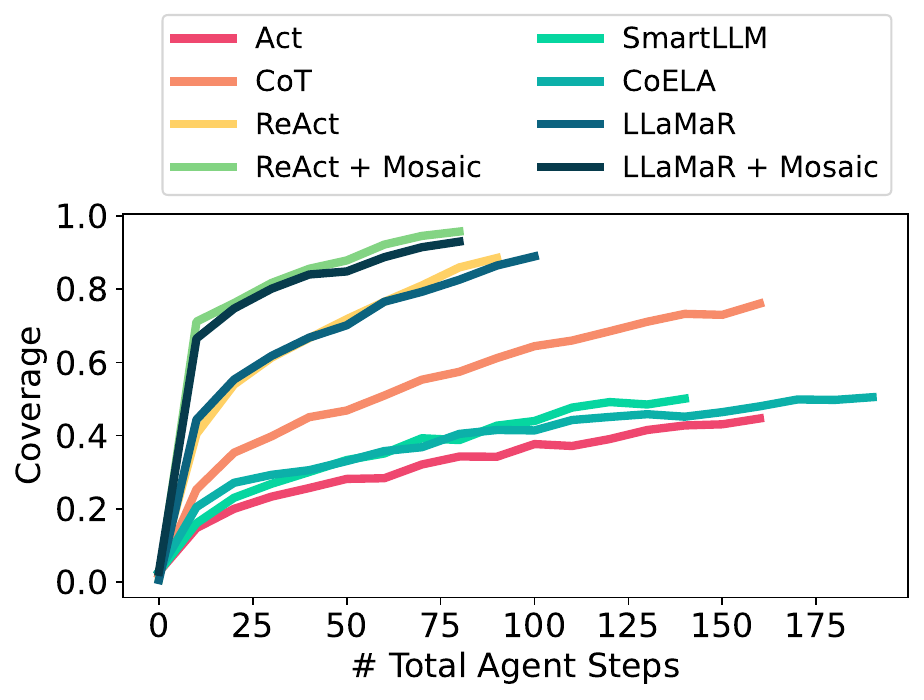}
         \caption{Coverage}
         \label{fig:sar-coverage}
     \end{subfigure}
     \hfill
     \begin{subfigure}[b]{0.3\textwidth}
         \centering
         \includegraphics[width=\textwidth]{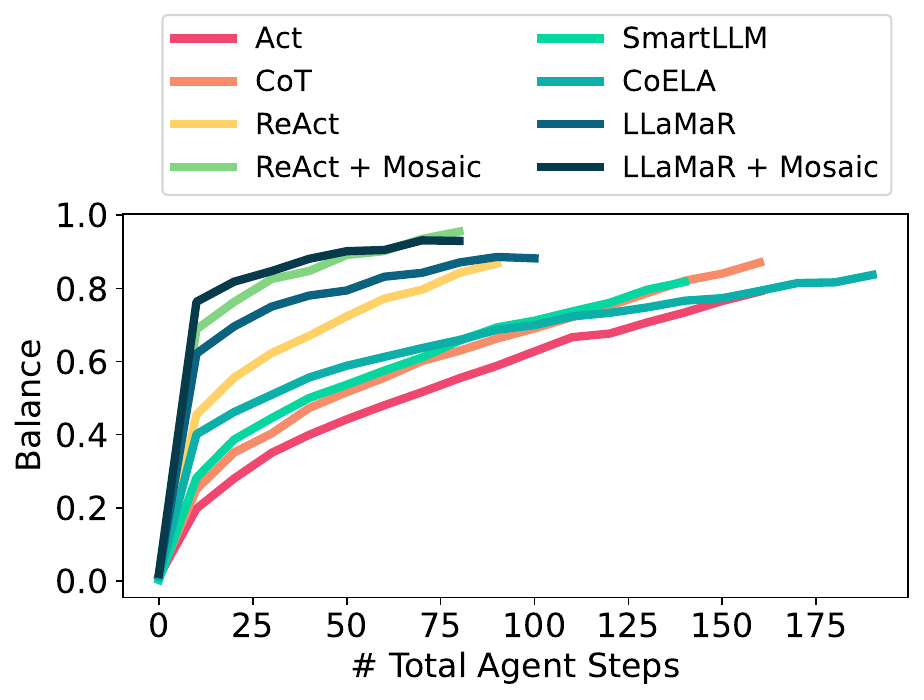}
         \caption{Balance}
         \label{fig:sar-balance}
     \end{subfigure}
        \caption{Transport Rate, Coverage, and Balance vs. Number of Agent Steps in SAR.}
        \label{fig:sar-plots}
\end{figure*}

Figures~\ref{fig:ai2thor-plots} and~\ref{fig:sar-plots} show the evolution of transport rate, balance, and coverage over planning steps for household rearrangement and search-and-rescue tasks.

\textbf{Balance correlates with transport efficiency.}
Across methods, improvements in balance closely track gains in transport rate. 
Planners that rapidly achieve high balance (such as \textsc{ReAct + Mosaic} and \textsc{LLaMaR + Mosaic}) also exhibit the fastest rise in transport rate. 
This indicates that equitable workload distribution is not merely a fairness metric, but a key enabler of effective rescue execution, allowing multiple agents to make simultaneous progress instead of interfering or idling.

\textbf{Coverage alone is insufficient without coordination.}
Coverage alone is insufficient without effective coordination like \projectname's. 
Although several methods exhibit steady increases in coverage over time, high coverage does not necessarily translate into high transport rates. 
In particular, planners without structured memory or with weak coordination mechanisms often achieve moderate-to-high coverage but remain limited in transport performance due to poor agent balance and inefficient task allocation.

In contrast, \projectname-augmented methods consistently convert coverage into successful transport by improving coordination once relevant regions are discovered. 
By enabling agents to share and reuse semantic and spatial context, \projectname promotes balanced engagement with rescue targets and reduces redundant exploration. 
This demonstrates that in AI2-THOR and SAR settings, effective coordination (rather than sheer exploration breadth) is the factor that determines whether discovered information leads to successful task completion.

\subsection{Generality across Models}
\label{subsubsec:models}

\begin{figure}[h]
     \centering
     \begin{subfigure}[b]{0.49\linewidth}
         \centering
         \includegraphics[width=\textwidth]{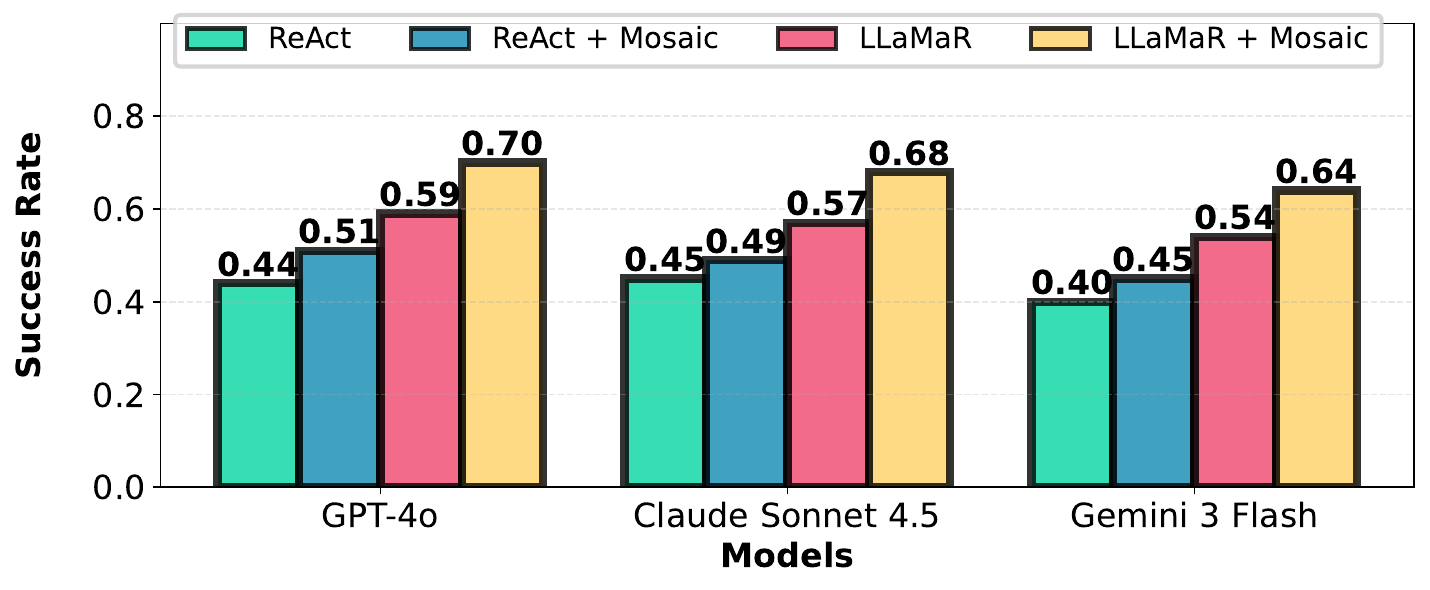}
         \caption{Success Rate}
         \label{fig:ai2thor-models-success-rate}
     \end{subfigure}
     \hfill
     \begin{subfigure}[b]{0.49\linewidth}
         \centering
         \includegraphics[width=\textwidth]{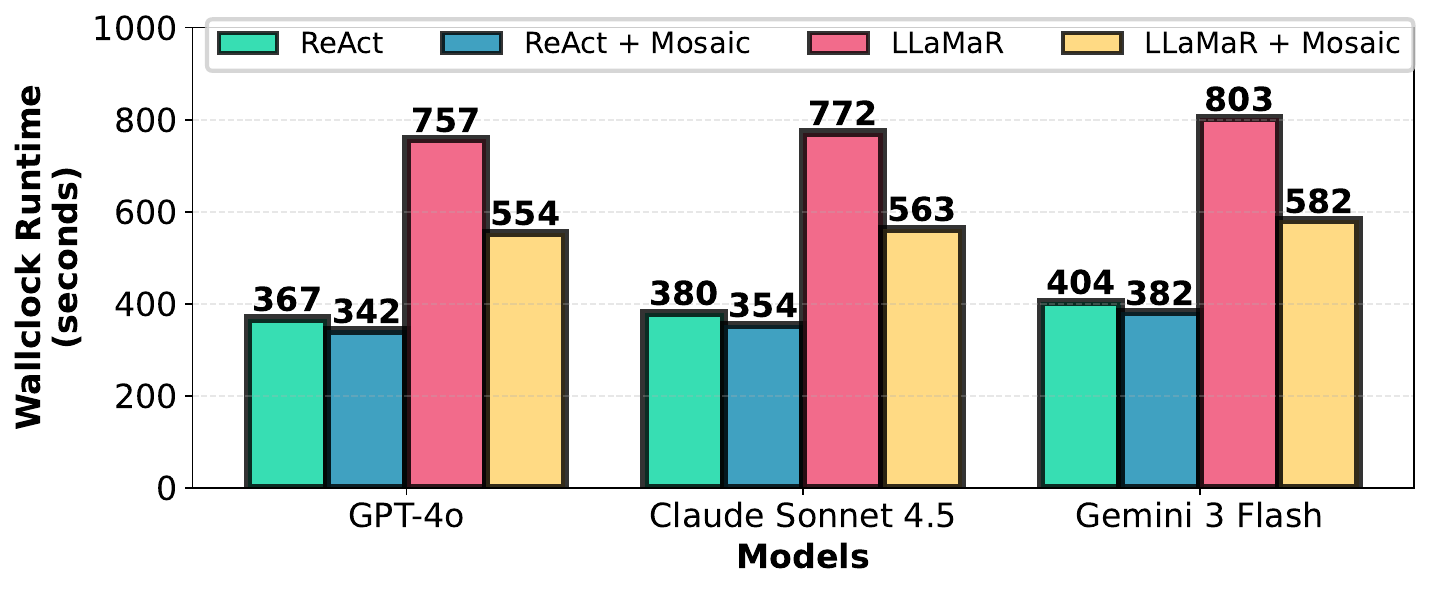}
         \caption{Runtime}
         \label{fig:ai2thor-models-wallclock-runtime}
     \end{subfigure}
    \caption{Evaluation of \projectname in AI2-THOR environment across three foundation models in terms of (a) success rate and (b) runtime. 
    Across all models and planning methods \textsc{ReAct}, \textsc{LLaMaR}, \projectname consistently improves task success while reducing runtime, demonstrating that its gains are model-agnostic and not tied to absolute model capability. }
        \label{fig:ai2thor-different-models}
\end{figure}

Figure~\ref{fig:ai2thor-different-models} reports success rate and runtime across three foundation models (GPT-4o, Claude Sonnet 4.5, and Gemini 3 Flash) under different planning methods. 
We analyze whether \projectname yields consistent improvements across models with varying success and inference cost.

\textbf{\projectname is generalizable across foundation models.}
Across all three foundation models, \projectname consistently improves success rate when paired with both \textsc{ReAct} and \textsc{LLaMaR}, indicating strong generality. 
For \textsc{ReAct}, \projectname increases success rate by +7\% points on GPT-4o (0.44~$\rightarrow$~0.51), +4\% points on Claude Sonnet 4.5 (0.45~$\rightarrow$~0.49), and +5\% points on Gemini 3 Flash (0.40~$\rightarrow$~0.45). 

When applied to \textsc{LLaMaR}, \projectname yields larger gains: +11\% points on GPT-4o (0.59~$\rightarrow$~0.70), +11\% points on Claude Sonnet 4.5 (0.57~$\rightarrow$~0.68), and +10\% points on Gemini 3 Flash (0.54~$\rightarrow$~0.64). 
Importantly, these improvements are achieved while reducing runtime in all cases (Figure~\ref{fig:ai2thor-models-wallclock-runtime}); for example, \textsc{LLaMaR} + \projectname reduces runtime by 26.9\% on GPT-4o (757.9s~$\rightarrow$~554.1s), 27.1\% on Claude Sonnet 4.5 (772s~$\rightarrow$~563s), and 27.5\% on Gemini 3 Flash (803s~$\rightarrow$~582s). 
These consistent trends suggest that \projectname acts as a planner-agnostic augmentation that generalizes across diverse foundation models.

\textbf{\projectname is robust even with weaker models.}
While absolute performance degrades as we move from GPT-4o and Claude Sonnet 4.5 to Gemini 3 Flash, the relative benefits of \projectname remain stable, demonstrating robustness under weaker models for the tasks of AI2THOR and SAR. 
For instance, baseline \textsc{ReAct} success drops by 4\% points from GPT-4o to Gemini 3 Flash (0.44 $\rightarrow$ 0.40), yet \projectname preserves comparable relative gains (+7\% points on GPT-4o vs. +5\% points Gemini 3 Flash) against the base planners. 
A similar pattern holds for \textsc{LLaMaR}, where baseline success decreases by 5\% points (0.59 $\rightarrow$ 0.54), but \projectname maintains nearly identical absolute improvements (+11\% points on GPT4o vs. +10\% points on Gemini 3 Flash).
Moreover, the ordering of methods remains invariant across all models, \textsc{ReAct} $<$ \textsc{ReAct} + \projectname $<$ \textsc{LLaMaR} $<$ \textsc{LLaMaR} + \projectname, indicating that \projectname's effectiveness is not tied to a specific foundation model's capability. 
These results suggest that \projectname improves planning robustness by mitigating execution failures and inefficiencies, rather than relying on increased model capacity.

\subsection{Performance under Observation Noise}
\label{adx:performance-under-observation-noise}

To evaluate robustness under realistic sensing imperfections, we define three noise regimes that progressively increase localization and perception errors. 
Low, medium, and high noise settings apply Gaussian position noise (0.1/0.3/0.5 grid units), rotation noise (5$^{\circ}$/10$^{\circ}$/15$^{\circ}$), and perception errors in terms of miss rate (visible objects not detected), false positive rate (non-existent objects hallucinated), and misclassification rate (objects detected with incorrect class), set to (2\%, 1\%, 1\%), (5\%, 2\%, 3\%), and (10\%, 5\%, 5\%) respectively. 
All regimes additionally include distance measurement errors of 5\%, 10\%, and 15\%, enabling a controlled evaluation of \projectname under increasingly adverse sensing conditions.

\begin{figure}[h]
    \centering
    \includegraphics[width=0.5\linewidth]{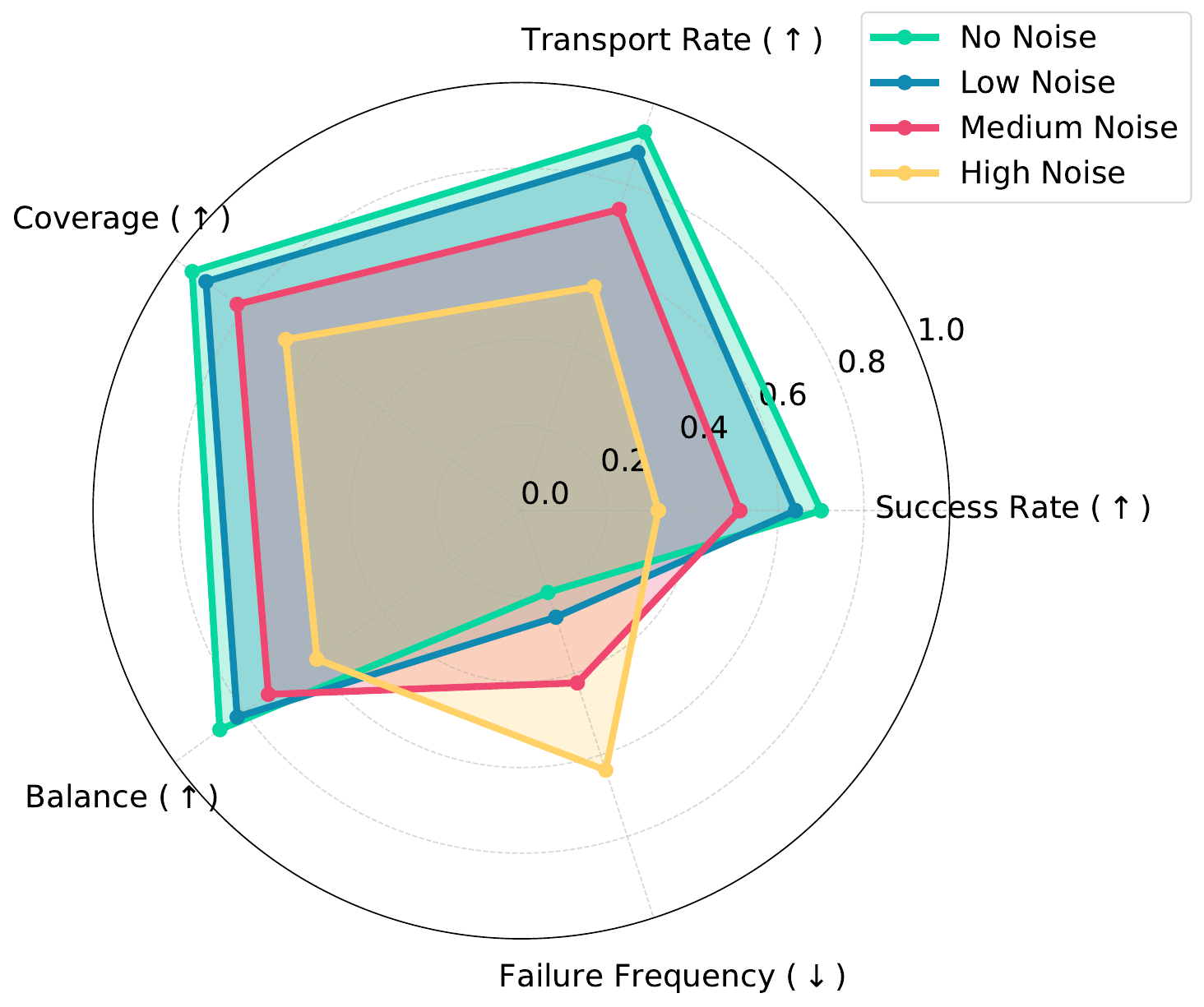}
    \caption{Key performance metrics under progressively increasing noise levels for AI2-THOR environment. 
    As noise intensifies, the system dynamically adapts by leveraging additional LLM calls and agent steps, enabling it to maintain functionality even in the high noise settings. 
    }
    \label{fig:mosaic-noise}
\end{figure}

Figure~\ref{fig:mosaic-noise} summarizes success metrics under increasing noise with \textsc{LLaMaR} + \projectname, while Table~\ref{tab:noisy-ai2thor} reports the complete set of success and efficiency metrics, providing a holistic view of performance and coordination cost.

\textbf{Increased noise triggers adaptive coordination and recovery behavior.}
As environmental noise increases, \projectname-based execution responds by allocating additional coordination and reasoning effort, as reflected across all efficiency metrics. 
From no noise to high noise, runtime increases by 88\% (554s $\rightarrow$ 1042s), alongside an 83\% increase in LLM calls (231 $\rightarrow$ 423), a 90\% increase in total agent steps (204 $\rightarrow$ 388), and a 77\% increase in total token usage (1.36M $\rightarrow$ 2.41M). 
Importantly, this additional coordination effort grows faster than the degradation in success rate (-54\%, 0.70 $\rightarrow$ 0.32), indicating that \projectname actively engages in coordination to preserve task execution under uncertainty rather than prematurely terminating or collapsing. 
This behavior suggests that higher noise elicits proactive synchronization and resilience mechanisms, allowing agents to continue coordinated execution despite increasingly unreliable dynamics.

\textbf{\projectname degrades gracefully under increasing noise.}
Through increased coordination effort, \projectname exhibits smooth performance degradation rather than catastrophic collapse.
Success rate decreases monotonically from 0.70 (no noise) to 0.64 (low), 0.51 (medium), and 0.32 (high), while intermediate task-quality metrics remain comparatively robust.
Even under high noise, \projectname maintains a transport rate of 0.55, coverage of 0.68, and balance of 0.59, compared to 0.93, 0.95, and 0.87 in the noiseless setting.
Meanwhile, failure rate increases gradually from 20.1\% $\rightarrow$ 26.2\% $\rightarrow$ 42.3\% $\rightarrow$ 63.7\%, indicating a progressive shift toward coordination breakdowns rather than abrupt loss of task engagement. 
These results suggest that \projectname preserves partial task structure and agent contribution under noise, with failures arising primarily in late-stage coordination.

\begin{table*}[h]
\centering
\setlength{\tabcolsep}{4pt} 
\caption{
Robustness of \textsc{LLaMaR} + \projectname under increasing sensor noise in AI2-THOR with 2 agents.
As sensing noise increases, \projectname degrades gracefully across all success metrics while adaptively increasing coordination effort, reflected in higher runtime, LLM calls, and agent steps. 
}
\label{tab:noisy-ai2thor}
\footnotesize
\begin{tabular}{lccccccccc}
\toprule
\textbf{\begin{tabular}[c]{@{}c@{}}\textsc{LLaMaR}\\+ \projectname\end{tabular}} & 
\textbf{\begin{tabular}[c]{@{}c@{}}Success\\ Rate ($\uparrow$)\end{tabular}} & 
\textbf{\begin{tabular}[c]{@{}c@{}}Transport\\ Rate ($\uparrow$)\end{tabular}} & 
\textbf{\begin{tabular}[c]{@{}c@{}}Coverage\\ ($\uparrow$)\end{tabular}} & 
\textbf{\begin{tabular}[c]{@{}c@{}}Balance\\ ($\uparrow$)\end{tabular}} & 
\textbf{\begin{tabular}[c]{@{}c@{}}Failure\\ Rate ($\downarrow$)\end{tabular}} & 
\textbf{\begin{tabular}[c]{@{}c@{}} Runtime\\ (Seconds, $\downarrow$)\end{tabular}} & 
\textbf{\begin{tabular}[c]{@{}c@{}}\# of LLM\\ Calls ($\downarrow$)\end{tabular}} & 
\textbf{\begin{tabular}[c]{@{}c@{}}\# of Agent\\ Steps ($\downarrow$)\end{tabular}} & 
\textbf{\begin{tabular}[c]{@{}c@{}}\# of Tokens\\ ($\times 10^3$, $\downarrow$)\end{tabular}} \\
\midrule
No Noise & 0.70 & 0.93 & 0.95 & 0.87 & 0.20 & \phantom{0}554.1 & 231.5 & 1364 & 204.1 \\
Low Noise & 0.64 & 0.88 & 0.91 & 0.82 & 0.26 & \phantom{0}612.5 & 258.1 & 1492 & 238.6 \\
Medium Noise & 0.51 & 0.74 & 0.82 & 0.73 & 0.42 & \phantom{0}745.3 & 314.9 & 1825 & 295.6 \\
High Noise & 0.32 & 0.55 & 0.68 & 0.59 & 0.64 & 1042.1 & 422.6 & 2410 & 388.1 \\
\bottomrule
\end{tabular}
\end{table*}

\subsection{Ablation Studies}
\label{subsec:ablation}

We analyze how ILP constraints and cost minimization contribute to performance, study the sensitivity of cost-function hyperparameters, and examine the runtime dynamics of the cost function.

\subsubsection{Role of Constraints and Cost Function in ILP}
\label{subsec:role-of-ILP}

Table~\ref{tab:main-results-ai2thor} shows that in the single-call setting, \textsc{ReAct}+\textsc{ASM} and \textsc{ReAct}+\textsc{ILP} achieve identical failure rates of 0.37 on AI2-THOR, yet differ in runtime (371s vs. 356s) and agent steps (230 vs. 211). 
This gap arises because the ILP framework improves efficiency beyond failure avoidance by optimizing action assignment and penalizing redundant behaviors, resulting in shorter execution trajectories even when failure rates are comparable.
These effects are amplified in the multi-call setting. 
For \textsc{LLaMaR}, adding ASM reduces the failure rate by 3\% points, yielding runtime reduction of 4.5\% and agent steps reduction of 3.9\%.
In contrast, adding ILP further lowers the failure rate by 8\% points and produces substantially larger efficiency improvements, reducing runtime  by 17.1\% and reducing agent steps by 18.8\%.
This demonstrates that coordinated action assignment and runtime-aware optimization are critical for translating improved feasibility into significant efficiency gains, particularly in multi-call planners due to the task decomposition.

To understand the effect of ILP further, Table \ref{tab:ablation-results-ai2thor} presents an ablation study on AI2-THOR that isolates the contributions of individual components in \projectname relative to the strongest baseline, \textsc{LLaMaR}.
Starting from \textsc{LLaMaR}, this table focuses on isolating the contributions of the ILP components by adding (i) the \emph{Feasibility Checker (FC)} only and (ii) \emph{Cost Function Minimization (CFM)} only.
For completeness and ease of comparison, we also report results for +ASM, full +ILP (FC + CFM), and the complete \projectname (ASM + ILP), which are analyzed in more detail in a separate table.

To construct the ablations, FC-only enforces all feasibility and coordination constraints but selects arbitrarily among feasible joint actions without optimization, while CFM-only optimizes the cost function but does not enforce hard feasibility constraints beyond basic action validity.
The full ILP combines both FC and CFM, while \projectname further augments ILP with semantic memory.

\begin{table*}[h]
\centering
\setlength{\tabcolsep}{4pt} 
\caption{Comparative evaluation of various components of \projectname against best -performing baseline method \textsc{LLaMaR} across key effectiveness and efficiency metrics in {AI2-THOR} environment.
\projectname integrates \emph{Agent-centric Semantic Memory (ASM)} with an \emph{Integer Linear Programming (ILP)} framework for coordinated multi-agent planning.
Within the ILP, the \emph{Feasibility Checker (FC)} enforces action validity and coordination constraints, while \emph{Cost Function Minimization (CFM)} optimizes among feasible joint actions.
Together, FC and CFM constitute the ILP module that regulates and optimizes multi-agent action selection.
Arrows ($\uparrow$/$\downarrow$) indicate whether higher or lower values are better. 
The numbers in subscript indicate the change relative to \textsc{LLaMaR}. 
For the first five columns (effectiveness metrics), subscript is the absolute difference, while for the last four columns (efficiency metrics) subscript is percent difference. 
Positive improvements are highlighted in \textcolor{Green}{green}, and drops are highlighted in \textcolor{red}{red}.
{Best results are highlighted in bold}.}
\label{tab:ablation-results-ai2thor}
\scriptsize
\begin{tabular}{lccccccccc}
\toprule & 
\multicolumn{5}{c}{Effectiveness Metrics} & \multicolumn{4}{c}{Efficiency Metrics} \\
\cmidrule(lr){2-6} \cmidrule(lr){7-10}
\textbf{Method} & 
\textbf{\begin{tabular}[c]{@{}c@{}}Success\\ Rate ($\uparrow$)\end{tabular}} & 
\textbf{\begin{tabular}[c]{@{}c@{}}Transport\\ Rate ($\uparrow$)\end{tabular}} & 
\textbf{\begin{tabular}[c]{@{}c@{}}Coverage\\ ($\uparrow$)\end{tabular}} & 
\textbf{\begin{tabular}[c]{@{}c@{}}Balance\\ ($\uparrow$)\end{tabular}} & 
\textbf{\begin{tabular}[c]{@{}c@{}}Failure\\ Rate ($\downarrow$)\end{tabular}} & 
\textbf{\begin{tabular}[c]{@{}c@{}} Runtime\\ (Seconds, $\downarrow$)\end{tabular}} & 
\textbf{\begin{tabular}[c]{@{}c@{}}\# of LLM\\ Calls ($\downarrow$)\end{tabular}} & 
\textbf{\begin{tabular}[c]{@{}c@{}}\# of Agent\\ Steps ($\downarrow$)\end{tabular}} & 
\textbf{\begin{tabular}[c]{@{}c@{}}\# of Tokens\\ ($\times 10^3$, $\downarrow$)\end{tabular}} \\
\midrule
\textsc{LLaMaR}     & 0.59 & 0.85 & 0.90 & 0.79 & 0.32 & 757.9 & 346.8 & 295.8 & 1472 \\
$\drsh$ + ASM & 0.61$_{\positive{+0.02}}$ & 0.87$_{\positive{+0.02}}$ & 0.91$_{\positive{+0.01}}$ & 0.82$_{\positive{+0.03}}$ & 0.29$_{\positive{-0.03}}$ &
723.8$_{\positive{-\phantom{0}4.5\%}}$ 
& 310.2$_{\positive{-10.6\%}}$ 
& 284.2$_{\positive{-\phantom{0}3.9\%}}$ 
& 1562$_{\negative{+6.1\%}}$\\
$\drsh$ + FC & 0.63 $_{\positive{+0.04}}$ & 0.88 $_{\positive{+0.03}}$ & 0.91 $_{\positive{+0.01}}$ & 0.84 $_{\positive{+0.05}}$ & 0.29 $_{\positive{-0.03}}$ & 702.0 $_{\positive{-\phantom{0}7.4\%}}$ & 285.3 $_{\positive{-17.7\%}}$ & 276.4 $_{\positive{-\phantom{0}6.6\%}}$ & 1475 $_{\negative{+0.2\%}}$ \\
$\drsh$ + CFM & 0.65$_{\positive{+0.06}}$ & 0.89 $_{\positive{+0.04}}$ & 0.92$_{\positive{+0.02}}$ & 0.84$_{\positive{+0.05}}$ & 0.26$_{\positive{-0.06}}$ & 665.1$_{\positive{-12.2\%}}$ &	271.8$_{\positive{-21.6\%}}$ & 258.3 $_{\positive{-12.7\%}}$ & 1447 $_{\positive{-1.7\%}}$\\
$\drsh$ + ILP & 0.67$_{\positive{+0.08}}$ & 0.90$_{\positive{+0.05}}$ & 0.93$_{\positive{+0.03}}$ & 0.85$_{\positive{+0.06}}$ & 0.24$_{\positive{-0.08}}$ & 
628.1$_{\positive{-17.1\%}}$ 
& 258.4$_{\positive{-25.5\%}}$ 
& 240.3$_{\positive{-18.8\%}}$ 
& 1420$_{\positive{-3.5\%}}$\\
$\drsh$ + \textsc{\projectname} & \textbf{0.69}$_{\positive{+0.10}}$ & \textbf{0.92}$_{\positive{+0.07}}$ & \textbf{0.95}$_{\positive{+0.05}}$ & \textbf{0.87}$_{\positive{+0.08}}$ & \textbf{0.20}$_{\positive{-0.12}}$ & \textbf{554.1}$_{\positive{-26.9\%}}$ 
& \textbf{231.5}$_{\positive{-33.3\%}}$ 
& \textbf{204.1}$_{\positive{-31.0\%}}$ 
& \textbf{1364}$_{\positive{-7.3\%}}$ \\

\bottomrule
\end{tabular}
\end{table*}

\textbf{Feasibility constraints primarily improve coordination and balance.}
Adding FC alone yields consistent gains in coordination-related metrics, highlighting the importance of enforcing joint-action feasibility.
Relative to \textsc{LLaMaR}, FC improves success and transport rates by 4\% and 3\% points, respectively, while increasing balance by 5\% points and reducing failure rate by 3\% points.
These gains translate into efficiency improvements, including a 17.7\% reduction in LLM calls and a 6.6\% reduction in agent steps, despite minimal change in token usage (+0.2\%).
This indicates that enforcing feasibility constraints alone already prevents conflicting or redundant actions, leading to better workload distribution and fewer coordination-induced failures.

\textbf{Cost-aware optimization drives efficiency and amplifies effectiveness.}
CFM-only further improves both effectiveness and efficiency by explicitly favoring low-cost joint actions among feasible candidates.
Compared to \textsc{LLaMaR}, CFM achieves larger gains than FC in success (6\% points), transport (4\% points), and coverage (2\% points), while reducing failure rate by 6\% points.
Efficiency improvements are also more pronounced, with 21.6\% fewer LLM calls, 12.7\% fewer agent steps, and a 12.2\% reduction in runtime.
These results show that cost-aware selection meaningfully differentiates among feasible joint actions, suppressing inefficient behaviors such as unnecessary navigation or repeated failed attempts, which feasibility constraints alone cannot distinguish.

\textbf{ILP and ASM deliver complementary benefits.}
Combining FC and CFM into the full ILP further amplifies these trends, yielding up to 8\% points in success and 17.1\% runtime reduction.
Finally, integrating ASM (\projectname) produces the strongest gains across all metrics, including a 10\% point increase in success, 7\% point increase in transport, and a 26.9\% runtime reduction, while also lowering token usage by 7.3\%.

Together, these results show that FC ensures valid coordination, CFM optimizes efficiency within the feasible space, and ASM improves global context, with each component contributing distinct and complementary benefits to multi-agent planning performance.

\subsubsection{Spatial and Temporal Penalties and their Hyperparameter Sensitivity}
\label{subsec:hyperparameter-sensitivity}

Table~\ref{tab:temporal-patterns} details the spatial and temporal inefficiency patterns used in the penalty term of Equation~\ref{eq:penal-cost-function}, along with their corresponding penalty definitions and weights. 
These patterns formalize common failure modes such as repetition, oscillation, regression, and stagnation, enabling structured penalization of inefficient behavior during planning.

\begin{table}[h]
\footnotesize
\setlength{\tabcolsep}{3pt} 
\begin{center}
\caption{Spatial and temporal inefficiency patterns used in the penalty component of the ILP cost function (Equation~\ref{eq:penal-cost-function}). 
Each pattern captures a type of redundant or inefficient behavior, and is weighted by a penalty $\lambda_p$ in Equation~\ref{eq:penal-cost-function}.}
\label{tab:temporal-patterns}
\begin{tabular}{llcc}
\toprule
\textbf{Pattern $p$} & \textbf{Description} & \textsf{Penalty}$_p$ & $\lambda_p$ \\
\midrule
\textsc{Cyclic} & \begin{tabular}[c]{@{}l@{}}Agent repeats a sequence of length $\ell$, $n_{\textsf{rep}}$ times\end{tabular} & $\ell \cdot n_{\textsf{rep}}$ & 2 \\
\rowcolor{gray!10}
\textsc{Failure} & \begin{tabular}[c]{@{}l@{}}Consecutive unsuccessful attempts of the same action\end{tabular} & $n_{\textsf{fail}}$ & 2 \\
\textsc{Oscillation} & \begin{tabular}[c]{@{}l@{}}Back-and-forth movement between locations\end{tabular} & $n_{\textsf{osc}}$ & 1 \\
\rowcolor{gray!10}
\textsc{Backtracking} & \begin{tabular}[c]{@{}l@{}}Spatial regressions or returning to previous positions\end{tabular} & $n_{\textsf{back}}$ & 1 \\
\textsc{Stagnation} & \begin{tabular}[c]{@{}l@{}}Consecutive idle or no-progress steps\end{tabular} & $n_{\textsf{idle}}$ & 2 \\
\bottomrule
\end{tabular}
\end{center}
\end{table}

\begin{table}[h]
\centering
\footnotesize
\caption{
Ablation study on the ILP cost function hyperparameters (\(\lambda_t\)) corresponding to different spatial–temporal inefficiency patterns (Table~\ref{tab:temporal-patterns}): Cyclic, Failure, Oscillation, Backtracking, and Stagnation.
Each planning configuration explores a different weighting strategy, illustrating trade-offs between task success rate and runtime in the AI2-THOR environment. 
The \textbf{Efficiency-Balanced} configuration serves as the baseline, achieving the best overall trade-off between effectiveness and efficiency.
}
\label{tab:cost-function-hyperparameters}
\begin{tabular}{llll}
\toprule
\begin{tabular}[c]{@{}l@{}}\textbf{Planning}\\ \textbf{Configuration}\end{tabular} & \textbf{Hyperparameters} & \begin{tabular}[c]{@{}l@{}}\textbf{Success}\\ \textbf{Rate ($\uparrow$)}\end{tabular} & \begin{tabular}[c]{@{}l@{}} \textbf{Runtime}\\\textbf{(s, $\downarrow$)}\end{tabular} \\\midrule
Efficiency-balanced & (2, 2, 1, 1, 2) & 0.69 & 554.1 \\
\rowcolor{gray!10} Aggressive temporal  suppression  & (3, 3, 2, 1, 1) & 0.65 & 575.8 \\
Spatial inefficiency-aware & (2, 2, 2, 2, 2) & 0.59 & 726.4 \\
\rowcolor{gray!10} Load-tolerant & (1, 1, 1, 1, 1) & 0.54 & 751.9 \\
Exploration-sensitive & (2, 2, 3, 1, 1) & 0.60 & 720.3 \\
\rowcolor{gray!10} Failure-dominant & (2, 4, 2, 1, 1) & 0.67 & 550.8 \\
\bottomrule
\end{tabular}
\end{table}

Building on these pattern definitions, Table~\ref{tab:cost-function-hyperparameters} presents an ablation study on the ILP cost function hyperparameters, highlighting how different weighting strategies for spatial-temporal inefficiency patterns affect planning performance. 
The \textbf{Efficiency-Balanced configuration achieves the highest success rate} while maintaining low runtime, demonstrating that a moderate, balanced weighting effectively filters redundant actions without over-constraining agents.
Configurations that heavily prioritize temporal penalties, such as Aggressive Temporal Suppression, reduce early inefficiencies but slightly increase runtime due to load imbalance, whereas Spatial Inefficiency-Aware and Load-Tolerant settings emphasize exploration or fairness at the cost of efficiency. 
The Failure-Dominant variant shows that emphasizing execution failures can match baseline success while slightly improving runtime. 
Overall, these results indicate that careful tuning of pattern-level penalties enables the ILP to balance feasibility, efficiency, and workload distribution, reinforcing the robustness of the proposed formulation.

\subsubsection{Cost Function Dynamics}
\label{subsec:cost-function-dynamics}

\begin{figure}[h]
    \centering
    \includegraphics[width=0.5\linewidth]{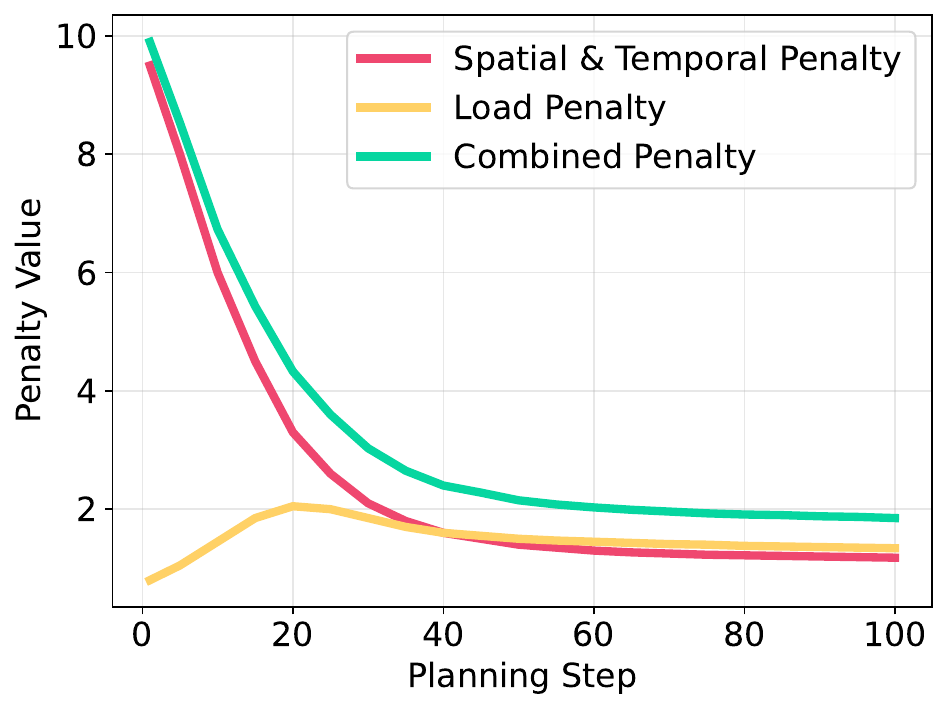}
    \caption{Evolution of ILP cost components over planning steps. 
    The spatial-temporal penalty decreases sharply early on, reflecting rapid elimination of inefficient actions, while the load penalty rises and falls, signaling emergent workload imbalance. 
    ILP cost function effectively prunes unproductive behaviors while promoting stable and balanced multi-agent coordination under partial observability of AI2-THOR.}
    \label{fig:cost-function-dynamics}
\end{figure}

Figure~\ref{fig:cost-function-dynamics} illustrates the evolution of the spatial-temporal penalty, load penalty, and their weighted combination over planning steps; we analyze these trends to understand how the ILP cost function formulation (Equation~\ref{eq:cost-function}) shapes coordination behavior over time.

\textbf{Early cost reduction is dominated by the removal of inefficient actions.}
The combined ILP cost decreases sharply in the early planning stage, dropping from 9.90 at step \#1 to 3.03 by step \#30, accounting for over 65\% of the total reduction observed by step \#100. 
This rapid decrease is primarily driven by the spatial–temporal penalty, which falls from 9.5 to 2.1 over the same interval. 
Such behavior indicates that the cost function is most effective at filtering out clearly inefficient joint actions (such as repeated failures, oscillations, and backtracking) early in the planning. 
The cost function formulation quickly prunes unproductive behaviors, which directly explains the substantial reductions in agent steps and execution failures observed in Table~\ref{tab:main-results-ai2thor}. 
This early pruning effect highlights the cost function's role as a proactive action regulator rather than a purely reactive correction mechanism.

\textbf{Load imbalance emerges as a meaningful coordination signal rather than an initial error.}
In contrast to the spatial–temporal penalty, the load penalty exhibits a non-monotonic trajectory, increasing from 0.80 at step \#1 to a peak of 2.05 at step \#19 before gradually decreasing to 1.34 by step \#100. 
The delayed increase and gradual correction of the load penalty demonstrate that imbalance is treated as an emergent signal that becomes meaningful only after sufficient execution history is available. 
By avoiding premature penalization, the ILP allows necessary action assignments early on while still promoting balanced utilization over longer horizons, consistent with the steady improvements in balance metrics reported in Table~\ref{tab:main-results-ai2thor}.

\textbf{Smooth late-stage decay indicates stable and well-regularized coordination.}
After approximately 40 planning steps, all cost components exhibit a slow and smooth decay, with the combined cost decreasing modestly from 2.40 at step \#40 to 1.85 at step \#100. 
This behavior suggests that the history-based penalties do not induce overcorrection or instability, but instead act as a stabilizing regularizer once agents converge to productive behaviors.
Importantly, the cost converges to a low but non-zero value, capturing unavoidable coordination overhead under partial observability and shared resources, rather than indicating failure to optimize.
The stability of the late-stage cost aligns with the observed gains in success and coverage, indicating that efficiency improvements do not come at the expense of task completion.

\subsubsection{Hyperparameter Selection for Action Candidates}
\label{subsec:hyperparameter-action-candidates}

Table~\ref{tab:action-candidate-counts} analyze the impact of action candidate selection on planning effectiveness and efficiency. 
The results show that using three action candidates per agent consistently achieves the best trade-off, maximizing task success, coverage, and coordination balance while substantially reducing runtime and LLM overhead. 
Increasing the candidate count beyond this point yields diminishing returns, with negligible effectiveness gains but no further efficiency improvements. 
Moreover, allocating candidates per agent significantly outperforms an equivalent global candidate budget, as shown in Table~\ref{tab:per-agent-action-ablation}, highlighting that per-agent candidates suggest higher quality actions based on each agent's local context.

\begin{table}[h]
\scriptsize
\centering
\caption{Effect of action candidate count per agent on effectiveness and efficiency in the AI2-THOR environment using \textsc{LLaMA} + \projectname. 
Increasing the candidate set improves both task success and coordination up to 3 candidates per agent, which emerges as the sweet spot.
With 3 action candidates per agent, we achieve the best overall effectiveness while substantially reducing runtime, LLM calls, agent steps, and token usage. 
Further increases yield diminishing returns, offering marginal gains in effectiveness with no meaningful efficiency improvements.}
\label{tab:action-candidate-counts}
\begin{tabular}{lccccccccc}
\toprule &
\multicolumn{5}{c}{Effectiveness Metrics} & \multicolumn{4}{c}{Efficiency Metrics} \\
\cmidrule(lr){2-6} \cmidrule(lr){7-10} 
\textbf{\begin{tabular}[l]{@{}l@{}}Action Candidate\\per Agent\end{tabular}} & \textbf{\begin{tabular}[c]{@{}c@{}}Success\\ Rate ($\uparrow$)\end{tabular}} & 
\textbf{\begin{tabular}[c]{@{}c@{}}Transport\\ Rate ($\uparrow$) \end{tabular}} & 
\textbf{\begin{tabular}[c]{@{}c@{}}Coverage\\($\uparrow$)\\ \end{tabular}} & 
\textbf{\begin{tabular}[c]{@{}c@{}}Balance\\($\uparrow$)\\\end{tabular}} & 
\textbf{\begin{tabular}[c]{@{}c@{}}Failure\\ Rate ($\downarrow$) \end{tabular}} & 
\textbf{\begin{tabular}[c]{@{}c@{}} Runtime\\ (Seconds, $\downarrow$)\end{tabular}} & 
\textbf{\begin{tabular}[c]{@{}c@{}}\# of LLM\\ Calls ($\downarrow$)\end{tabular}} & 
\textbf{\begin{tabular}[c]{@{}c@{}}\# of Agent\\ Steps ($\downarrow$) \end{tabular}} & 
\textbf{\begin{tabular}[c]{@{}c@{}}\# of Tokens\\ ($\times 10^3$, $\downarrow$)\end{tabular}} \\ \midrule
1 & 0.62 & 0.84 & 0.86 & 0.78 & 0.29 & 705.4 & 284.2 & 275.2 & 1556 \\
2 & 0.64 & 0.87 & 0.89 & 0.84 & 0.24 & 657.2 & 267.7 & 286.9 & 1425 \\
\rowcolor{gray!10} 3 & 0.69 & 0.92 & 0.95 & 0.87 & 0.20 & 554.1 & 231.5 & 204.1 & 1364 \\
4 & 0.68 & 0.91 & 0.92 & 0.88 & 0.21 & 551.6 & 234.9 & 211.2 & 1355 \\
5 & 0.67 & 0.91 & 0.94 & 0.85 & 0.20 & 552.5 & 229.3 & 201.5 & 1339 \\
\bottomrule
\end{tabular}
\end{table}

\begin{table}[h]
\scriptsize
\centering
\caption{Comparison of per-agent versus global action candidate selection in the AI2-THOR environment using \textsc{LLaMA} + \projectname. 
Allocating 3 action candidates per agent significantly outperforms an equivalent global budget of nine candidates in both effectiveness and efficiency. 
This highlights the importance of  per-agent candidate allocation over a larger unstructured candidate pool which doesn't take into account the agent's local context.}
\label{tab:per-agent-action-ablation}
\begin{tabular}{lccccccccc}
\toprule &
\multicolumn{5}{c}{Effectiveness Metrics} & \multicolumn{4}{c}{Efficiency Metrics} \\
\cmidrule(lr){2-6} \cmidrule(lr){7-10} 
\textbf{\begin{tabular}[l]{@{}l@{}}Action Candidate\\per Agent\end{tabular}} & \textbf{\begin{tabular}[c]{@{}c@{}}Success\\ Rate ($\uparrow$)\end{tabular}} & 
\textbf{\begin{tabular}[c]{@{}c@{}}Transport\\ Rate ($\uparrow$) \end{tabular}} & 
\textbf{\begin{tabular}[c]{@{}c@{}}Coverage\\($\uparrow$)\\ \end{tabular}} & 
\textbf{\begin{tabular}[c]{@{}c@{}}Balance\\($\uparrow$)\\\end{tabular}} & 
\textbf{\begin{tabular}[c]{@{}c@{}}Failure\\ Rate ($\downarrow$) \end{tabular}} & 
\textbf{\begin{tabular}[c]{@{}c@{}} Runtime\\ (Seconds, $\downarrow$)\end{tabular}} & 
\textbf{\begin{tabular}[c]{@{}c@{}}\# of LLM\\ Calls ($\downarrow$)\end{tabular}} & 
\textbf{\begin{tabular}[c]{@{}c@{}}\# of Agent\\ Steps ($\downarrow$) \end{tabular}} & 
\textbf{\begin{tabular}[c]{@{}c@{}}\# of Tokens\\ ($\times 10^3$, $\downarrow$)\end{tabular}} \\ \midrule
\rowcolor{gray!10} Per-Agent Candidates (3 per agent) & 0.69 & 0.92 & 0.95 & 0.87 & 0.20 & 554.1 & 231.5 & 204.1 & 1364 \\
Global Candidate Pool (9 total) & 0.54 & 0.72 & 0.76 & 0.71 & 0.36 & 898.7 & 394.6 & 384.2 & 1787 \\
\bottomrule
\end{tabular}
\end{table}

\subsection{Planning Step Budget}
\label{subsec:planning-step-budget}

Table~\ref{tab:planning-step-budget} summarizes the trade-offs between effectiveness and efficiency as the planning step budget increases.

\textbf{A planning budget of 100 steps emerges as a sweet spot, offering strong effectiveness gains without disproportionate efficiency costs}.
Across all methods, increasing the budget from 50 to 100 steps leads to substantial improvements in success rate by +0.21--0.25, transport rate by +0.20--0.26, coverage by +0.09--0.15, and balance by +0.14--0.17, alongside a notable reduction in failure rate of -0.13--0.19. 
These gains indicate that 100 steps provide sufficient planning horizon for resolving most coordination and execution challenges. 
Beyond this point, improvements become marginal: 
moving from 100 to 150 or 200 steps yields only incremental effectiveness gains (i.e., success rate improves only by 0.04--0.05\% points from 100 to 150 planning steps and only by 0.01--0.02\% points from 150 to 200 planning steps), suggesting diminishing returns.

\begin{table}[h]
\scriptsize
\centering
\setlength{\tabcolsep}{3pt} 
\caption{Performance metrics for various planning step budgets. 
While increasing the planning step limit initially boosts success rates and environmental coverage, the gains show diminishing returns and plateauing efficiency beyond a budget of 100 steps.}
\label{tab:planning-step-budget}
\begin{tabular}{llccccccccc}
\toprule & &
\multicolumn{5}{c}{Effectiveness Metrics} & \multicolumn{4}{c}{Efficiency Metrics} \\
\cmidrule(lr){3-7} \cmidrule(lr){8-11} 
\textbf{\begin{tabular}[l]{@{}l@{}}Planning Step\\Budget\end{tabular}} & \textbf{2 Agent} & \textbf{\begin{tabular}[c]{@{}c@{}}Success\\ Rate ($\uparrow$)\end{tabular}} & 
\textbf{\begin{tabular}[c]{@{}c@{}}Transport\\ Rate ($\uparrow$) \end{tabular}} & 
\textbf{\begin{tabular}[c]{@{}c@{}}Coverage\\($\uparrow$) \end{tabular}} & 
\textbf{\begin{tabular}[c]{@{}c@{}}Balance\\($\uparrow$)\end{tabular}} & 
\textbf{\begin{tabular}[c]{@{}c@{}}Failure\\ Rate ($\downarrow$) \end{tabular}} & 
\textbf{\begin{tabular}[c]{@{}c@{}} Runtime\\ (Seconds, $\downarrow$)\end{tabular}} & 
\textbf{\begin{tabular}[c]{@{}c@{}}\# of LLM\\ Calls ($\downarrow$)\end{tabular}} & 
\textbf{\begin{tabular}[c]{@{}c@{}}\# of Agent\\ Steps ($\downarrow$) \end{tabular}} & 
\textbf{\begin{tabular}[c]{@{}c@{}}\# of Tokens\\ ($\times 10^3$, $\downarrow$)\end{tabular}} \\ \midrule
\multirow{4}{*}{50} & ReAct & 0.23 & 0.42 & 0.71 & 0.54 & 0.58 & 235.9 & \phantom{0}85.0 & 150.3 & \phantom{0}412.9 \\
 & $\drsh$ + \projectname & 0.30 & 0.51 & 0.76 & 0.59 & 0.52 & 222.3 & \phantom{0}79.1 & 140.5 & \phantom{0}470.6 \\
 & LLaMaR & 0.35 & 0.61 & 0.81 & 0.65 & 0.45 & 492.6 & 225.4 & 192.3 & \phantom{0}956.8 \\
 & $\drsh$ + \projectname & 0.44 & 0.72 & 0.86 & 0.72 & 0.38 & 360.2 & 150.5 & 132.7 & \phantom{0}886.6 \\ \midrule
\multirow{4}{*}{100} & ReAct & 0.44 & 0.68 & 0.86 & 0.71 & 0.39 & 367.5 & 135.4 & 235.4 & \phantom{0}649.6 \\
 & $\drsh$ + \projectname & 0.51 & 0.73 & 0.90 & 0.76 & 0.36 & 342.2 & 121.7 & 216.1 & \phantom{0}724.8 \\
 & LLaMaR & 0.59 & 0.85 & 0.92 & 0.79 & 0.32 & 757.9 & 346.8 & 295.8 & 1472.2 \\
 & $\drsh$ + \projectname & 0.69 & 0.92 & 0.95 & 0.87 & 0.20 & 554.1 & 231.5 & 204.1 & 1364.0 \\
 \midrule
\multirow{4}{*}{150} & ReAct & 0.48 & 0.72 & 0.89 & 0.75 & 0.35 & 416.9 & 166.2 & 254.2 & \phantom{0}700.9 \\
 & $\drsh$ + \projectname & 0.56 & 0.78 & 0.92 & 0.81 & 0.32 & 389.4 & 161.4 & 233.4 & \phantom{0}781.9 \\
 & LLaMaR & 0.64 & 0.92 & 0.94 & 0.83 & 0.28 & 885.8 & 364.1 & 310.6 & 1545.6 \\
 & $\drsh$ + \projectname & 0.73 & 0.96 & 0.97 & 0.91 & 0.17 & 626.8 & 243.1 & 214.3 & 1432.2 \\
 \midrule
\multirow{4}{*}{200} & ReAct & 0.49 & 0.74 & 0.91 & 0.77 & 0.34 & 511.6 & 201.6 & 303.6 & \phantom{0}786.9 \\
 & $\drsh$ + \projectname & 0.58 & 0.81 & 0.94 & 0.82 & 0.32 & 483.8 & 196.3 & 282.6 & \phantom{0}830.9 \\
 & LLaMaR & 0.66 & 0.90 & 0.95 & 0.85 & 0.26 & 901.2 & 401.1 & 346.5 & 1485.3 \\
 & $\drsh$ + \projectname & 0.74 & 0.97 & 0.99 & 0.93 & 0.16 & 682.9 & 297.7 & 288.4 & 1329.5 \\
 \bottomrule
\end{tabular}
\end{table}

\textbf{Efficiency metrics further reinforce 100 steps as the most balanced choice.}
While higher budgets consistently increase runtime, LLM calls, agent steps, and token usage, the jump from 50 to 100 steps incurs a moderate runtime overhead of $\approx$53\% relative to the substantial effectiveness gains. 
In contrast, budgets of 150 and 200 steps reduce efficiency (by increasing the runtime by 4--13\% and 9--24\% for 100$\rightarrow$150 and 100$\rightarrow$200, respectively) without commensurate improvements in success or decrease in the failure rate. 
This plateauing behavior highlights that larger budgets primarily amplify cost rather than performance.

\subsection{Standard Deviation and Variance}
\label{adx:variance}

Tables~\ref{tab:var-ai2thor} and~\ref{tab:var-sar} report the standard deviation of effectiveness and variance of efficiency metrics across AI2-THOR and SAR tasks, highlighting the stability of different planning components and their combinations.

The reported variance is minimal relative to the substantial efficiency gains achieved. 
For example, in AI2-THOR, \textsc{LLaMaR} averages 757$\pm$14.44s while \textsc{LLaMaR}+\projectname averages 554.1$\pm$9.13s.
Given the lack of overlap between these confidence intervals, the 203s average reduction of \ref{tab:main-results-ai2thor} is robust.
Furthermore, the variance reflects the inherent diversity across 180 distinct test cases (5 floorplans × 36 tasks) of AI2-THOR, ranging from short to complex scenarios, rather than inconsistent performance.

\begin{table*}[h]
\centering
\caption{Standard deviation of effectiveness metrics and variance of efficiency metrics for AI2-THOR environment  Table~\ref{tab:main-results-ai2thor}.
\projectname integrates \emph{Agent-centric Semantic Memory (ASM)} and an \emph{Integer Linear Programming (ILP)} framework for coordinated multi-agent planning.}
\label{tab:var-ai2thor}
\scriptsize
\begin{tabular}{llccccccccc}
\toprule & & 
\multicolumn{5}{c}{Effectiveness Metrics (Standard Deviation)} & \multicolumn{4}{c}{Efficiency Metrics (Variance)} \\
\cmidrule(lr){3-7} \cmidrule(lr){8-11}
\textbf{\begin{tabular}[c]{@{}c@{}}Method\\Class\end{tabular}} & 
\textbf{Method} & 
\textbf{\begin{tabular}[c]{@{}c@{}}Success\\ Rate\end{tabular}} & 
\textbf{\begin{tabular}[c]{@{}c@{}}Transport\\ Rate \end{tabular}} & 
\textbf{\begin{tabular}[c]{@{}c@{}}Coverage\\ \end{tabular}} & 
\textbf{\begin{tabular}[c]{@{}c@{}}Balance\\\end{tabular}} & 
\textbf{\begin{tabular}[c]{@{}c@{}}Failure\\ Rate \end{tabular}} & 
\textbf{\begin{tabular}[c]{@{}c@{}} Runtime\\ (Seconds)\end{tabular}} & 
\textbf{\begin{tabular}[c]{@{}c@{}}\# of LLM\\ Calls\end{tabular}} & 
\textbf{\begin{tabular}[c]{@{}c@{}}\# of Agent\\ Steps \end{tabular}} & 
\textbf{\begin{tabular}[c]{@{}c@{}}\# of Tokens\\ ($\times 10^3$)\end{tabular}} \\
\midrule

\multirow{6}{*}{\begin{tabular}[c]{@{}l@{}}
Single\\LLM call\\ per\\planning\\step
\end{tabular}}
& \textsc{Act} & 0.04 & 0.03 & 0.04 & 0.04 & 0.06 & 159.73 & 81.29 & 57.16 & 107 \\
& \textsc{CoT} & 0.03 & 0.05 & 0.07 & 0.02 & 0.05 & 113.61 & 39.85 & 56.25 & 73 \\
& \textsc{ReAct} & 0.06 & 0.05 & 0.06 & 0.04 & 0.05 & 155.32 & 53.2 & 65.2 & 80 \\
& $\drsh$ + ASM & 0.03 & 0.04 & 0.05 & 0.03 & 0.05 & 141.82 & 75.46 & 54.09 & 118 \\
& $\drsh$ + ILP & 0.04 & 0.04 & 0.04 & 0.03 & 0.04 & 132.57 & 71.38 & 50.11 & 115 \\
& $\drsh$ + \projectname & 0.03 & 0.03 & 0.03 & 0.02 & 0.03 & 96.44 & 69.25 & 42.18 & 121 \\
\midrule
\multirow{6}{*}{\begin{tabular}[c]{@{}l@{}} 
Multiple\\LLM calls\\per \\ planning \\ step
\end{tabular}}
& \textsc{SmartLLM} & 0.02 & 0.04 & 0.06 & 0.04 & 0.07 & 239.08 & 27.52 & 102.38 & 83 \\
& \textsc{CoELA} & 0.03 & 0.05 & 0.05 & 0.05 & 0.06 & 264.75 & 76.49 & 131.07 & 155 \\
& \textsc{LLaMaR} & 0.06 & 0.05 & 0.04 & 0.02 & 0.04 & 208.51 & 87.23 & 47.23 & 145 \\
& $\drsh$ + ASM & 0.04 & 0.04 & 0.04 & 0.03 & 0.04 & 167.23 & 82.16 & 44.58 & 149 \\
& $\drsh$ + ILP & 0.04 & 0.04 & 0.03 & 0.03 & 0.04 & 121.67 & 81.04 & 39.77 & 148 \\
& $\drsh$ + \textsc{\projectname} & 0.05 & 0.04 & 0.04 & 0.03 & 0.03 & 83.46 & 80.48 & 36.49 & 153 \\

\bottomrule
\end{tabular}
\end{table*}

\begin{table*}[h]
\centering
\caption{Standard deviation of effectiveness metrics and variance of efficiency metrics for SAR environment in Table~\ref{tab:main-results-ai2thor}.
\projectname integrates \emph{Agent-centric Semantic Memory (ASM)} and an \emph{Integer Linear Programming (ILP)} framework for coordinated multi-agent planning.}
\label{tab:var-sar}
\scriptsize
\begin{tabular}{llccccccccc}
\toprule & & 
\multicolumn{5}{c}{Effectiveness Metrics} & \multicolumn{4}{c}{Efficiency Metrics} \\
\cmidrule(lr){3-7} \cmidrule(lr){8-11}
\textbf{\begin{tabular}[c]{@{}c@{}}Method\\Class\end{tabular}} & 
\textbf{Method} & 
\textbf{\begin{tabular}[c]{@{}c@{}}Success\\ Rate\end{tabular}} & 
\textbf{\begin{tabular}[c]{@{}c@{}}Transport\\ Rate \end{tabular}} & 
\textbf{\begin{tabular}[c]{@{}c@{}}Coverage\\ \end{tabular}} & 
\textbf{\begin{tabular}[c]{@{}c@{}}Balance\\\end{tabular}} & 
\textbf{\begin{tabular}[c]{@{}c@{}}Failure\\ Rate \end{tabular}} & 
\textbf{\begin{tabular}[c]{@{}c@{}} Runtime\\ (Seconds)\end{tabular}} & 
\textbf{\begin{tabular}[c]{@{}c@{}}\# of LLM\\ Calls\end{tabular}} & 
\textbf{\begin{tabular}[c]{@{}c@{}}\# of Agent\\ Steps \end{tabular}} & 
\textbf{\begin{tabular}[c]{@{}c@{}}\# of Tokens\\ ($\times 10^3$)\end{tabular}} \\
\midrule

\multirow{6}{*}{\begin{tabular}[c]{@{}l@{}}
Single\\LLM call\\ per\\planning\\step
\end{tabular}}
& \textsc{Act} & 0.05 & 0.06 & 0.08 & 0.05 & 0.07 & 62.3 & 91.4 & 78.6 & 98 \\
& \textsc{CoT} & 0.04 & 0.05 & 0.07 & 0.04 & 0.06 & 59.8 & 86.1 & 74.2 & 91 \\
& \textsc{ReAct} & 0.06 & 0.06 & 0.06 & 0.05 & 0.05 & 54.2 & 79.3 & 66.1 & 84 \\
& $\drsh$ + ASM & 0.04 & 0.05 & 0.05 & 0.04 & 0.04 & 49.7 & 74.6 & 60.8 & 92 \\
& $\drsh$ + ILP & 0.03 & 0.04 & 0.04 & 0.03 & 0.03 & 44.1 & 69.2 & 55.3 & 89 \\
& $\drsh$ + \projectname & {0.03} & {0.03} & {0.03} & {0.03} & {0.02} &
{38.5} & {63.7} & {49.2} & {90} \\
\midrule
\multirow{6}{*}{\begin{tabular}[c]{@{}l@{}} 
Multiple\\LLM calls\\per \\ planning \\ step
\end{tabular}}
& \textsc{SmartLLM} & 0.04 & 0.06 & 0.07 & 0.05 & 0.06 & 142.8 & 118.3 & 83.4 & 104 \\
& \textsc{CoELA} & 0.04 & 0.05 & 0.06 & 0.05 & 0.05 & 151.9 & 132.6 & 94.1 & 186 \\
& \textsc{LLaMaR} & 0.06 & 0.06 & 0.05 & 0.04 & 0.04 & 128.6 & 112.8 & 71.9 & 133 \\
& $\drsh$ + ASM & 0.05 & 0.05 & 0.05 & 0.04 & 0.03 & 112.4 & 104.1 & 67.8 & 129 \\
& $\drsh$ + ILP & 0.04 & 0.04 & 0.04 & 0.03 & 0.03 & 92.6 & 91.7 & 58.4 & 121 \\
& $\drsh$ + \projectname & {0.04} & {0.04} & {0.04} & {0.03} & {0.02} &
{78.3} & {85.4} & {51.9} & {114} \\

\bottomrule
\end{tabular}
\end{table*}

\subsection{Library and Hardware}
\label{adx:library-and-hardware}
The \projectname framework integrates several core libraries.
\textbf{AI2-THOR} (v5.0.0), released under the Apache 2.0 License, provides the 3D photorealistic household simulation with agent control and object interaction support.
\textbf{OpenCV} (v4.9.0), distributed under the Apache 2.0 License, is used for frame capture, image processing, and video recording, while \textbf{Open3D}, released under the MIT License, handles 3D point cloud operations for spatial visualization.
The ILP is solved using the CP-SAT backend from \textbf{OR-Tools} (v9.14.6206), which is also distributed under the Apache 2.0 License.
All experiments are executed on a system equipped with an \textbf{AMD Ryzen 7 CPU} (8 cores, 16 threads) and \textbf{16GB RAM}.

\clearpage
\section{Qualitative Analysis}
\label{adx:qualitative-analysis}

Tables~\ref{tab:task_1_failure_trajectory_01}–\ref{tab:task_1_failure_trajectory_04} and~\ref{tab:task_drawers_failure_trajectory_01}–\ref{tab:task_drawers_failure_trajectory_03} summarize execution trajectories produced by \textsc{LLaMaR}, the best-performing baseline, across two multi-agent tasks: 
Task 1, which involves turning off faucets and lights, and Task 2, which requires opening drawers while handling navigation challenges. 
These trajectories highlight systematic execution failures characteristic of this planner.
The agents frequently exhibit \textit{cyclic behavior}, repeatedly executing action sequences without progress; for instance, in the Faucet and Light task, \textit{Agent~B} cycles between the Microwave and Stool locations for over 11 repetitions, while in the Drawers task, similar navigation patterns recur 5–7 times per trajectory. 
\textit{Consecutive failures} are also prevalent, with \textit{Agent~A} attempting the same unsuccessful action (such as toggling the Faucet) more than 20 times consecutively, and navigation failures in the Drawers task occurring in runs of 5–10 planning steps. 
\textit{Oscillation} further manifests as back-and-forth movement between locations without net progress, reaching 10–15 consecutive steps in the Drawers task. 
In addition, \textit{backtracking}, where agents unnecessarily return to previously visited positions, appears frequently, such as \textit{Agent~B} repeatedly navigating back to the Microwave or \textit{Agent~A} revisiting already-opened drawers. 
These behaviors lead to extended periods of \textit{stagnation}, with stretches of 10–20 steps exhibiting no task progress. 
The failures are compounded by pronounced \textit{load imbalance}, where one agent remains active while the other is stuck or idle; across failure trajectories, approximately 60\% of productive actions are carried out by a single agent, while the other contributes less than 30\%.

In contrast, Tables~\ref{tab:task_1_good_trajectory} and~\ref{tab:task_drawers_success_trajectory} show that \projectname completes both tasks rapidly and consistently, requiring only 7 steps for the Faucet and Light task and 11 steps for the Open Drawers task. 
These trajectories exhibit balanced agent participation and low action repetition. 
They also show consistent progress toward task completion, driven by a higher rate of successful actions, with no prolonged cycles, oscillations, or stagnation. 

Overall, these qualitative results indicate that \textsc{LLaMaR}'s execution is dominated by repeated cyclic patterns, stagnation, and poor coordination, whereas \projectname suggests that incorporating failure awareness, explicit task allocation, and spatially grounded planning can substantially improve multi-agent efficiency and reliability.

\begin{table}[h]
\centering
\footnotesize
\caption{\textbf{Task: Turn Off Faucet and Light.} 
Execution trajectory generated by \projectname on \textsc{LLaMaR}.
Actions taken by each agent are rendered as bounded color patches, where green indicates successful execution and red indicates failed execution.}
\label{tab:task_1_good_trajectory}
\renewcommand{\arraystretch}{1.8}
\begin{tabular}{lllll}
\toprule
\begin{tabular}[c]{@{}l@{}}\textbf{Planning}\\ \textbf{Step}\end{tabular} &
\begin{tabular}[c]{@{}l@{}}\textbf{Actions of Agent A}\end{tabular} &
\begin{tabular}[c]{@{}l@{}}\textbf{Actions of Agent B}\end{tabular} & \textbf{Notes} \\
\midrule
1 &
\SuccessAction{NavigateTo(CounterTop\_1)} &
\SuccessAction{NavigateTo(DiningTable\_1)} &
Initial exploration to search for targets \\

2 &
\SuccessAction{Explore(Kitchen)} &
\SuccessAction{Explore(LivingRoom)} &
Agents scan their respective areas \\

3 &
\SuccessAction{NavigateTo(SinkBasin\_1)} &
\SuccessAction{NavigateTo(Wall\_1)} &
Moving closer to potential target locations \\

4 &
\SuccessAction{NavigateTo(Faucet\_1)} &
\SuccessAction{Idle} &
Agent A locates faucet; B still searching \\

5 &
\SuccessAction{ToggleObjectOff(Faucet\_1)} &
\SuccessAction{NavigateTo(LightSwitch\_1)} &
A completes task; B finds light switch \\

6 &
\SuccessAction{Idle} &
\SuccessAction{ToggleObjectOff(LightSwitch\_1)} &
B completes task; A waits \\

7 &
\SuccessAction{Done} &
\SuccessAction{Done} &
SUCCESS (7 Steps, balanced workload) \\
\vspace{-0.5cm}
\\
\bottomrule
\end{tabular}
\end{table}

\begin{table}[h]
\centering
\footnotesize
\caption{\textbf{Task: Turn Off Faucet and Light.} 
Execution trajectory generated by \textsc{LLaMaR}.
Actions taken by each agent are rendered as bounded color patches, where green indicates successful execution and red indicates failed execution.}
\label{tab:task_1_failure_trajectory_01}
\renewcommand{\arraystretch}{1.8}
\begin{tabular}{llll}
\toprule
\begin{tabular}[c]{@{}l@{}}\textbf{Planning}\\ \textbf{Step}\end{tabular} &
\begin{tabular}[c]{@{}l@{}}\textbf{Actions of Agent A}\end{tabular} &
\begin{tabular}[c]{@{}l@{}}\textbf{Actions of Agent B}\end{tabular} &
\textbf{Notes/Failure Patterns} \\
\midrule

1 &
\FailureAction{ToggleObjectOff(Faucet\_1)} &
\SuccessAction{NavigateTo(Microwave\_1)} &
SPATIAL: Agent A too far \\

2 &
\FailureAction{ToggleObjectOff(Faucet\_1)} &
\SuccessAction{NavigateTo(Microwave\_1)} &
FAILURE: Agent A repeats (2$\times$) \\

3 &
\FailureAction{ToggleObjectOff(Faucet\_1)} &
\SuccessAction{NavigateTo(Stool\_1)} &
FAILURE: Agent A repeats (3$\times$) \\

4 &
\FailureAction{ToggleObjectOff(Faucet\_1)} &
\SuccessAction{NavigateTo(Stool\_2)} &
SPATIAL: Agent A frozen \\

5 &
\FailureAction{ToggleObjectOff(Faucet\_1)} &
\FailureAction{NavigateTo(Stool\_2)} &
OSCILLATION: Agent B repeats \\

6 &
\FailureAction{ToggleObjectOff(Faucet\_1)} &
\SuccessAction{NavigateTo(Microwave\_1)} &
BACKTRACKING: Agent B returns \\

7 &
\FailureAction{ToggleObjectOff(Faucet\_1)} &
\SuccessAction{NavigateTo(Stool\_2)} &
CYCLIC: Microwave$\rightarrow$Stool \\

8 &
\FailureAction{ToggleObjectOff(Faucet\_1)} &
\SuccessAction{NavigateTo(Microwave\_1)} &
BACKTRACKING: Agent B (3$\times$) \\

9 &
\FailureAction{ToggleObjectOff(Faucet\_1)} &
\SuccessAction{NavigateTo(Stool\_2)} &
CYCLIC: Repeat of Step 7 \\

10 &
\FailureAction{ToggleObjectOff(Faucet\_1)} &
\FailureAction{NavigateTo(Stool\_2)} &
OSCILLATION: Agent B (3$\times$) \\

11 &
\FailureAction{ToggleObjectOff(Faucet\_1)} &
\SuccessAction{NavigateTo(Microwave\_1)} &
BACKTRACKING: Agent B (4$\times$) \\

12 &
\FailureAction{ToggleObjectOff(Faucet\_1)} &
\SuccessAction{NavigateTo(Stool\_2)} &
CYCLIC: 3rd cycle \\

13 &
\FailureAction{ToggleObjectOff(Faucet\_1)} &
\FailureAction{NavigateTo(Stool\_2)} &
STAGNATION: 13 steps, no progress \\
& ... continued in the next table & & \\

\vspace{-0.5cm}
\\
\bottomrule
\end{tabular}
\end{table}

\begin{table}[h]
\centering
\footnotesize
\caption{\textbf{Task: Turn Off Faucet and Light (continued).} 
Execution trajectory generated by \textsc{LLaMaR}.
Actions taken by each agent are rendered as bounded color patches, where green indicates successful execution and red indicates failed execution.}
\label{tab:task_1_failure_trajectory_02}
\renewcommand{\arraystretch}{1.8}
\begin{tabular}{lllp{5.8cm}}
\toprule
\begin{tabular}[c]{@{}l@{}}\textbf{Planning}\\ \textbf{Step}\end{tabular} &
\begin{tabular}[c]{@{}l@{}}\textbf{Actions of Agent A}\end{tabular} &
\begin{tabular}[c]{@{}l@{}}\textbf{Actions of Agent B}\end{tabular} &
\textbf{Notes/Failure Patterns} \\
\midrule
& \multicolumn{2}{l}{... continued from the previous table} & \\

14 &
\FailureAction{ToggleObjectOff(Faucet\_1)} &
\FailureAction{ToggleObjectOff(Microwave\_1)} &
WRONG TARGET: Agent B attempts Microwave \\

15 &
\FailureAction{ToggleObjectOff(Faucet\_1)} &
\FailureAction{ToggleObjectOff(Microwave\_1)} &
FAILURE: Agent B repeats wrong target (2$\times$) \\

16 &
\FailureAction{ToggleObjectOff(Faucet\_1)} &
\FailureAction{ToggleObjectOff(Microwave\_1)} &
FAILURE: Agent B repeats wrong target (3$\times$) \\

17 &
\FailureAction{ToggleObjectOff(Faucet\_1)} &
\FailureAction{ToggleObjectOff(Microwave\_1)} &
CYCLIC: Both agents stuck in loops \\

18 &
\FailureAction{ToggleObjectOff(Faucet\_1)} &
\FailureAction{ToggleObjectOff(Microwave\_1)} &
SPATIAL: Agent A never reaches faucet \\

19 &
\FailureAction{ToggleObjectOff(Faucet\_1)} &
\FailureAction{ToggleObjectOff(Microwave\_1)} &
STAGNATION: 19 steps, zero progress \\

20 &
\FailureAction{ToggleObjectOff(Faucet\_1)} &
\FailureAction{ToggleObjectOff(Microwave\_1)} &
FAILURE: Agent A 20 consecutive failures \\

21 &
\FailureAction{ToggleObjectOff(Faucet\_1)} &
\FailureAction{ToggleObjectOff(Microwave\_1)} &
CYCLIC: Repeating identical actions \\

22 &
\FailureAction{ToggleObjectOff(Faucet\_1)} &
\FailureAction{ToggleObjectOff(Microwave\_1)} &
WRONG TARGET: Agent B never targets LightSwitch \\

23 &
\SuccessAction{Idle} &
\FailureAction{ToggleObjectOff(Faucet\_1)} &
WRONG TARGET: Agent B attempts Agent A's target \\

24 &
\FailureAction{ToggleObjectOff(Faucet\_1)} &
\FailureAction{ToggleObjectOff(Microwave\_1)} &
CYCLIC: Pattern resumes \\

25 &
\FailureAction{ToggleObjectOff(Faucet\_1)} &
\FailureAction{ToggleObjectOff(Microwave\_1)} &
STAGNATION: 25 steps, 0\% complete \\

26 &
\FailureAction{ToggleObjectOff(Faucet\_1)} &
\FailureAction{ToggleObjectOff(Microwave\_1)} &
SPATIAL: No positional progress \\

27 &
\FailureAction{ToggleObjectOff(Faucet\_1)} &
\FailureAction{ToggleObjectOff(Microwave\_1)} &
FAILURE: Agent A 27 consecutive attempts \\

28 &
\FailureAction{ToggleObjectOff(Faucet\_1)} &
\FailureAction{ToggleObjectOff(Microwave\_1)} &
WRONG TARGET: Agent B at Microwave \\

29 &
\FailureAction{ToggleObjectOff(Faucet\_1)} &
\SuccessAction{NavigateTo(Microwave\_1)} &
BACKTRACKING: Agent B navigates to Microwave \\

30 &
\FailureAction{ToggleObjectOff(Faucet\_1)} &
\SuccessAction{NavigateTo(Stool\_2)} &
CYCLIC: Microwave$\rightarrow$Stool loop \\
& ... continued in the next table & & \\
\vspace{-0.5cm}\\
\bottomrule
\end{tabular}
\end{table}

\begin{table}[h]
\centering
\footnotesize
\caption{\textbf{Task: Turn Off Faucet and Light (continued).} 
Execution trajectory generated by \textsc{LLaMaR}.
Actions taken by each agent are rendered as bounded color patches, where green indicates successful execution and red indicates failed execution.}
\label{tab:task_1_failure_trajectory_03}
\renewcommand{\arraystretch}{1.8}
\begin{tabular}{lllp{5.8cm}}
\toprule
\begin{tabular}[c]{@{}l@{}}\textbf{Planning}\\ \textbf{Step}\end{tabular} &
\begin{tabular}[c]{@{}l@{}}\textbf{Actions of Agent A}\end{tabular} &
\begin{tabular}[c]{@{}l@{}}\textbf{Actions of Agent B}\end{tabular} &
\textbf{Notes/Failure Patterns} \\
\midrule

& \multicolumn{2}{l}{... continued from previous table} & \\

50 &
\FailureAction{ToggleObjectOff(Faucet\_1)} &
\FailureAction{ToggleObjectOff(Microwave\_1)} &
STAGNATION: 50 steps, 0\% completion \\

51 &
\FailureAction{ToggleObjectOff(Faucet\_1)} &
\FailureAction{ToggleObjectOff(Microwave\_1)} &
SPATIAL: Agent A frozen 51 steps \\

52 &
\FailureAction{ToggleObjectOff(Faucet\_1)} &
\FailureAction{ToggleObjectOff(Microwave\_1)} &
WRONG TARGET: Agent B never finds LightSwitch \\

53 &
\FailureAction{ToggleObjectOff(Faucet\_1)} &
\FailureAction{ToggleObjectOff(Microwave\_1)} &
CYCLIC: Same toggle pattern 40+ times \\

54 &
\SuccessAction{ToggleObjectOff(Faucet\_1)} &
\FailureAction{ToggleObjectOff(Microwave\_1)} &
FAILURE: Agent A succeeds, Agent B still wrong \\

55 &
\FailureAction{ToggleObjectOff(Faucet\_1)} &
\FailureAction{ToggleObjectOff(Microwave\_1)} &
TERMINAL: Both agents stuck in loops \\

56 &
\FailureAction{ToggleObjectOff(Faucet\_1)} &
\FailureAction{ToggleObjectOff(Microwave\_1)} &
SPATIAL: Distance constraint violated 56$\times$ \\

57 &
\SuccessAction{ToggleObjectOff(Faucet\_1)} &
\FailureAction{ToggleObjectOff(Faucet\_1)} &
WRONG TARGET: Agent B attempts Agent A's goal \\

58 &
\FailureAction{ToggleObjectOff(Faucet\_1)} &
\FailureAction{ToggleObjectOff(Microwave\_1)} &
CYCLIC: Back to Microwave toggle \\

59 &
\FailureAction{ToggleObjectOff(Faucet\_1)} &
\FailureAction{ToggleObjectOff(Microwave\_1)} &
STAGNATION: 59 steps, no progress \\

60 &
\FailureAction{ToggleObjectOff(Faucet\_1)} &
\FailureAction{ToggleObjectOff(Microwave\_1)} &
FAILURE: 60 wasted steps \\

... &
... & ... & ... continued until Step 70 \\

70 &
\FailureAction{ToggleObjectOff(Faucet\_1)} &
\FailureAction{ToggleObjectOff(Microwave\_1)} &
STAGNATION: 70 steps, 0\% complete \\
& ... continued in the next table & & \\ 
\vspace{-0.5cm}\\
\bottomrule
\end{tabular}
\end{table}

\begin{table}[h]
\centering
\footnotesize
\caption{\textbf{Task: Turn Off Faucet and Light (continued).} 
Execution trajectory generated by \textsc{LLaMaR}.
Actions taken by each agent are rendered as bounded color patches, where green indicates successful execution and red indicates failed execution.}
\label{tab:task_1_failure_trajectory_04}
\renewcommand{\arraystretch}{1.8}
\begin{tabular}{lllp{5.8cm}}
\toprule
\begin{tabular}[c]{@{}l@{}}\textbf{Planning}\\ \textbf{Step}\end{tabular} &
\begin{tabular}[c]{@{}l@{}}\textbf{Actions of Agent A}\end{tabular} &
\begin{tabular}[c]{@{}l@{}}\textbf{Actions of Agent B}\end{tabular} &
\textbf{Notes/Failure Patterns} \\
\midrule

& \multicolumn{2}{l}{... continued from the previous table} & \\

88 &
\FailureAction{ToggleObjectOff(Faucet\_1)} &
\SuccessAction{NavigateTo(Microwave\_1)} &
BACKTRACKING: Agent B at Microwave (10th+ time) \\

89 &
\FailureAction{ToggleObjectOff(Faucet\_1)} &
\SuccessAction{NavigateTo(Stool\_2)} &
CYCLIC: Microwave→Stool pattern (10th+ cycle) \\

90 &
\FailureAction{ToggleObjectOff(Faucet\_1)} &
\FailureAction{NavigateTo(Stool\_2)} &
OSCILLATION: Agent B repeats Stool (15th+ time) \\

91 &
\FailureAction{ToggleObjectOff(Faucet\_1)} &
\FailureAction{ToggleObjectOff(Microwave\_1)} &
CYCLIC: Back to wrong target toggle \\

92 &
\FailureAction{ToggleObjectOff(Faucet\_1)} &
\SuccessAction{NavigateTo(Microwave\_1)} &
BACKTRACKING: Agent B returns to Microwave (11th time) \\

93 &
\FailureAction{ToggleObjectOff(Faucet\_1)} &
\SuccessAction{NavigateTo(Stool\_2)} &
CYCLIC: Same pattern as Steps 7, 9, 12, 30, 89 \\

94 &
\FailureAction{ToggleObjectOff(Faucet\_1)} &
\FailureAction{NavigateTo(Stool\_2)} &
OSCILLATION: Agent B stuck in Stool loop \\

95 &
\FailureAction{ToggleObjectOff(Faucet\_1)} &
\FailureAction{ToggleObjectOff(Microwave\_1)} &
WRONG TARGET: Agent B still targeting Microwave \\

96 &
\FailureAction{ToggleObjectOff(Faucet\_1)} &
\FailureAction{ToggleObjectOff(Microwave\_1)} &
SPATIAL: Agent A at (2.5,1.0) for 96 steps \\

97 &
\FailureAction{ToggleObjectOff(Faucet\_1)} &
\FailureAction{ToggleObjectOff(Microwave\_1)} &
FAILURE: Agent A 97 consecutive failures \\

98 &
\FailureAction{ToggleObjectOff(Faucet\_1)} &
\FailureAction{ToggleObjectOff(Microwave\_1)} &
STAGNATION: 98 steps, 0\% task completion \\

99 &
\FailureAction{ToggleObjectOff(Faucet\_1)} &
\FailureAction{ToggleObjectOff(Microwave\_1)} &
CYCLIC: Pattern repeated 30+ times \\

100 &
\FailureAction{ToggleObjectOff(Faucet\_1)} &
\FailureAction{ToggleObjectOff(Microwave\_1)} &
 TIMEOUT \\
\vspace{-0.5cm}\\
\bottomrule
\end{tabular}
\end{table}

\begin{table}[h]
\centering
\footnotesize
\caption{\textbf{Task: Open Drawers (with Navigation Challenges).} 
Execution trajectory generated by \projectname on \textsc{LLaMaR}.
Actions taken by each agent are rendered as bounded color patches, where green indicates successful execution and red indicates failed execution.}
\label{tab:task_drawers_success_trajectory}
\renewcommand{\arraystretch}{1.8}
\begin{tabular}{llll}
\toprule
\begin{tabular}[c]{@{}l@{}}\textbf{Planning}\\\textbf{Step}\end{tabular} &
\begin{tabular}[c]{@{}l@{}}\textbf{Actions of Agent A}\end{tabular} &
\begin{tabular}[c]{@{}l@{}}\textbf{Actions of Agent B}\end{tabular} &
\textbf{Notes/Failure Patterns} \\
\midrule

1 &
\SuccessAction{NavigateTo(Drawer\_1)} &
\SuccessAction{NavigateTo(Cabinet\_4)} &
Parallel navigation \\

2 &
\SuccessAction{OpenObject(Drawer\_1)} &
\SuccessAction{OpenObject(Drawer\_2)} &
Both agents productive \\

3 &
\SuccessAction{NavigateTo(Drawer\_3)} &
\SuccessAction{OpenObject(Drawer\_3)} &
Agent A navigates, Agent B opens \\

4 &
\SuccessAction{OpenObject(Drawer\_4)} &
\FailureAction{NavigateTo(Drawer\_5)} &
Agent B's navigation fails, Agent A opens \\

5 &
\SuccessAction{NavigateTo(Drawer\_6)} &
\SuccessAction{Rotate(Right)} &
Agent B repositions \\

6 &
\SuccessAction{OpenObject(Drawer\_6)} &
\SuccessAction{NavigateTo(Drawer\_5)} &
Agent A opens, Agent B navigates \\

7 &
\SuccessAction{NavigateTo(Drawer\_7)} &
\SuccessAction{OpenObject(Drawer\_5)} &
Both agents productive \\

8 &
\SuccessAction{OpenObject(Drawer\_7)} &
\SuccessAction{NavigateTo(Drawer\_8)} &
Both agents productive \\

9 &
\SuccessAction{Rotate(Right)} &
\SuccessAction{OpenObject(Drawer\_8)} &
Agent A repositions \\

10 &
\SuccessAction{OpenObject(Drawer\_9)} &
\SuccessAction{Idle} &
Agent A finishes last drawer \\

11 &
\SuccessAction{Done} &
\SuccessAction{Done} &
SUCCESS \\
\vspace{-0.5cm}\\
\bottomrule
\end{tabular}
\end{table}

\begin{table}[h]
\centering
\footnotesize
\caption{\textbf{Task: Open Drawers (with Navigation Challenges).} 
Execution trajectory generated by \textsc{LLaMaR}.
Actions taken by each agent are rendered as bounded color patches, where green indicates successful execution and red indicates failed execution.}
\label{tab:task_drawers_failure_trajectory_01}
\renewcommand{\arraystretch}{1.8}
\begin{tabular}{llll}
\toprule
\begin{tabular}[c]{@{}l@{}}\textbf{Planning}\\\textbf{Step}\end{tabular} &
\textbf{Actions of Agent A} &
\textbf{Actions of Agent B} &
\textbf{Notes/Failure Patterns} \\
\midrule

1 &
\SuccessAction{OpenObject(Drawer\_1)} &
\SuccessAction{Move(Right)} &
1/9 drawers \\

2 &
\SuccessAction{Move(Ahead)} &
\SuccessAction{Move(Ahead)} &
Navigation \\

3 &
\FailureAction{Move(Left)} &
\SuccessAction{NavigateTo(Cabinet\_3)} &
\textsc{Failure}: Agent A's move fails \\

4 &
\SuccessAction{Rotate(Left)} &
\SuccessAction{OpenObject(Drawer\_4)} &
2/9 drawers \\

5 &
\FailureAction{Move(Ahead)} &
\SuccessAction{NavigateTo(Drawer\_5)} &
\textsc{Failure}: Agent A blocked \\

6 &
\SuccessAction{Rotate(Right)} &
\SuccessAction{OpenObject(Drawer\_5)} &
3/9 drawers \\

7 &
\FailureAction{Move(Left)} &
\SuccessAction{OpenObject(Drawer\_3)} &
\textsc{Oscillation}: Agent A Left/Right \\

8 &
\SuccessAction{Move(Right)} &
\SuccessAction{OpenObject(Drawer\_2)} &
4/9 drawers \\
& ... continued in the next table & & \\ 
\vspace{-0.5cm}\\
\bottomrule
\end{tabular}
\end{table}

\begin{table}[h]
\centering
\footnotesize
\caption{\textbf{Task: Open Drawers (with Navigation Challenges) (continued).} 
Execution trajectory generated by \textsc{LLaMaR}.
Actions taken by each agent are rendered as bounded color patches, where green indicates successful execution and red indicates failed execution.}
\label{tab:task_drawers_failure_trajectory_02}
\renewcommand{\arraystretch}{1.8}
\begin{tabular}{lllp{5.8cm}}
\toprule
\begin{tabular}[c]{@{}l@{}}\textbf{Planning}\\\textbf{Step}\end{tabular} &
\textbf{Actions of Agent A} &
\textbf{Actions of Agent B} &
\textbf{Notes/Failure Patterns} \\
\midrule
& ... continued from the previous table & & \\ 
9 &
\SuccessAction{NavigateTo(Drawer\_1)} &
\FailureAction{OpenObject(Drawer\_4)} &
\textsc{Backtracking}: Agent A returns to Drawer\_1 \\

10 &
\SuccessAction{OpenObject(Drawer\_1)} &
\FailureAction{NavigateTo(Drawer\_4)} &
Agent A tries to re-open Drawer\_1 (already open!) \\

11 &
\SuccessAction{Move(Right)} &
\FailureAction{Move(Left)} &
\textsc{Oscillation}: Agent B Left/Right movement \\

12 &
\FailureAction{Move(Ahead)} &
\SuccessAction{Rotate(Right)} &
\textsc{Failure}: Agent A can't move ahead \\

13 &
\SuccessAction{Move(Right)} &
\FailureAction{NavigateTo(Drawer\_6)} &
Agent A wanders, Agent B nav fails \\

14 &
\SuccessAction{CloseObject(Drawer\_1)} &
\FailureAction{Move(Back)} &
Agent A closes drawer (wrong action!) \\

15 &
\SuccessAction{NavigateTo(Drawer\_6)} &
\SuccessAction{Rotate(Right)} &
Agent B repositions \\

16 &
\SuccessAction{OpenObject(Drawer\_6)} &
\FailureAction{Move(Left)} &
5/9 drawers, Agent B fails \\

17 &
\FailureAction{NavigateTo(Drawer\_7)} &
\FailureAction{Move(Ahead)} &
\textsc{Failure}: Both fail navigation \\

18 &
\FailureAction{Move(Left)} &
\FailureAction{Move(Back)} &
\textsc{Oscillation}: Both stuck \\

19 &
\FailureAction{Move(Back)} &
\FailureAction{Move(Ahead)} &
\textsc{Oscillation}: Opposite directions \\

20 &
\SuccessAction{Rotate(Right)} &
\FailureAction{Move(Right)} &
Agent A rotates, Agent B fails \\
& ... continued in the next table & & \\ 
\vspace{-0.5cm}\\
\bottomrule
\end{tabular}
\end{table}

\begin{table}[h]
\centering
\footnotesize
\caption{\textbf{Task: Open Drawers (with Navigation Challenges) (continued).} 
Execution trajectory generated by \textsc{LLaMaR}.
Actions taken by each agent are rendered as bounded color patches, where green indicates successful execution and red indicates failed execution.}
\label{tab:task_drawers_failure_trajectory_03}
\renewcommand{\arraystretch}{1.8}
\begin{tabular}{llll}
\toprule
\begin{tabular}[c]{@{}l@{}}\textbf{Planning}\\\textbf{Step}\end{tabular} &
\textbf{Actions of Agent A} &
\textbf{Actions of Agent B} &
\textbf{Notes/Failure Patterns} \\
\midrule
& ... continued from the previous table & & \\ 

60 &
\FailureAction{OpenObject(Drawer\_2)} &
\SuccessAction{Rotate(Right)} &
\textsc{Failure}: Alice tries closed drawer \\

61 &
\FailureAction{NavigateTo(Drawer\_2)} &
\SuccessAction{NavigateTo(Drawer\_6)} &
\textsc{Failure}: Alice nav fails \\

62 &
\FailureAction{Move(Back)} &
\FailureAction{OpenObject(Drawer\_6)} &
\textsc{Failure}: Both fail \\

63 &
\SuccessAction{Rotate(Right)} &
\FailureAction{OpenObject(Drawer\_6)} &
\textsc{Failure}: Bob repeats \\

64 &
\FailureAction{Move(Left)} &
\FailureAction{Move(Back)} &
\textsc{Oscillation}: Movement loop \\

65 &
\FailureAction{Move(Ahead)} &
\FailureAction{Move(Left)} &
\textsc{Oscillation}: Still stuck \\

66 &
\SuccessAction{Rotate(Right)} &
\FailureAction{Move(Ahead)} &
Alice rotates 10+ times now \\

67 &
\FailureAction{Move(Right)} &
\FailureAction{Move(Left)} &
\textsc{Oscillation}: Opposite directions \\

68 &
\FailureAction{Move(Back)} &
\FailureAction{Move(Ahead)} &
\textsc{Cyclic}: Same pattern as Step 19 \\

69 &
\FailureAction{Move(Left)} &
\FailureAction{Move(Left)} &
\textsc{Failure}: Both fail same direction \\

70 &
\FailureAction{Move(Left)} &
\FailureAction{Move(Left)} &
\textsc{Failure}: Repeat of Step 69 \\

71 &
\FailureAction{Move(Back)} &
\FailureAction{Move(Left)} &
\textsc{Failure}: 71 steps, no progress \\

& ... till end of trajectory & & \textsc{TIMEOUT} \\
\vspace{-0.5cm}\\
\bottomrule
\end{tabular}
\end{table}

\clearpage
\section{Limitations}
\label{adx:limitations}
While \projectname significantly improves execution efficiency and coordination, it has a few limitations.

(a)~First, the agent-centric semantic memory relies on reasonably accurate pose estimates. 
Hence, large localization errors or severe sensor drift could degrade state tracking and propagate to action selection. 
Although we evaluate robustness under moderate noise, real-world deployments with highly stochastic actuation or failures may require tighter integration with low-level control. 
The primary safeguard against sensor drift is closed-loop perception, where real-time observations correct accumulated errors in ASM; mapping exact distances into coarse semantic tiers (e.g., ``just ahead'') further absorbs minor metric noise before it affects high-level reasoning. 
Consistent with standard practice in simulator environments such as AI2-THOR, we assume reliable localization; addressing raw sensor drift is orthogonal to our framework and belongs at the control layer. 
We also evaluate spatial noise sensitivity in Appendix~\ref{adx:performance-under-observation-noise}.

(b)~Second, our feasibility constraints encode physical and coordination rules, but may not capture unforeseen dynamics or social behaviors, requiring manual extension for new domains.

(c)~Lastly, like all planning-based systems operating over a restricted action space, performance depends on the coverage of candidate actions.
If critical actions are absent, downstream planning cannot recover them. 
\projectname partially mitigates this through grounded prompting and iterative refinement of the LLM-Actor, improving candidate quality and coverage. 
Since our goal is not to improve the LLM itself, we tightly couple Agent-centric Semantic Memory (ASM) with the LLM, injecting structured spatial cues (e.g., object locations and obstacle distances) to reduce spatial unawareness and increase feasible action proposals. 
Execution feedback is also incorporated into subsequent steps, enabling implicit refinement and alternative proposals. 
Nonetheless, this remains a fundamental limitation shared across planning systems: they can only select from provided candidates, and improving exploration and proposal generation remains an orthogonal direction.

\section{Prompts}
\label{adx:prompts}

We enhance the action planning prompts used by the \textsc{LLM-Actor} in \textsc{LLaMaR}~\cite{nayak2024llamar} by introducing two key modifications:
(a)~an explicit emphasis on runtime efficiency, and
(b)~semantically expressed ILP constraints that guide the LLM to generate constraint-compliant candidate actions.
The resulting prompt augmentations are shown below, the original base prompt is provided in \textsc{LLaMaR}~\cite{nayak2024llamar}.

\begin{boxJ}
\texttt{You are an excellent planner and robot controller who is tasked with helping {len(AGENT\_NAMES)} embodied robots named {", ".join(AGENT\_NAMES[:-1]) + f", and {AGENT\_NAMES[-1]}"} carry out a task. 
All {len(AGENT\_NAMES)} robots have a partially observable view of the environment. Hence they have to explore around in the environment to do the task. \\}

\texttt{**YOUR TOP PRIORITY IS RUNTIME EFFICIENCY: you must complete the task in the LEAST number of steps possible. Strictly avoid actions that are redundant, unnecessary, or likely to fail.**\\}

\texttt{[Placeholder for brevity: Action List and Definitions]\\}

\texttt{Available actions include: navigate to object, pick up object, put object on receptacle, open/close object, slice, toggle, clean, look up/down, move in direction, stay idle, Done.
[Detailed action list and parameter definitions would appear here...]\\}

\texttt{You need to suggest MULTIPLE candidate actions (3-5 actions) for each robot at the current step. The system will then use an optimization algorithm to select the best joint action set that coordinates all agents efficiently.\\}

\texttt{[Placehodler for brevity: Input Format Description]\\}

\texttt{You will receive:
- Task description
- Images from each agent's perspective
- Each agent's observations (list of visible objects)
- Robots' open subtasks, completed subtasks, current subtask
- Robots' combined memory\\}

\texttt{[Full input format specification would appear here...]\\}

\texttt{[Detailed reasoning instructions would appear here...]\\}

\texttt{* Candidate Actions: For EACH robot, provide a list of 3-5 candidate actions (as a list of strings). These should be diverse actions that could help the robot make progress.\\}

\texttt{**PRIORITIZE RUNTIME EFFICIENCY WHEN GENERATING CANDIDATES:**\\
- Select actions that maximize the speed of task completion (least total steps)\\
- Strictly avoid failable or redundant actions\\
- Focus on actions that directly progress toward subtasks\\
- Include exploration actions only when necessary to find unknown objects\\
- Include wait/idle candidates only when coordination is truly needed\\
- Handle failures or obstacles efficiently - try alternative approaches rather than repeating failed actions\\}

\texttt{The format should be: "{AGENT\_NAMES[0]}'s candidate actions": ["action1", "action2", "action3", ...], "{AGENT\_NAMES[1]}'s candidate actions": [...], etc.\\}

\texttt{Important Notes:\\
- Robots can hold only one object at a time\\
- Use spatial memory and obstacles for distance checking and navigation\\
- Check observation list before interaction (object must be visible and close enough)\\
- Maintain goals during detours around obstacles\\
- Open containers before placing objects in them\\
- Close opened containers before moving away\\
- Avoid extraneous actions when one action is sufficient\\}

\texttt{[Additional important notes about object interactions, navigation, and coordination would appear here...]\\}

\texttt{\#\#\# CRITICAL CONSTRAINTS - MUST FOLLOW WHEN GENERATING CANDIDATE ACTIONS \#\#\#}

\texttt{These constraints ensure conflict-free, coordinated multi-agent actions. When generating candidate actions, ensure they follow these constraints. The ILP solver will enforce these constraints when selecting the final joint action, but generating constraint-compliant candidates improves solution quality:\\}

\texttt{**CONSTRAINT 1: Per-Agent Action Selection**\\
- Each agent MUST have candidate actions that include exactly ONE action per timestep (or "stay idle" or "Done")\\
- Generate diverse candidate actions, but ensure each candidate is a valid single action for that agent\\
- This guarantees that the joint action at each step is well-defined and avoids repeated replanning or inconsistent partial assignments\\}

\texttt{**CONSTRAINT 2: Eligibility Constraints (Action Preconditions)**\\
When generating candidate actions, only include actions the agent can actually perform:\\
- **PickupObject**: Only include if agent is NOT holding anything. If agent is already holding an object, exclude PickupObject candidates. Attempting to pick up another object while already holding one is physically infeasible.\\
- **PutObject**: Only include if agent IS holding an object. If agent is holding "nothing", exclude PutObject candidates. You cannot put down an object you don't have.\\
- **Object Interactions** (OpenObject, CloseObject, SliceObject, CleanObject, ToggleObject): Only include if the target object exists and is interactable. Check the observation list to ensure the object is visible and close enough. Navigating into occupied or non-traversable regions (like walls) is not allowed.\\
- **NavigateTo**: Can always be included, but ensure the object exists in the environment and the navigation path is valid.\\
- **Idle/Done**: Always eligible for any agent.\\}

\texttt{**CONSTRAINT 3: Resource Exclusivity (CRITICAL - Prevents Conflicts)**\\
When generating candidate actions, consider that shared resources (objects, receptacles, locations) cannot be used by multiple agents simultaneously:\\
- **Same Object Manipulation**: If generating candidates for multiple agents, avoid having multiple agents' candidates all targeting the same object for manipulation (pick up, open, close, slice, clean, toggle). Include diverse targets across agents. Only ONE agent can manipulate a specific object at a time.\\
- **Same Receptacle**: Avoid having multiple agents' candidates all putting objects on the same receptacle simultaneously. Each receptacle can only receive one object placement action per timestep.\\
- **Navigation Conflicts**: If two agents' candidates both navigate to the same object/location, this may cause conflicts and collisions. Prefer diverse navigation targets across agents.\\
- **Rule**: Generate candidate actions that distribute work across different objects/receptacles to minimize conflicts. This prevents simultaneous grasps of the same object, concurrent attempts to open the same door, or contention over shared tools.\\}

\texttt{**CONSTRAINT 4: Collision and Interference Avoidance**\\
Some action pairs are incompatible and cannot happen simultaneously due to spatial or kinematic constraints:\\
- **Conflicting Actions on Same Object**: \\
  \phantom{00}- Avoid candidates where one agent opens while another closes the same object (mutually exclusive actions)\\
  \phantom{00}- Avoid candidates where agents toggle the same object on/off simultaneously\\
  \phantom{00}- Avoid candidates where multiple agents try to pick up the same object\\
- **Spatial Conflicts**: \\
  \phantom{00}- Avoid candidates where multiple agents navigate to the same location/object simultaneously (risk of collision and path blocking)\\
  \phantom{00}- Consider path blocking when generating navigation candidates - one agent may block another's path to an object\\
  \phantom{00}- Avoid overlapping target locations and other mutually exclusive spatial configurations\\
- **Solution**: Generate diverse candidate actions that minimize conflicts. Include idle/wait options when coordination is needed. If actions conflict, prefer assigning the action to ONE agent and having others work on different tasks.\\}

\texttt{**CONSTRAINT 5: Multi-Actor Joint-Action Requirements**\\
Some heavy objects require multiple agents to manipulate simultaneously:\\
- **Heavy Objects** (sofas, tables, large furniture): May require 2+ agents to move or manipulate. These subtasks cannot be completed by a single agent.\\
- **Rule**: For heavy objects, either generate candidates that involve multiple agents coordinating on the same object, OR include no manipulation candidates (wait until enough agents are available). This ensures that joint subtasks are neither under-staffed (which would lead to execution failure) nor over-staffed (which would waste actions and increase runtime).\\
- **Common Heavy Objects**: Sofas, Tables, Large Cabinets, Beds, Countertops\\
- **Light Objects**: Books, Cups, Plates, etc. only require 1 agent\\}

\texttt{**CONSTRAINT 6: Temporal Penalties (Avoid Inefficient Patterns)**\\
When generating candidate actions, avoid patterns that lead to inefficiency and increase runtime:\\
- **Cyclic Patterns**: Avoid candidates that repeat recent action sequences (e.g., PickupObject(Book\_1) $\rightarrow$ PutObject(Sofa\_1) $\rightarrow$ PickupObject(Book\_1) $\rightarrow$ PutObject(Sofa\_1)). These waste steps without making progress.\\
- **Failure Loops**: If an action failed 2+ times consecutively, include alternative approaches in candidates instead of repeating the same failed action. Repeating failed actions wastes time.\\
- **Oscillation**: Avoid candidates that create back-and-forth movement (e.g., navigating to Object\_1, then Object\_2, then back to Object\_1). This increases total path length.\\
- **Backtracking**: Avoid candidates that return to locations/objects recently visited. This wastes motion and increases runtime.\\
- **Stagnation**: Avoid consecutive idle candidates when there's work to be done. Agents should be making progress when possible.\\}

\texttt{**CONSTRAINT 7: Load Balancing**\\
Generate candidate actions that distribute work evenly across agents:\\
- Include candidates that allow idle agents to help with tasks\\
- Avoid candidates where one agent does all the work while others are idle\\
- Include candidates that balance workload (e.g., if one agent is carrying objects, include candidates for other agents to help with navigation or other tasks)\\
- This ensures efficient parallelization and reduces the total time to complete the task\\}

\texttt{**BEFORE OUTPUTTING CANDIDATE ACTIONS - CONSTRAINT CHECKLIST:**\\
1. Each agent has 3-5 diverse candidate actions
2. No agent has PickupObject candidates if they're already holding something\\
3. No agent has PutObject candidates if they're holding "nothing"\\
4. Candidate actions distribute work across different objects (minimize same-object conflicts)\\
5. No conflicting candidates (e.g., one agent opening while another closes same object)\\
6. Heavy objects have coordination candidates OR no manipulation candidates\\
7. Candidates avoid cyclic patterns, failure loops, and oscillation\\
8. Candidates allow for load balancing across agents\\}

\texttt{* NOTE: DO NOT OUTPUT ANYTHING EXTRA OTHER THAN WHAT HAS BEEN SPECIFIED\\
Let's work this out in a step by step way to be sure we have the right answer.}
\end{boxJ}

We also incorporate a dedicated prompt component that instructs the LLM to reason explicitly over agent-centric spatial memory and obstacle-relative information for navigation and coordination.

\begin{boxK}
    \texttt{* Memory: You will receive **AGENT-CENTRIC MEMORY** and **OBSTACLES** with relative locations and distances.\\}
    
    \texttt{**AGENT-CENTRIC MEMORY** shows object locations: "Book\_1 - Alice: Far left (4 moves), Bob: Just ahead (1 move)"
    \phantom{00}- **OBSTACLES** shows immediate surroundings with distances: "Ahead: Wall\_1 (2 moves), Chair\_1 (1 move); Left: Clear"\\}
      
    \texttt{**Use these for all navigation decisions!**\\
    The coordinate system is NOT meaningful - only use relative descriptions.\\}
      
    \texttt{Your task is to record task progress and key decisions:\\
    **DON'T repeat spatial info** (already provided). Instead, focus on:\\
    \phantom{00}-What agents are HOLDING: "Alice holding Book\_1, Bob holding nothing"\\
    \phantom{00}-TASK PROGRESS: "COMPLETED: Laptop\_1 on Sofa. REMAINING: Book\_1, RemoteControl\_1 need transport"\\
    \phantom{00}-RECENT ACTIONS \& OUTCOMES: Track what worked and what failed, adapt strategies\\
    \phantom{00}-NAVIGATION STRATEGY: "Alice navigating around Chair to reach Laptop" (use OBSTACLES distances to plan)\\
    \phantom{00}-MAINTAIN GOALS DURING DETOURS: Always state the ultimate goal when taking workarounds\\
    \phantom{00}-CROSS-AGENT COORDINATION: Agents share discoveries and help each other navigate\\
    **KEY PRINCIPLE**: Focus on TASK DECISIONS using provided spatial context and obstacle distances}
\end{boxK}


\end{document}